\documentclass[a4paper,fleqn,longmktitle]{cas-sc}

\usepackage[authoryear,longnamesfirst]{natbib}

\def\tsc#1{\csdef{#1}{\textsc{\lowercase{#1}}\xspace}}
\tsc{WGM}
\tsc{QE}

\newcommand{\modu}[1]{\left\vert#1 \right\vert}
\usepackage{caption}
\usepackage{subcaption}

\begin{document}
\let\WriteBookmarks\relax
\def\floatpagepagefraction{1}
\def\textpagefraction{.001}

\shorttitle{Quasi-coplanar co-orbital asteroids in the solar system}    

\shortauthors{Di Ruzza, Pousse, Alessi}  

\title [mode = title]{On the co-orbital asteroids in the solar system: medium-term timescale analysis of the quasi-coplanar objects}  



%

\author[1]{Sara {Di Ruzza}}[orcid=0000-0003-4858-3535]

\affiliation[1]{organization={Universit\`a degli Studi di Palermo,
Dipartimento di Matematica e Informatica},
addressline={Via Archirafi 34}, 
postcode={90123},
  city={Palermo}, 
  country={Italy}}


\ead{sara.diruzza@unipa.it}




\author[2]{Alexandre Pousse}[orcid=0000-0003-3996-5232]
\affiliation[2]{organization={Istituto di Matematica Applicata e Tecnologie Informatiche ``E. Magenes",
Consiglio Nazionale delle Ricerche
(IMATI-CNR)},
 addressline={Via Alfonso Corti 12},
postcode={20133},
city={Milano}, 
country={Italy}}


\ead{poussealexandre@gmail.com}




\author[2]{Elisa Maria Alessi}[orcid=0000-0001-6693-0014]
\cormark[1]


\ead{elisamaria.alessi@cnr.it}



\cortext[1]{Corresponding author}



\begin{abstract}
   The focus of this work is the current distribution of asteroids 
    in co-orbital motion with Venus, Earth and Jupiter, 
    under a quasi-coplanar configuration 
    and for a medium-term timescale of the order of 900 years. 
    A co-orbital trajectory is a heliocentric orbit trapped in a 1:1 mean-motion resonance 
    with a given planet. 
    As such, to model it this work considers the Restricted Three-Body Problem 
    in the circular-planar case with the help of averaging techniques.
	The domain of each co-orbital regime, 
		that is, 
			the quasi-satellite motion, 
			the horseshoe motion 
			and the tadpole motion,
	can be neatly defined by means of an integrable model 
	and a simple bi-dimensional map, 
	that is invariant with respect to the mass parameter of the planet, 
	and turns out to be a remarkable tool 
	to investigate the distribution of the co-orbitals objects of interest.	
	The study is based on the data corresponding 
	to the ephemerides computed by the JPL Horizons system for asteroids 
	with a sufficient low orbital inclination with respect to the Sun-planet orbital plane.	
	These objects are cataloged according to their current dynamics, 
	together with the transitions that occur 
	in the given time frame from a given type of co-orbital motion
	to another. 	
	The results provide a general catalog of co-orbital asteroids 
	in the solar system, the first one to our knowledge, 
	and an efficient mean to study transitions.	
\end{abstract}




\maketitle

\section{Introduction} 
\label{sec:intro}


	The co-orbital motion is a particular type of dynamics 
 	according to which two bodies 
		(e.g., an asteroid and a planet) 
	gravitate towards a more massive body 
	    (e.g., Sun) 
    with the same period. 
%
	Many objects susceptible to be at least temporary co-orbitals have been observed 
	in the solar system. 
	Earth, Venus, Mars and Jupiter are the planets 
	with the largest number of documented co-orbital objects, 
	while Saturn, Neptune and Uranus possess at least one object of this type
\citep[see, e.g.,][]{2006Ga}.
%
	A major example is given by the Trojans of Jupiter. 
	These asteroids are observed since the 1900s on tadpole-shaped (TP)  trajectories 
\citep{1906W}
	in the neighborhood of the two Lagrange's equilateral equilibrium configuration $L_4$ and $L_5$
	in the synodic reference system.
%
	Another illustrative case is given by the moons Janus and Epimetheus 
	that orbit at Saturn on quasi-coplanar and quasi-circular trajectories.
	As their orbital periods are slightly different, 
	the inner moon catches up with the outer one, 
	and their mutual gravitational influence leads to a swapping of the orbits 
	every 4 years, 
	as confirmed by Voyager 1 
\citep{1985Ak}. 
	In  the synodic reference system, they follow horseshoe-shaped (HS) trajectories.
	Several near-Earth asteroids have been observed with a similar behavior. 
	For instance, 2013 BS45 orbits the Sun in about one year 
	on a low eccentric quasi-planar trajectory and
	experiences regular close encounters with the Earth every 80 years
\citep{2013dede}.
%
	If, instead, the asteroid orbits around the Sun in such a way 
	that the Sun's and the planet's gravitational accelerations 
	act to create an asteroid retrograde motion with respect 
	to the planet in the rotating frame, 
	we speak of quasi-satellite (QS) orbit. 
	The first confirmed body in QS motion was 2002 VE68 harbored by Venus 
\citep{2004Mk}.

	On the other hand, co-orbital trajectories, 
	the possible transitions between them, 
	together with escaping/trapping mechanisms, 
	are considered
	as novel solutions for space retrieval and exploration missions.
	For instance, the JAXA-CNES MMX mission, 
	aimed at the exploration of the Martian moons, 
	plans to place a probe on a QS orbit at Phobos 
\citep{2021Nakamura}. 
%
	Moreover, ESA is planning to exploit 
	the neighborhood of $L_5$ of the Sun-Earth system to monitor the solar activity 
	and the related space weather issues 
\citep{ESAVigil}. 
%
	As an additional example, 
	the Chinese space agency is intended to visit the near-Earth asteroid Kamo'oalewa, 
	which is exceptional due to its long-term QS state 
\citep{2019Nat,2020Ch}.

	As it was implicitly assumed in the description above, 
	since the masses of the considered objects 
		(e.g., asteroids or spacecraft) are
	significantly smaller than the mass of the two other bodies 
	involved in the problem (i.e., Sun and planet),
	the Restricted Three-Body Problem (RTBP)
	is the classical model 
	for describing the co-orbital motion.
	In particular, the corresponding area of the phase space
	can be tackled through the perturbation theory,
	with the help of averaging techniques.
%
	The averaged problem of the 1:1 mean-motion resonance 
	has the advantage to be integrable 
	in the circular-planar case, thus
	providing a complete understanding of the co-orbital resonance. In this context, the domain of each dynamical regime in the phase space  can be neatly defined 
\citep{2002NeThFe,2022PoAl}.

	The objective of this work is to show 
	how the integrable approximation obtained 
	with the help of the classical tools of dynamical systems theory
	and perturbation techniques 
	can be applied to describe the dynamics of real natural objects 
	in the solar system, 
	and to see what we can learn from their behavior 
	to improve the analytical description.

	To the best of our knowledge, 
	there exist only official catalogues of Trojan asteroids 
	of Earth, Mars, Jupiter, Uranus and Neptune 
\citep{MPC},
	while for other co-orbital motions there exist only works that focus specifically on given objects
\citep[see, e.g., ][]{2004Mk, 2004BrInCo, 2007KN, 2010Wa, 2012WK, 2012DD, 2016DD}.

	Contrarily to the classical studies 
    that investigate the stability on secular timescales,
	from several millennia to million years
\citep[see, e.g.,][for studies on HS and TP, respectively]{2012CuHaHo,2006RoGa}
	the co-orbital motion is investigated here in a medium timescale, 
	of the order of several centuries.
%
	More precisely, the analysis is based on the ephemerides 
	computed by the JPL Horizons system \citep{NASAHor}, covering a time span of the order of 900 years.
	This constraint in time 
	together with the features of the co-orbital motion 
	imposes to restrict the study to the co-orbitals of Venus, Earth and Jupiter.
	
	In addition, in order to remain close to the domain of validity 
	of the integrable approximation,
	only the asteroids
	whose orbit can be considered, within a certain threshold, on the orbital planet of the given planet
	 will be analyzed.
	They will be referred as asteroids in
	``quasi-planar'' co-orbital motion with the given planet.
	
	By using the data obtained on the basis of real observations,
	the theory will be linked to the experimental data,
	showing whether
    the integrable approximation
	is a relevant model in order to describe the co-orbital asteroids 
	in the considered time frame. 
	Moreover, some conclusions on the future directions to be taken to improve the averaged model, especially in the three-dimensional case, will be drawn.
	In particular, one of the goals is to highlight 
		the  transitions from a given type of dynamics to another,
		showing the ones that can be ascribable to the integrable approximation
		within its limit of validity, 
		and those that are not.

	This work is organized as follows. 
	Without too much details,  the perturbation scheme that leads to the averaged problem is explained
    in Sec.~\ref{sec:av}, together with the tool developed in the circular-planar case. 
	In Sec.~\ref{sec:JPL}, the data considered are described. 
	In Sec.~\ref{sec:results}, the results obtained 
	for each Sun-planet system taken into account are given. 
	In Sec.~\ref{sec:concl}, conclusions and future directions are highlighted.


\section{The averaged problem in the circular-planar case}
\label{sec:av}


	For the sake of clarity, 
	we will summarize here the reasoning that leads 
	to the integrable approximation of the co-orbital resonance
	and refer the reader to 
\citet{2022PoAl}
	for the full derivation and analysis.

	The model is based on the RTBP,
	that is, 
	the study of the motion of a massless body 
		(a particle, an asteroid or a spacecraft), 
	affected by the gravitational attraction of the Sun and a planet.
%
	In that framework,
	the motion of the two massive bodies follows a solution of the two-body problem,
	and, in the circular case, 
	the planet and the Sun move on a circular orbit around their barycenter.
%
	The following Hamiltonian function provides 
	the dynamics of the problem in the heliocentric reference frame:
	\begin{equation}
	\begin{aligned}
	\mathcal{H}\left(\mathbf{r}, \dot{\mathbf{r}}, \lambda_p\right)
	&=
		\frac{\|\dot{\mathbf{r}}\|^{2}}{2}
	-	\frac{\mu}{\|\mathbf{r}\|}
	-	\frac{(\mu + \mu_p)\varepsilon}{\left\|\mathbf{r}-\mathbf{r}_p\left(\lambda_p\right)\right\|}\\
	&+	(\mu + \mu_p) \varepsilon 
	\mathbf{r} \cdot\mathbf{r}_p\left(\lambda_p\right)
\label{eq:Ham}
	\end{aligned}
	\end{equation}
	where
		$\mu$ and $\mu_p$ are the respective mass parameter of the Sun and the planet,
		\begin{equation*}
			\varepsilon := \frac{\mu_p}{\mu + \mu_p},
		\end{equation*}
	    is a dimensionless parameter characterizing 
	    the mass ratio of the Sun-planet system,
	    the vectors $\mathbf{r}, \dot{\mathbf{r}}\in\mathbb{R}^2$ 
	    are canonical variables
    	corresponding to the position and the velocity of the particle 
	    and
    	$||\cdot||$ is the Euclidean norm associated with the scalar product 
	    denoted $\,\cdot\,$. 
	    In the planar case, 
	    the motion of the particle takes place on the orbital plane of the planet.
	The heliocentric vector $\mathbf{r}_p(\lambda_p)$ 
	denotes the position of the planet 
	for a given value of the mean longitude $\lambda_p = M_p + \varpi_p$,
	whose evolution is proportional to time
	with an angular velocity -- generally known as mean-motion --
	that reads
	\begin{equation*}
		n_p := \sqrt{\frac{\mu + \mu_p}{a_p^3}}
	\end{equation*}
	being $a_p$, $M_p$ and $\varpi_p$ 
	    its semi-major axis,
	    its mean anomaly
	    and its longitude of the periaster, respectively.
	
	The problem is usually analyzed in the synodic reference system,
		that rotates together with the planet. 
%
	In this system, the planet is located 
	    at $a_p(1-\varepsilon, 0)$, 
		while the Sun at $a_p(-\varepsilon, 0)$. 
%
	In the synodic reference system, it is well known that there exist 5 equilibrium points, 
		called the {\it Lagrangian points}, 
		denoted as $L_j$, $(j=1, \dots 5)$ 
	and that it can be defined one first integral of the problem, 
		called the {\it Jacobi constant}. 
%
	The dynamics of interest for this work is connected 
		to some relevant families of periodic orbits that exist in the synodic reference frame 
\citep{1967Sz}.
%
	For instance, 
		the HS motion is associated with 
		the hyperbolic invariant manifolds of the Lyapunov family $\mathscr{L}_3$ 
		that stems from $L_3$ 
\citep{2006BaOl},
		while the TP motion arises from the neighborhood of 
		$\mathscr{L}_j^s$ and $\mathscr{L}_j^l$, that are
			the two Lyapunov families of $L_j$ for $j = 4,5$,
			generally known as the short and long periodic families, due to their associated timescale
		in the neighborhood of the equilibrium point.
%
	The QS regime is remarkable for the fact 
	    that it is not connected to the Lagrange points, 
		but stems from 
		a family of simple-periodic symmetrical retrograde satellite orbits
		generally known as the family $f$ 
\citep[see, e.g., ][for more details]{2017PoRoVi}.
	According to the Poincar\'e classification 
		these families of periodic orbits are the continuation, 
		from the limit case $\varepsilon = 0$, 
		of heliocentric Kepler orbits in 1:1 mean-motion resonance with the planet.
%
	Consequently, a natural choice of perturbative treatment 
		in order to recover these structures
		is to consider $\varepsilon$ as a small parameter.

    For $\varepsilon$ small enough, 
    the Hamiltonian in Eq.\,\eqref{eq:Ham}
    can be split as
	\begin{equation*}
    \mathcal{H}\left(\mathbf{r}, \dot{\mathbf{r}}, \lambda_p\right)
	=	                            \mathcal{H}_{\mathrm{K}}
	                                \left(\mathbf{r}, \dot{\mathbf{r}} \right)
	+(\mu + \mu_p)\varepsilon    \mathcal{H}_{\mathrm{P}}
		                            \left(\mathbf{r}, \lambda_p\right),
	\end{equation*}
	where $\mathcal{H}_{\mathrm{K}}$ corresponds to the Kepler motion 
		around the Sun, 
	while $\mathcal{H}_{\mathrm{P}}$ 
	models the perturbations 
	which derive from the gravitational effect of the planet.
	In the unperturbed problem, 
	that is for $\varepsilon = 0$,
    the particle lies on an ellipse that can be defined 
    by the orbital elements $(a, e, \varpi)$ (semi-major axis, eccentricity, longitude of the periaster, respectively).
    The position on its ellipse can be obtained through several angles:
        the true anomaly $v$, 
        the mean anomaly $M$ 
        or the mean longitude $\lambda = M + \varpi$, 
    a fictitious angle proportional to time such that 
    \begin{equation*}
        \dot{\lambda} = \sqrt{\frac{\mu + \mu_p}{a^3}}.
    \end{equation*}
%
    Therefore, the commensurability associated with 
    the 1:1 mean-motion resonance 
    occurs for a  semi-major axis equals to $a_p$.

    In order to focus on the co-orbital resonance,
    one can define the resonant angle
    \begin{equation*}
        \theta := \lambda - \lambda_p,
    \end{equation*}
    as well as its conjugated dimensionless action
    \begin{equation*}
        u := \sqrt{\frac{a}{a_p}} - 1,        
    \end{equation*}
    whose modulus measures the distance to the commensurabilty,
    such that $\dot{\theta}$ is equal to zero at the resonance 
    in the unperturbed problem, that is, for $u=0$.
%
    Hence, studying the co-orbital resonance 
	in that perturbative frame consists in 
	understanding how the perturbation $\mathcal{H}_{\mathrm{P}}$ 
	transforms the unperturbed phase space 
	in the neighborhood of $u=0$.
    All the variables $(\theta, u, e, \varpi)$ might vary 
    and this makes the motion very tricky to understand;
    nevertheless, 
	for $\varepsilon$ and $|u|$ small enough, 
	the angular variables 
	    $\theta$ and $\varpi$ 
	    evolve at different rates with respect to the 
        ``fast" angle $\lambda_p$.
%
    More precisely,
	    $\theta$ undergoes ``semi-fast" variations
	    corresponding to the resonant dynamics
	    while $\varpi$ experiences a ``secular" drift. 
	A classical way to exploit this feature
	is to replace the original Hamiltonian by one 
	in which the fast oscillations are removed. 
%
    This process defines the averaged problem in the co-orbital resonance.
%
    In a more practical way, 
    the averaged Hamiltonian
    is obtained by replacing the mean longitude $\lambda$ by 
    $\theta + \lambda_p$,
    and implementing an averaging with respect to the fast angle $\lambda_p$.   
%
    For further details
    the reader is referred to the work of 
\citet{2002NeThFe},
        that develops a semi-analytical method
        in order to compute the averaged Hamiltonian 
        as well as its derived equations of motion,
    and thus provides an integrator for co-orbital trajectories.

    Finally, an important feature occurs in the circular-planar case:
    the invariance of $\mathcal{H}$ 
	under the group of rotations SO(2) in the plane
	implies that the averaged Hamiltonian does not depend on $\varpi$
	while the quantity
	\begin{equation*}
	    \Gamma := \sqrt{a}(1 - \sqrt{1 - e^2})
	\end{equation*}
    is a first integral.
%
    Hence
	for various $\Gamma$, seen as a parameter, 
	the averaged Hamiltonian
	$\overline{H}^{\Gamma}(\theta, u)$ is integrable
	and allows to understand the global dynamics of 
	the 1:1 mean-motion resonance 
	through a simple description of phase portraits 
	in the resonant variables $(\theta, u)$.


\subsection{Phase portraits description and $(\theta,e)$-map of the co-orbital resonance}



\begin{figure}
	\begin{center}
	\includegraphics[scale = 0.63]{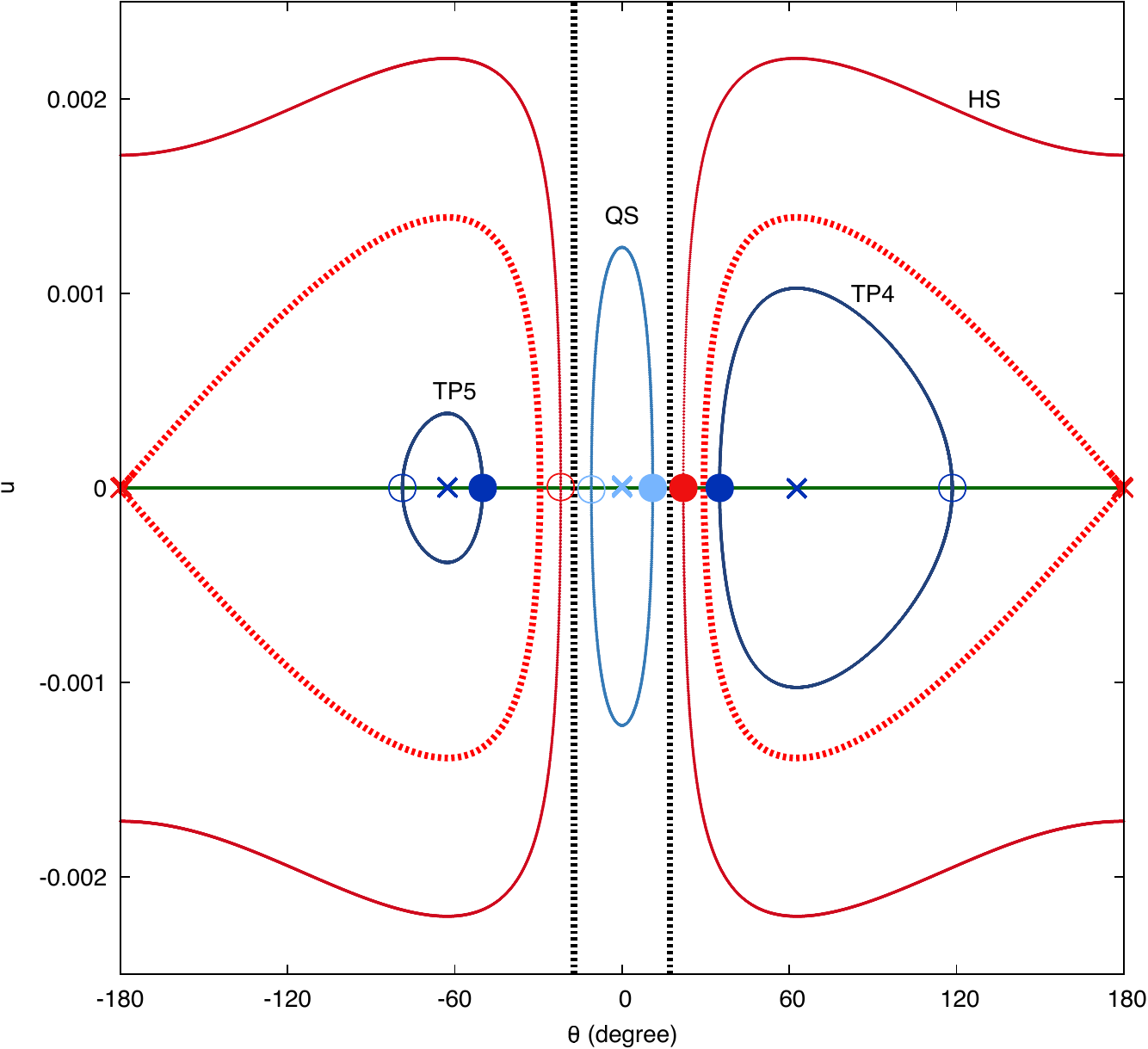}
	\caption{Phase portrait of the averaged Hamiltonian 
	for a Sun-Earth system 
	and  $\Gamma  = \sqrt{a_p}(1 - \sqrt{1 - e^2})$ with $e= 0.15$.
	The black dashed lines represent 
	    the collision with the planet.
	The blue, sky blue and red trajectories are level curves 
	    associated with TP orbits of $L_4$ 
	    and $L_5$, QS and HS motion, respectively.
	The blue and sky blue crosses are elliptic equilibrium points.
	More precisely, they correspond, respectively, 
	    to periodic orbits of the family 
	    $\mathscr{L}_j^s$ for $j = 4,5$ and $f$.
	The red crosses are a hyperbolic equilibrium point 
	    associated with a periodic orbit of $\mathscr{L}_3$
	    and from which emerges a separatrix (the red dashed curve) 
	    that bounds the two domains of TP motion and the one of the HS motion.
    The green line corresponds to the section $u=0$.
    Each periodic solution of the phase portrait
        crosses the section $u=0$ in two points,
    represented by colored disks and circles on the phase portrait.
	}
\label{fig:PhaseP}
	\end{center}
	\end{figure}

	For a given $0 \leq \Gamma <\sqrt{a_p} $, 
	the averaged Hamiltonian is integrable 
		with one degree of freedom 
		and a phase portrait can be computed.
%
    The co-orbital trajectories are solutions
    located in the neighborhood of $u=0$
	and such that $\theta$ oscillates around a given value.
%
    Only three types of co-orbital dynamics 
	exist in the circular-planar case:
    the quasi-satellite regime where $\theta$ librates around zero 
		for $\Gamma>0$,
    the tadpole motion of $L_j$ with $j=4,5$ 
		where $\theta$ experiences a periodic oscillation 
		around a given $\theta_j(\Gamma)$ satisfying 
		${23.9^\circ< (-1)^j\theta_j(\Gamma)< 180^\circ}$,
	and the horseshoe trajectories 
		where $\theta$ oscillates around $180^\circ$ 
		with a large amplitude that decreases as long as $\Gamma$ increases.

	For instance, Fig.\,\ref{fig:PhaseP} displays a phase portrait 
	in the case of the Sun-Earth system 
	with $\Gamma  = \sqrt{a_p}(1 - \sqrt{1 - e^2})$ and $e= 0.15$.
%
	The phase portrait has four equilibrium points
	represented on the figure by crosses.
%
	The elliptic equilibrium points located in the neighborhood of 
	$(\theta, u) = ((-1)^j 60^\circ, 0)$ for $j=4,5$
	stand for periodic orbits of $\mathscr{L}_4^s$ and $\mathscr{L}_5^s$,
    while
	the elliptic equilibrium point located close to the origin 
	is associated with a periodic orbit of the family $f$.
%
	Therefore, 
		the trajectories
		that librate around them, are respectively,
		the TP of $L_4$, the TP of $L_5$,
		and the QS. 
%
	The hyperbolic equilibrium point located 
	at $\theta = 180^\circ$ and $u\simeq 0$
		corresponds to a periodic orbit
		that belongs to $\mathscr{L}_3$.
%
	A separatrix, represented by the red dashed curve,
    emerges from it
    and 
       approximates the corresponding
       stable and unstable invariant manifolds.
    Besides, the separatrix
        bounds the two TP regions,
	    giving rise to the HS trajectories
	    that encompass it.
%
	Finally, the two black dashed lines	    
	    embody  the collision with the Earth
	    where the averaged Hamiltonian is not defined,
	    and delimit the domains of QS and HS motion.

	\begin{figure*}
	\begin{center}
	\begin{minipage}{0.1cm}
    (a)
    \end{minipage}	
    \begin{minipage}{12cm}
        \centering
        \includegraphics[scale = 0.425]{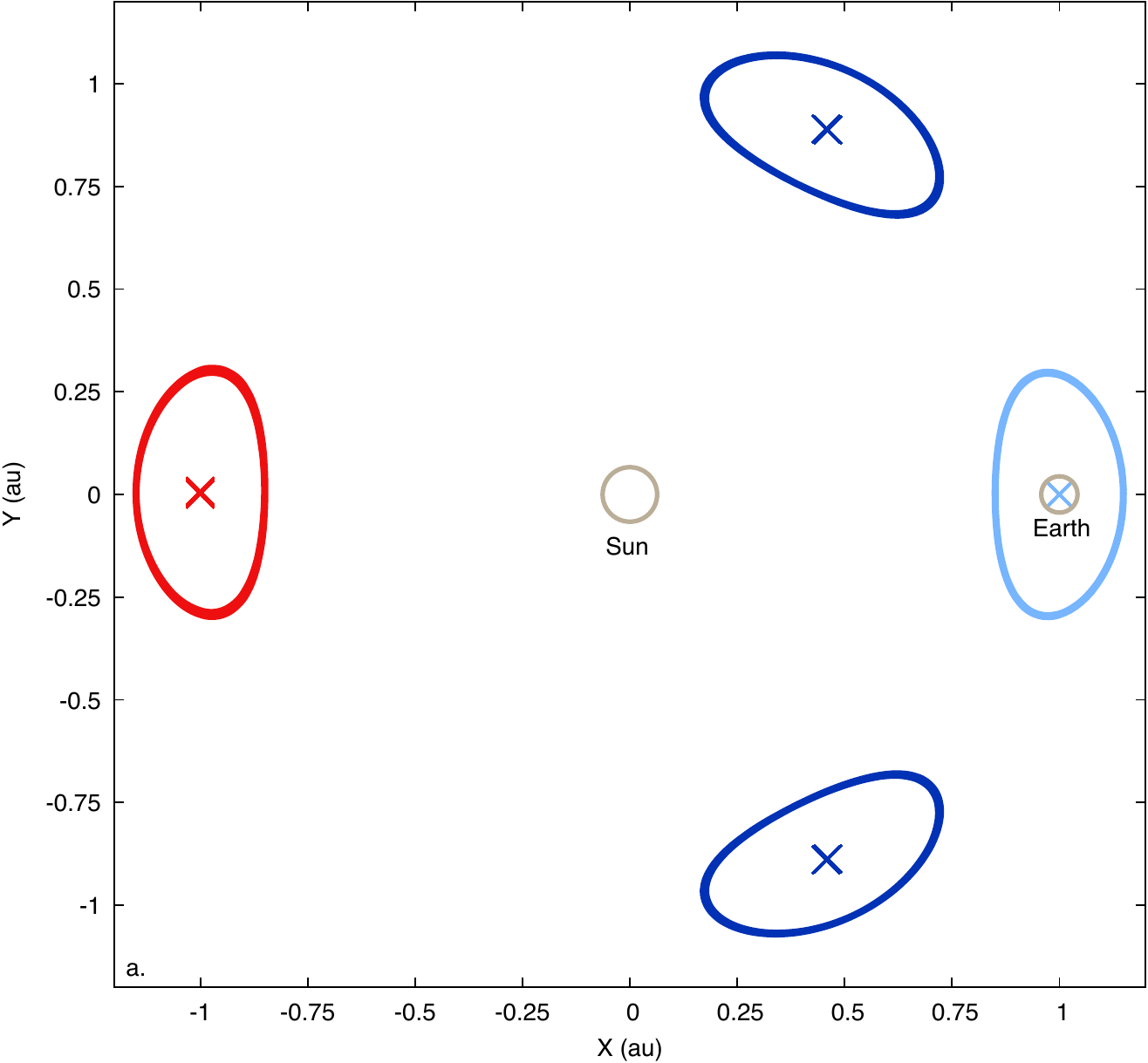}
	    \includegraphics[scale = 0.425]{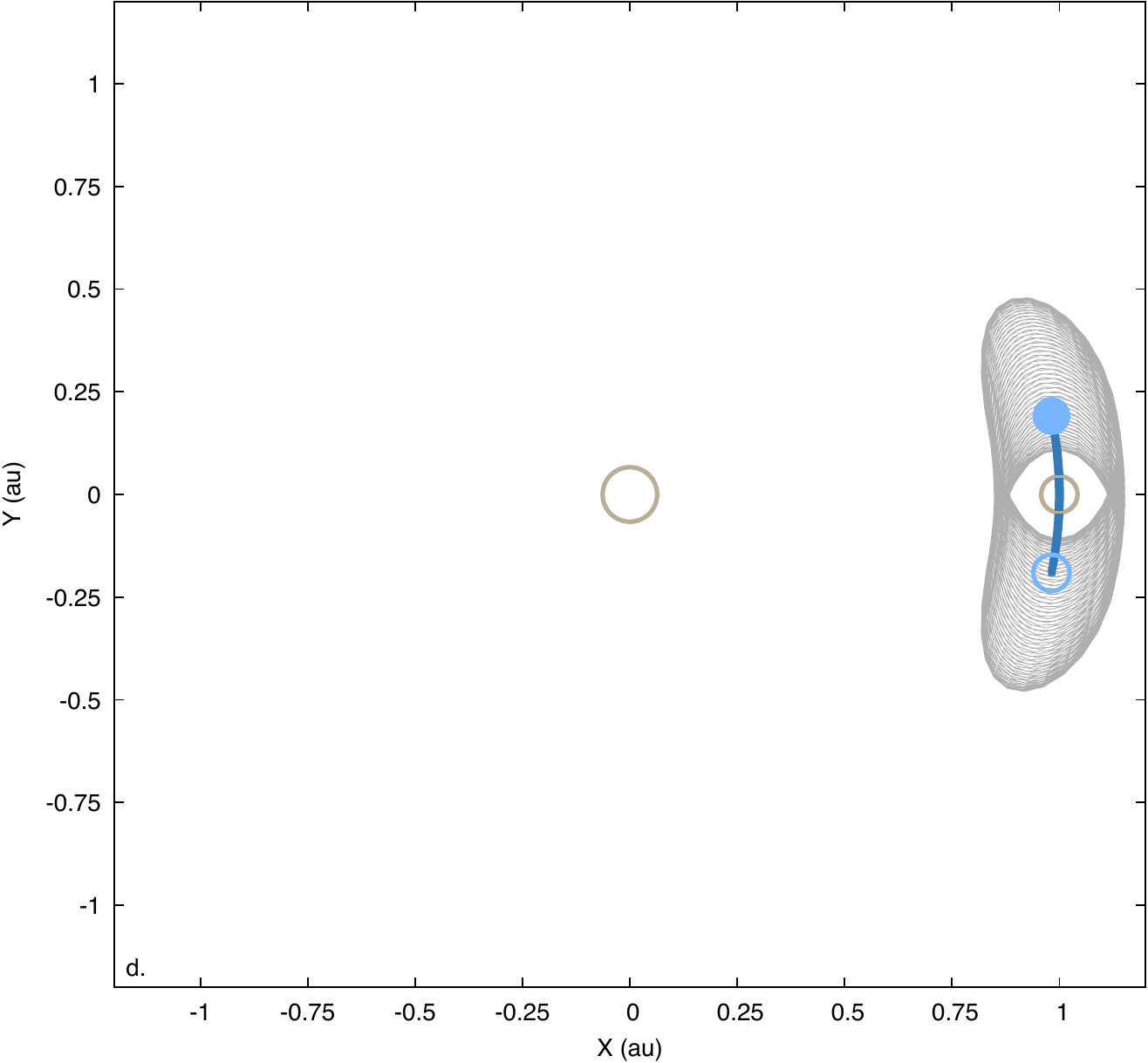}
    \end{minipage}
	\begin{minipage}{0.1cm}
    (b)
    \end{minipage}\\
	\begin{minipage}{0.1cm}
    (c)
    \end{minipage}	
    \begin{minipage}{12cm}
        \centering
        \includegraphics[scale = 0.425]{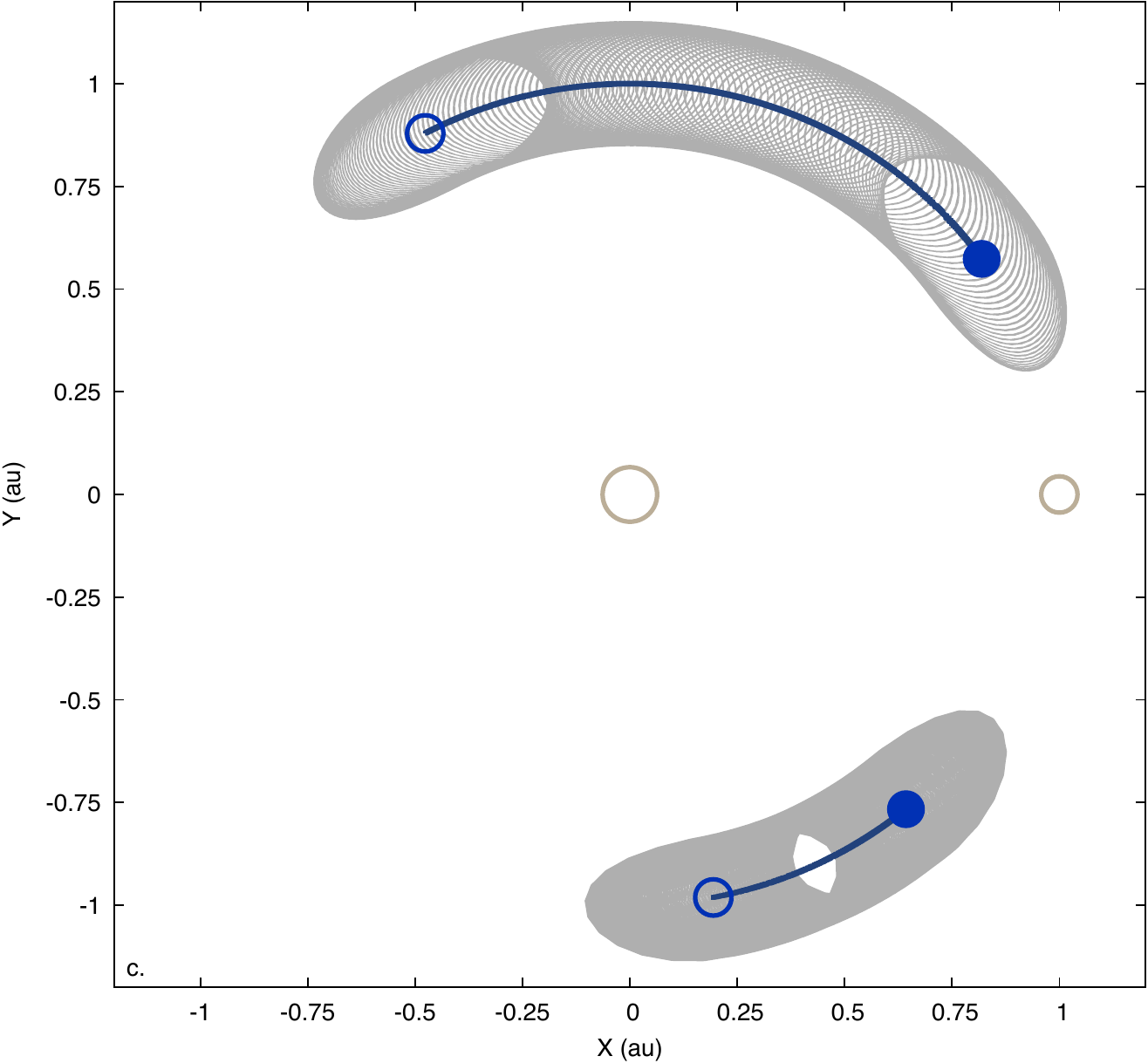}
	    \includegraphics[scale = 0.425]{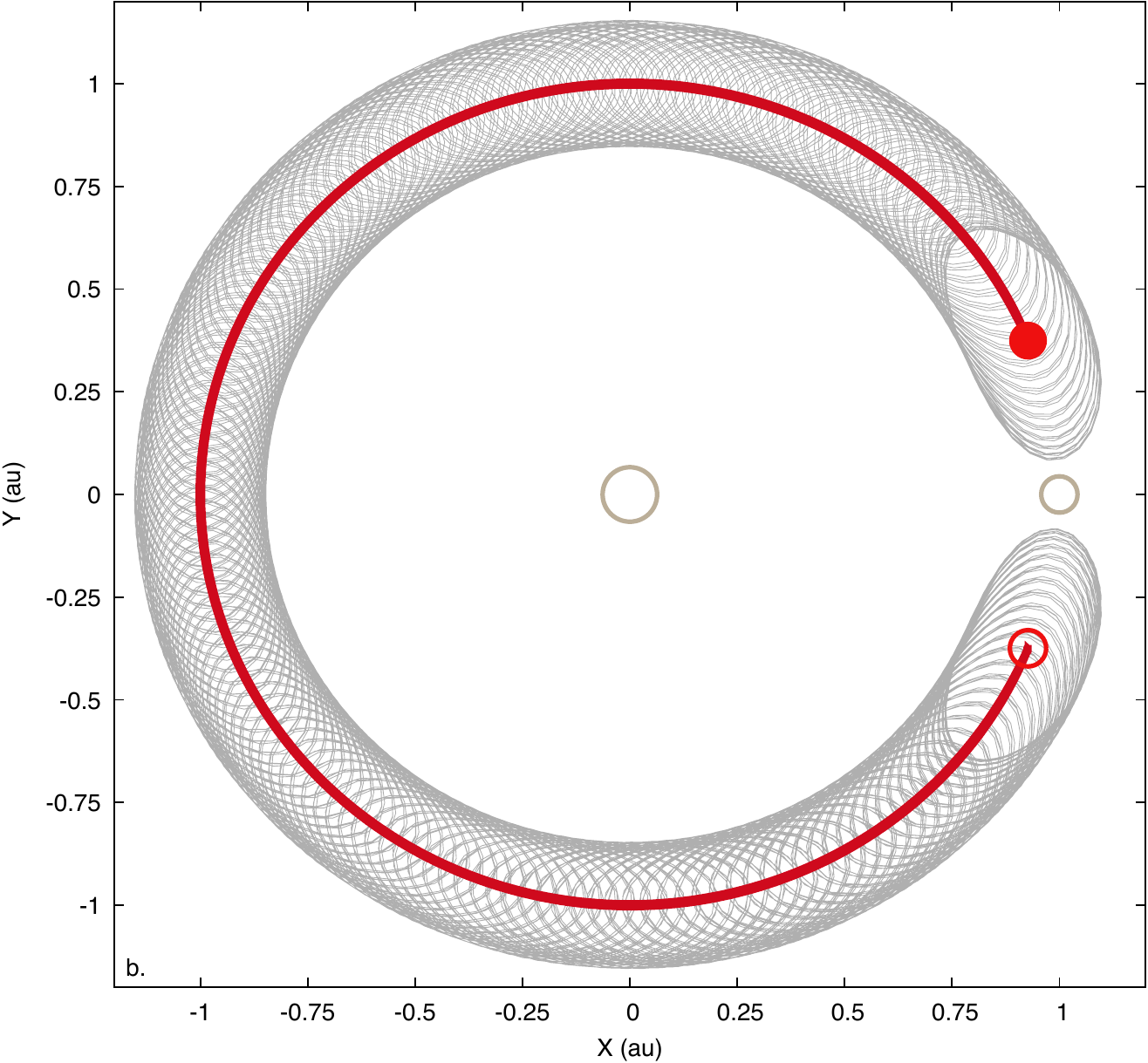}
    \end{minipage}
	\begin{minipage}{0.1cm}
    (d)
    \end{minipage}

    \end{center}
	\caption{Trajectories in the synodic reference frame propagated
	        with the integrator of the averaged problem,
	        for 500 years.
	   The Sun is at the origin and the Earth is located at $(1,0)$.
	   Panel (a): Periodic orbits and respective guiding center 
	        (red, blue, sky blue crosses) 
	   associated with the four equilibrium points 
	   of the phase portrait of Fig.\,\ref{fig:PhaseP}.
	   Panels (b) (c) and (d): trajectories (grey dots) corresponding respectively 
	   to the QS, TP and HS orbits 
	   of Fig.\,\ref{fig:PhaseP}.
	   The thick colored curves embody the evolution of their respective guiding center while the colored disks and circles represent the crossings with the section $u=0$.}
\label{fig:Traj}
	\end{figure*}

	In the case of quasi-circular orbits, 
	that is for $\Gamma = 0$, 
	the resonant angle $\theta$ coincides with the angular separation 
	between the particle and the planet, denoted  $\phi$,
	while the heliocentric norm, denoted $R$, is equal to $a = a_p(1 + u)^2$.
    Hence,
        in the synodic reference frame 
	    for which the Sun is at the origin
	    and the planet is located in $(a_p,0)$,
		the motion of the particle
		is equivalent to the following trajectory:
	\begin{equation}	
		(X(t),Y(t)) = a_p(1 + u(t))^2(\cos \theta(t), \sin \theta(t) ).
\label{eq:XY}
	\end{equation}
%
	For $\Gamma>0$, Eq.\,\eqref{eq:XY} 
	does not provide the motion of the particle
	in the synodic reference frame, 
	but the dynamics of its guiding center,
	a concept that is used in literature 
	to describe the resonant behavior of a co-orbital trajectory.

	To take into account 
	the degree of freedom $(e, \varpi)$ is necessary
	to model the trajectory in the synodic reference frame. 
%
	Due to the conservation of $\Gamma$,
		$e(t)$ undergoes semi-fast oscillations 
		around a value $e_0 = \sqrt{1 - (1 - \Gamma)^2}$
		which corresponds to the eccentricity for $u=0$,
	while $\varpi(t)$ experiences 
	a secular drift
	combined with the oscillations generated by the resonant behavior.
%
	Besides, the transformation from 
	the orbital elements 
	to the synodic reference frame 
	introduces a fast oscillation generated by the angle $\lambda_p(t)$.

  	Hence, in first approximation,
  	    the motion of the particle can be written as follows:
    \begin{equation*}
        (X(t), (Y(t))) =   R(t)(\cos \phi(t), \sin \phi(t) )
    \end{equation*}
    with
	\begin{equation*}	
	    R(t) = a_p(1 - e_0\cos M(t)), \quad 
		\phi(t) = \theta(t) + e_0 F(M(t))
	\end{equation*}
	where 
		$M(t) = M(0) + (n_p - g)t$ 
		    is the mean anomaly of the particle, 
	    $g$ is the frequency of the secular drift in $\varpi$,
		and $F$ is a $2\pi$-periodic function 
		derived from the difference between the true anomaly $v$ and $M$.	

	As a consequence,
	    a periodic solution $(\theta(t), u(t))$ of a given phase portrait
	    provides a quasi-periodic trajectory in the synodic reference frame
	    that results from the combination 
	    of a fast oscillation of frequency $n_p-g$,
	    whose amplitude is related to the eccentricity,
	    and a slower oscillation 
	    generated by the resonant behavior
	    embodied by the guiding center.
 %
    For instance, Fig.\,\ref{fig:Traj}b-d display, respectively, 
	    the TP, HS and QS orbits of Fig.\,\ref{fig:PhaseP} 
	    (grey dots)
	        as well as the evolution of their guiding center 
	    (colored curves).
%
    Finally, as depicted in Fig.\,\ref{fig:Traj}a,
    for an equilibrium point of Fig.\,\ref{fig:PhaseP},
        the corresponding trajectory in the synodic reference frame
            (colored curves)
	    is the one of a periodic orbit
	    while its guiding center is a fixed point
	        (colored crosses).


\begin{figure}
\begin{center}
    	\includegraphics[scale = 0.63]{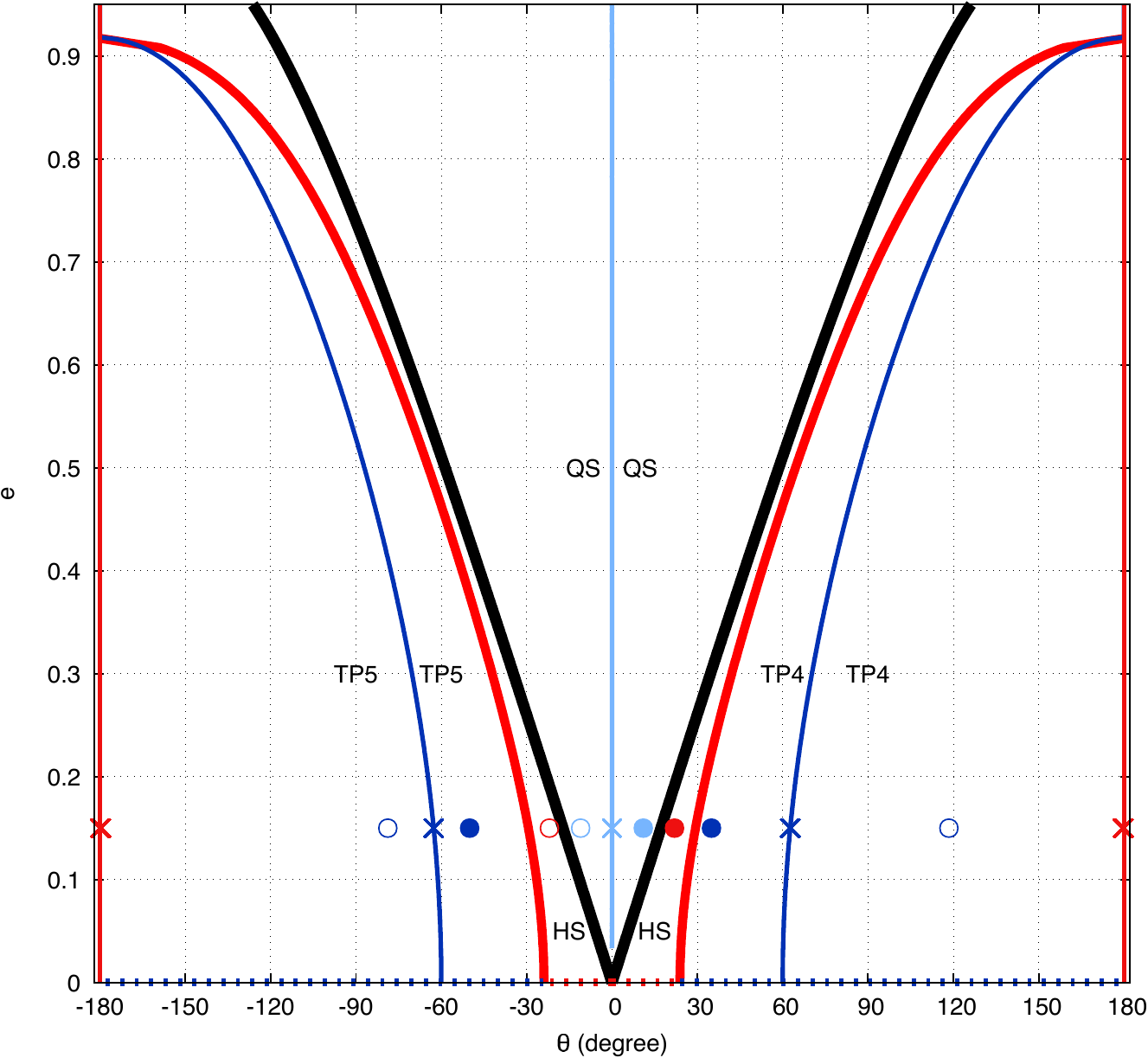}
    	\caption{The $(\theta,e)$-map of the co-orbital motion 
    	defined by the section $u = 0$. 
    	The black and red thick curves stand, respectively, 
    	for the singularity of collision 
    	and the crossing of the separatrices that originate from $L_3$ (thick red curve). 
    	They divide the map in three regions. 
    	The QS and TP domains are centered, respectively, 
    	on the family $f$ (sky blue curve) 
    	and the short periodic families $\mathscr{L}_j^s$ (blue curves). 
    	The HS region is split in two parts, 
    	between the separatrix and the curve associated with the collision.
    	The crosses, colored disks and circles denote 
    	the crossings of the solutions $(\theta(t), u(t))$ plotted in 
    	the phase portrait of Fig\,\ref{fig:PhaseP}.}
\label{fig:Map}
\end{center}
\end{figure}

	The topology of the phase portraits evolves by increasing $\Gamma$.
%
    Inside the collision curve, the size of the QS domain increases 
    until it dominates the phase portrait 
    for high values of $\Gamma$. 
%
	Outside the collision curve,  
	the two elliptic equilibrium points get closer to the hyperbolic fixed point, 
		therefore the TP domains shrink and vanish when the three merge.
%
    More generally, for all $0\leq\Gamma<\sqrt{a_p}$,
		each domain of dynamics extends quasi-symmetrically 
		with respect to the axis $u = 0$
		and is neatly defined by the collision curve or the separatrix.
%
	Therefore, their crossings with the section $u=0$, that is $a = a_p$, 
		provides a synthetic way to understand the global evolution of 
		the different domains of dynamics by varying $\Gamma$, 
		or equivalently, by varying the eccentricity of the orbit.	

	Figure \ref{fig:Map} displays the $(\theta, e)$-map 
	    of the co-orbital resonance in the case of a Sun-Earth system.
%
	The black and red thick curves
	$\modu{\theta} = \Theta_{col}(e)$
	and 
	$\modu{\theta} = \Theta_{sep}(e)$,
		depict, respectively, 
		the crossings of the collision curve 
		and of the separatrix 
		with the section $u=0$, for all $0 \leq e<1$.
	Note that $\Theta_{col}(e)$ can be approximated by $2e$
	    up to high values of eccentricities
	    while, according to the classical result
\citep{1977Ga}, 
    $\Theta_{sep}(0) \simeq 23.9^\circ$.
%
    In this framework,
        each type of dynamics belong to one of the following domains:
   \begin{equation}
    \begin{aligned}
    \mathcal{D}_{QS} 	&= \{\modu{\theta}<\Theta_{col}(e)\},\\
	\mathcal{D}_{HS}	&= \{ \Theta_{col}(e) < \modu{\theta}< \Theta_{sep}(e)\},\\      
	\mathcal{D}_{TP_j} 	&= \{ \Theta_{sep}(e) < (-1)^j\theta< 180^\circ\}.
	\end{aligned}
\label{eq:dom}	
	\end{equation}
%
    Hence, for a given eccentricity $e$ 
    when the particle is at the resonance,
    a periodic solution $(\theta(t), u(t))$
	    intersects the section in exactly two values of $\theta$, 
	    that belong to one of the three domains of Eq.\,\eqref{eq:dom},
	    on both sides of a curve
        that represents a family of equilibrium points,
        that is, $\mathscr{L}_4^s$ or $\mathscr{L}_5^s$ for the TP,
        $f$ for the QS 
        and $\mathscr{L}_3$ for the HS.
%
    These two crossings $(\theta_j,e)_{j=1,2}$ 
    are identified by colored disks and circles on
    Fig.\,\ref{fig:PhaseP},
    and Fig.\,\ref{fig:Map},
	as well as on Fig.\,\ref{fig:Traj}b-d.

    As a consequence, 
        the averaged problem in the circular-planar case, through the $(\theta, e)$-map,
        provides a simple method to identify
        the type of motion of co-orbital trajectories.
%
 	Besides the tool given by the $(\theta,e)$-map of Fig.\,\ref{fig:Map}
 	    does not apply only to the Sun-Earth system,
 	    as it is invariant with respect to $\varepsilon$.
%
    This remarkable feature derives to the fact that the following rescaling 
        $\sqrt{\varepsilon}^{-1}\overline{H}^{\Gamma}(\theta, \sqrt{\varepsilon}u)$
        makes the averaged Hamiltonian independent of $\varepsilon$.
%
    Hence, as long as $\varepsilon$ can be considered as a small parameter,
        that is the case for all the Sun-planet systems of the solar system,
        the $(\theta,e)$-map of Fig.\,\ref{fig:Map} is reliable 
        to identify the type of dynamics
        of the co-orbital motion.
        

\subsection{On the validity limit of the model}


    The averaged problem in circular-planar case
        has some limits of validity
        which might constrain its application to real objects in the solar system.
%
    The first one derives from the perturbative treatment.
%
    Indeed,    
        according to the perturbation theory, 
        the averaging process coincides 
        with the existence of a symplectic transformation
        which maps the original Hamiltonian to the averaged Hamiltonian 
        plus a remainder that is supposed to be small 
        and thus dropped in the averaged problem.
%
    As long as the planet and the particle are getting closer,
        the remainder can increase 
        and can no longer be considered negligible
        with respect to the averaged perturbation in $\overline{H}^{\Gamma}$.
%
    Through a rigorous treatment, 
\cite{2022PoAl} 
    proved that the model 
    and the $(\theta,e)$-map 
    are reliable as long as 
    the considered trajectories lay outside the Hill's sphere of the planet. 
%
    Outside the Hill's sphere, 
        the dropped remainder generates a perturbation
        whose size increases as long as the distance to the planet decreases,
        which may destabilize the quasi-periodic character
        of the co-orbital trajectories described in the averaged problem.
%
    In the framework of the $(\theta,u)$-map, 
        it is expected 
        that the crossings of a co-orbital trajectory slowly drift
        such that transitions from a given domain to another are possible,
        especially for crossings located close to the collision curve
        or in the neighborhood of the separatrices.
%
    Nevertheless, since these instabilities occur in the framework of the
    RTBP in circular-case,
        they conserve the Jacobi constant.

    Another limit of validity of the model  
        is given by the RTBP in circular-planar case itself.
%
    Real objects are not co-planar with the planet,
    therefore it is necessary to define an arbitrary threshold
    in order to distinguish the quasi-coplanar trajectories.
%
    According to \cite{1999Na} and \cite{2002NeThFe},
        the effect of the inclination may generate 
        transitions, escapes
        and also new types of co-orbital dynamics, 
        as for instance
        the compound orbits 
        (see Sect.\,\ref{sec:Ven} for a discussion on this dynamics).

    Finally,
        other effects may have a role on the dynamics of real objects,
        such as the eccentricity of the planet,
        possible close encounters with other massive objects
        or non-gravitational forces as the solar radiation pressure.
%
    In these situations, the Jacobi constant is no longer conserved.
%
    For possible close encounters with other planets,
    an indication can be obtained 
    by assuming that all the planets of the solar system move in the same plane on circular orbits
    and estimating the eccentricities corresponding to the crossing of the orbit
    of the inner planets 
        (at perihelion), 
    and of the outer planets
        (at aphelion),
    with respect to the considered Sun-planet system.
%
    In principle,
        an asteroid that crosses the orbit of another planet
        will do it twice a year and the corresponding
     gravitational effect cannot be neglected over long timescales.

\begin{table}
\caption{For each Sun-planet system, 
                $\varepsilon$ is the mass parameter, 
                $T_{rev}$ is the period of revolution of the planet, 
                $T_{lib}$ is the period of libration of $\theta$ 
                in the neighborhood of $L_4$ or $L_5$, 
                and $N_{lib}$ the number of librations possibly performed
                in 900 years.           
                All the periods are reported in years. 
                The physical values are taken from 
\citet{NASApar} }
\label{librations}
\begin{tabular*}{\tblwidth}{@{}LL@{}LL@{}LL@{}LL@{}LL@{}}
\toprule
 Planet      & $\varepsilon$  & $T_{rev}$  & $T_{lib}$ & $N_{lib}$ \\ 
\midrule
Mercury     & 1.66012055 $\times$ 10$^{-7}$     &   0.24            & 227.51            & 3.95 \\
Venus       &   2.44783229 $\times$ 10$^{-6}$   &   0.61            & 151.34            & 5.95 \\
Earth       &   3.00348059 $\times$ 10$^{-6}$   &   1.00            & 222.09            & 4.05 \\
Mars        &   3.22715504 $\times$ 10$^{-7}$   &   1.88            & 1274.36           & 0.71 \\
Jupiter     & 9.53881152 $\times$ 10$^{-4}$     &   11.87           & 147.88            &  6.09 \\
Saturn      &  2.85803962 $\times$ 10$^{-4}$    &   29.47           &  671.02           & 1.34 \\
Uranus      &  4.36605898 $\times$ 10$^{-5}$    &   84.05           & 4896.06           & 0.18 \\
Neptune     & 5.15111841 $\times$ 10$^{-5}$     &   164.89          & 8842.75           & 0.10 \\
\bottomrule
\end{tabular*}
\end{table}

\begin{table}
\caption{  Range of semi-major axis (in au) 
                    for the possible asteroids to be analyzed.
                The orbit parameters are taken from
 \citet{NASAPla}.}
\label{sma}
\begin{tabular*}{\tblwidth}{@{}LL@{}LL@{}LL@{}LL@{}}
\toprule
Planet      & $a_p$   & $[a_{\min},a_{\max}]$\\
\midrule
Venus       & 0.72332102    & $a_p  \pm 2.7 \% $\\
Earth       & 1.00000018    & $a_p  \pm  2.9 \% $\\
Jupiter     & 5.20248019    & $a_p  \pm 19.7 \% $\\
\bottomrule
\end{tabular*}
\end{table}



\section{JPL Horizons medium-term ephemerides}\label{sec:JPL}


    The test bench for the integrable model
    presented above is given 
    by the ephemerides of asteroids 
    computed by the JPL Horizons system \citep{NASAHor}. 
    
    The analysis starts from the choice of the Sun-planet systems 
    that can be addressed and of the small bodies 
    that may follow a quasi-coplanar co-orbital motion in such systems.
%
    The asteroids are first filtered 
        according to their semi-major axis and inclination,
        to fulfill the condition 
            of being in 1:1 mean-motion resonance with the given planet,
        and to orbit on a quasi-coplanar configuration 
        with respect to the Sun-planet reference plane. 


\subsection{On the selection of the Sun-planet systems and the asteroids}


    The maximum time span covered by the ephemerides computed 
    by the JPL Horizons system 
    for small bodies is 900 years, 
    from 1599-12-10 23:59 
    to 2500-12-31 23:58 \citep{NASAJPLHor-c}. 
%
    According to that, 
        it is necessary to estimate for each planet 
        if the characteristic period
        associated with the resonant behavior 
        can be revealed in this time span.
    The period of libration of $\theta$ in the neighborhood of $L_4$ or $L_5$,
        that is 
\citep{1977Ga}
   \begin{equation*}
\label{eq:period_libration}
        T_{lib} =\frac{2\pi}{n_p}\sqrt{\frac{27}{4}\varepsilon}^{-1}
                \left(1 + \mathcal{O}(\varepsilon)\right),
    \end{equation*}
    is introduced as a period of reference to this end.
   Only the planets such that an asteroid can perform 
    at least two librations
        in 900 years
        are considered.
    
    In Table~\ref{librations}, 
        the number of years corresponding to one libration
     is reported for each Sun-planet system
        as well as the number of librations in the considered time frame.
%
    From the data shown, 
    the Sun-planet systems 
    that fulfill the given criterion are the ones
    associated with Mercury, Venus, Earth and Jupiter.
%
    However, since the orbit of Mercury is significantly eccentric ($e_p \simeq 0.2$),
        the case of the Sun-Mercury system has been removed from the study.

    Regarding the choice of the asteroids to be considered,
    it is necessary to define an admissible range of semi-major axis
    located in the neighborhood of the resonance, that is $a = a_p$.
    To this end, 
        we use the thresholds given by the separatrices
        that stem from $L_1$ and $L_2$ in the averaged problem
        for quasi-circular orbits 
            (i.e., $\Gamma = 0$)
        and bound the co-orbital region  \cite[see][]{2022PoAl}.
%
    Following the same reasoning as in \cite{2013RoPo},
        it can be proved that the co-orbital motion is located 
        in the following subset of the phase space in terms of resonant action: 
    \begin{equation*}
        |u| <\varepsilon^{1/3} \,.
    \end{equation*}
%
   For the planets considered in this study,
        it corresponds to the range of semi-major axis $[a_{\min},a_{\max}]$ 
        shown in Table \ref{sma}.

    Finally, 
        the condition of co-planarity required by the model 
        is considered as fulfilled for all the asteroids that orbit the Sun
        with an inclination with respect to the ecliptic plane, denoted $I^{ecl}$,
        that satisfies
    \begin{equation*}
        I^{ecl} - I_{p}^{ecl} < 10^\circ,
    \end{equation*}
    where $I_{p}^{ecl}$ denotes the inclination of the planet to the ecliptic.
    From a geometric point of view,
        the two inclinations do not subtract,
        unless $\Omega^{ecl}$ and $\Omega_p^{ecl}$,
        respectively the longitude of the node of the asteroid and of the planet
        with respect to the ecliptic reference system, 
        coincide.
    The above assumption
        is a technical choice,
        in order to simplify the filtering.
    Notice that only asteroids with a prograde motion are studied.


\subsection{Methodology}


    The asteroids 
    whose motion is analyzed 
    are taken from the website of the
\citet{NASASDB}, 
    as the ones satisfying the conditions defined above 
    at the date of 2021-03-21 00.00.00 (JD 2459294.50). 
%
    Then, through the API service of the JPL Horizons system, 
    the heliocentric osculating Cartesian coordinates are computed
    in the IAU76/J2000 ecliptic reference system for each of such asteroids 
    and for the nominal planet for the maximum timescale available 
    that depends on the planet and the specific object. 

    In order to apply 
    the model and take advantage of the $(\theta,e)$-map,
    the orbital elements must be defined with respect to the planet orbital plane
    rather that the ecliptic.
%
    Thus, a set of three orthogonal rotations are applied 
    to the position and velocity vectors 
    of the object to change its coordinates 
    from the ecliptic to the orbital reference frame of the planet. 

    Let $X^{ecl}$ and $X$ 
    be the three-dimensional state vector of the asteroid 
    in the ecliptic and in the orbital reference frame of the planet, respectively. 
%
    Then, the following coordinate change is applied:
    \begin{equation*}
        X = \mathcal R_z(-\omega_{p}^{ecl}) \mathcal R_x(-I_{p}^{ecl})  \mathcal R_z(-\Omega_{p}^{ecl}) X^{ecl}
    \end{equation*}
 where 
        $\omega_{p}^{ecl}$,         is 
        the argument of periaster of the planet, 
    while
    the matrix $\mathcal R_k(\alpha)$ is the orthogonal rotation 
    about the $k$-axis of an angle $\alpha$.

    Finally, from $X$, 
    the orbital elements of the asteroid and of the planet are computed
    in the new reference system 
\citep{1971BMW}.
%
    Notice that in this way, 
    osculating orbital elements are computed
    in the osculating orbital plane of the planet.

    Once computed the orbital elements in the suitable reference system, 
    then the algorithm looks for the possible intersections of the 
    variable $u=\sqrt{a/a_p}-1$ with zero. 
%
    If this condition -- assumed to be the {\it co-orbital condition} --  occurs, 
    then also the corresponding $(\theta,e)$ are computed and stored. 
%
    Concerning $\theta$, it is computed 
    as the difference between the two osculating longitudes in the orbital plane of the planet 
    as
    \begin{equation*}
        \theta=(M+\varpi)-(M_p+\varpi_p)
    \end{equation*}
    with the longitude of the periaster $\varpi = \Omega + \omega$
    where $\omega$ and $\Omega$
    denotes respectively 
    the argument of periaster
    and the longitude of the node in the orbital plane of the planet.


\section{Analysis}\label{sec:results}


    This section describes
    the co-orbital objects found for each planet, 
    showing common features both in case their number is small, 
        as for the case of Venus, 
    and in case their number is very high,
        as for Jupiter. 

\begin{figure}
\begin{center}
\, \, \, (a)\\
\includegraphics[width=0.4\textwidth]{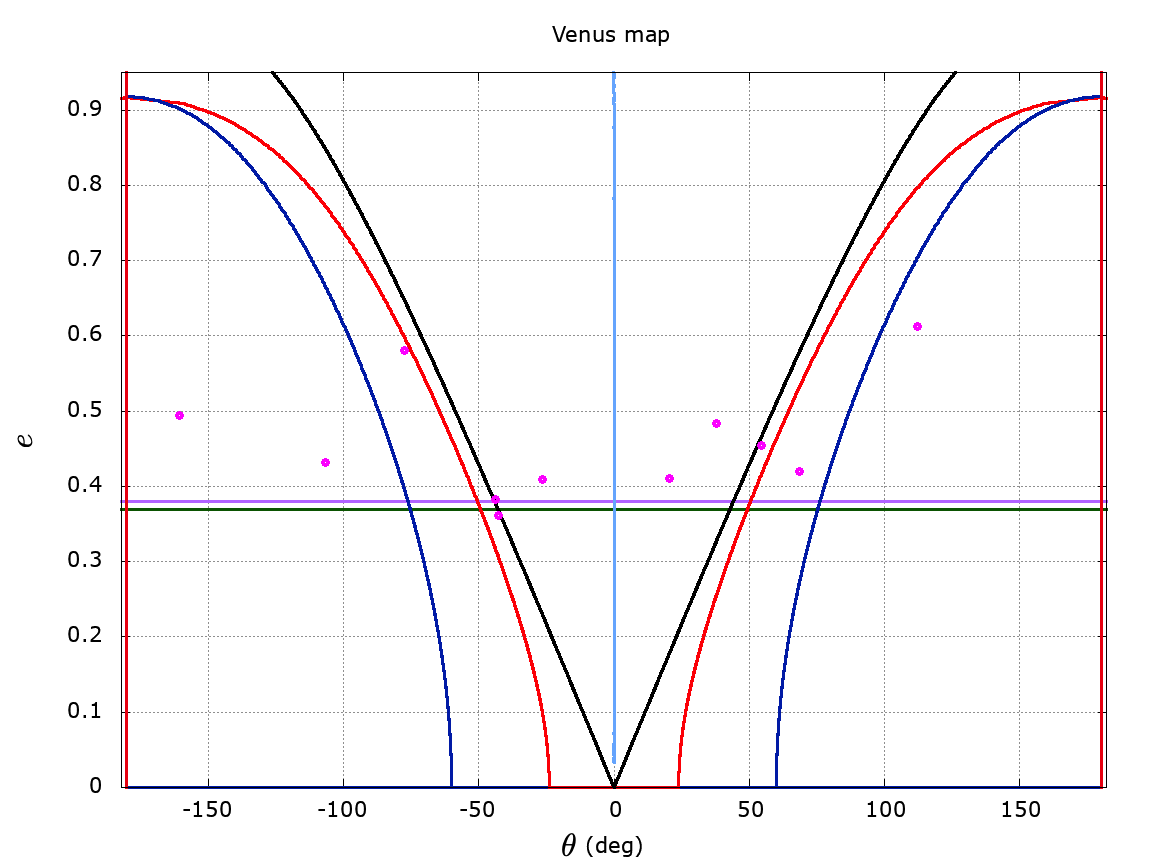}\\
\, \, \, (b)\\
\includegraphics[width=0.4\textwidth]{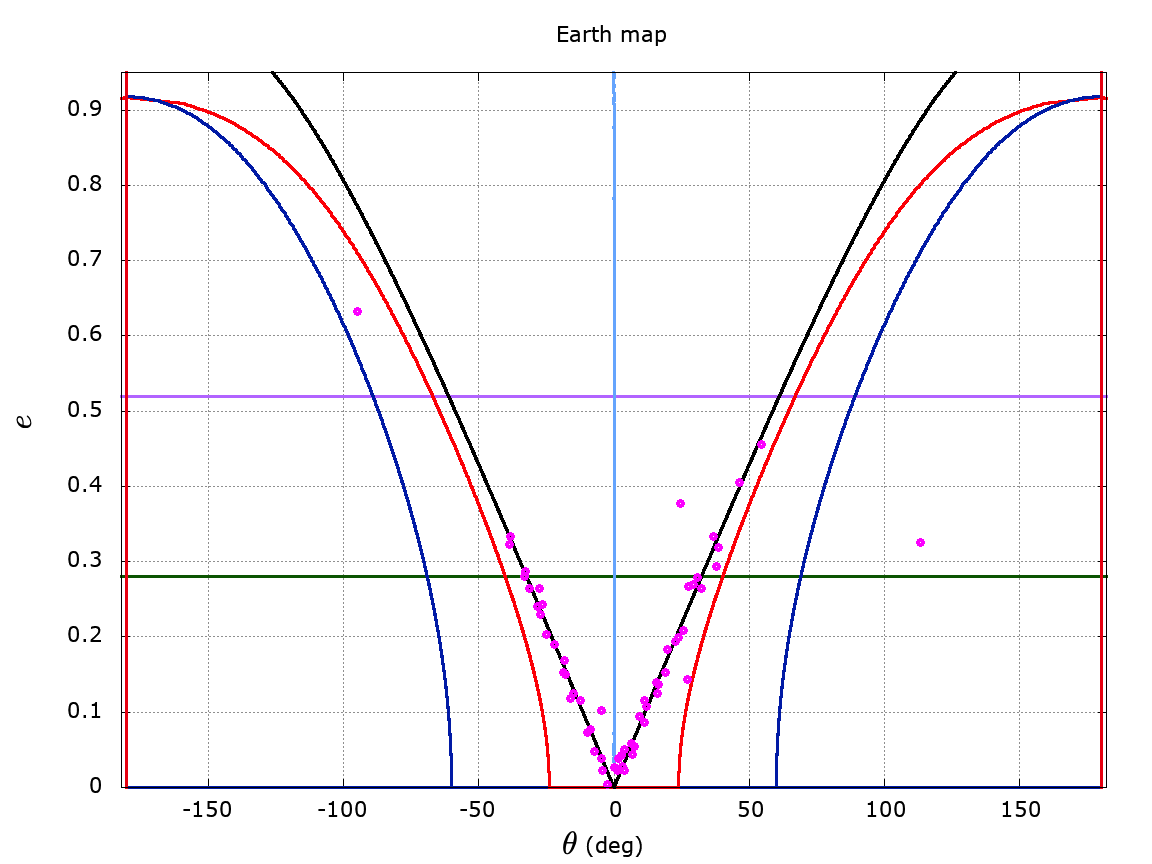}\\
\, \, \, (c)\\
\includegraphics[width=0.4\textwidth]{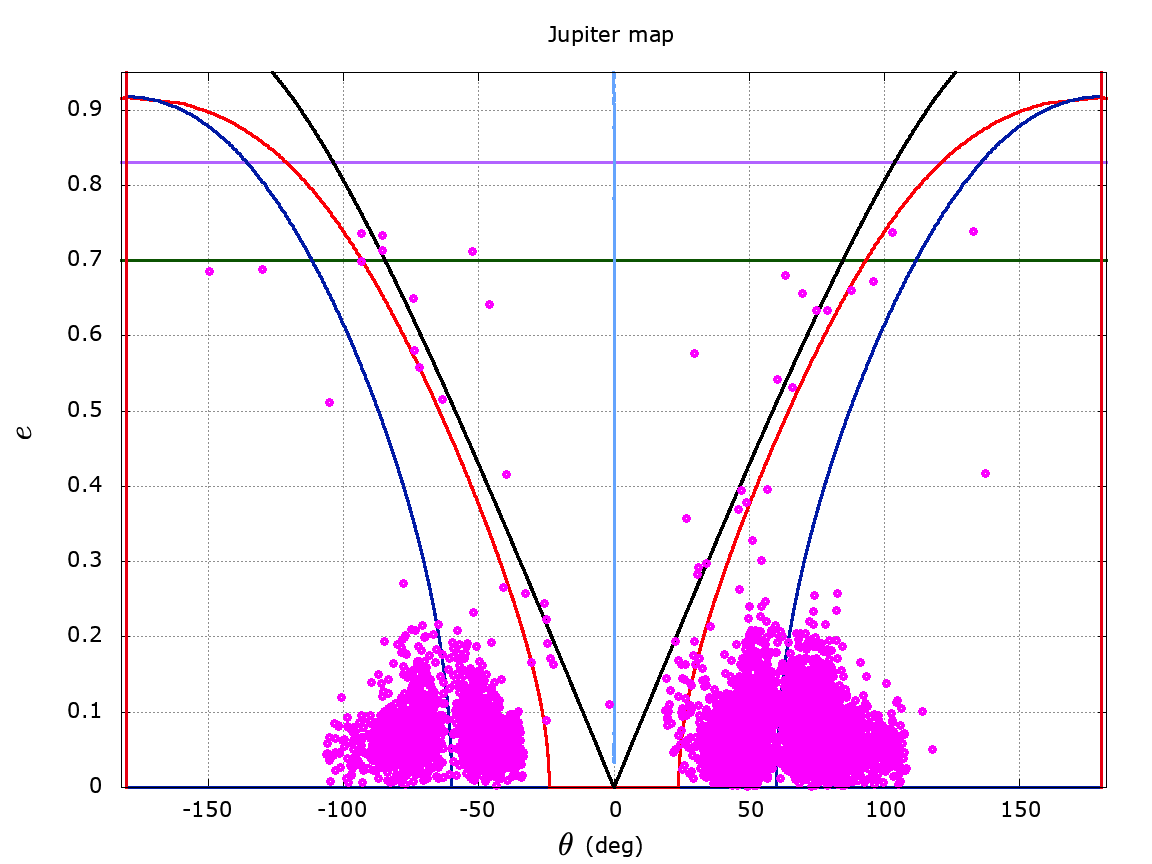}	
\end{center}
    \caption{The co-orbital objects found on quasi-coplanar orbits are depicted 
        in magenta in the $(\theta,e)$-map, 
        considering the values of the closest co-orbital configuration 
        with respect to the current date. 
        (a) Venus, 
        (b) Earth, 
        (c) Jupiter.}
\label{fig:planet_maps}
\end{figure}

    Figure \ref{fig:planet_maps}a-c displays
    the $(\theta,e)$-maps for the three planets, 
    respectively, Venus, Earth and Jupiter. 
%
    The points in magenta represent 
    the distribution of co-orbital asteroids 
    in the $(\theta,e)$-map
    at the current date,
    while
    the two horizontal lines stand for the eccentricities of an object 
    in co-orbital motion with the considered planet $P$ 
    when it crosses the orbit of an inner and an outer
    planet 
        (respectively in green and purple) 
    with respect to $P$.

    Let us start with the analysis of Venus 
        (Fig.~\ref{fig:planet_maps}a). 
%
    The horizontal lines in purple and green, 
        corresponding to 
        the eccentricity values of $e=0.38$ and $e=0.37$, 
    are associated with the crossing
    with the Earth's and Mercury's orbit, respectively. 
%
    Most of the objects found are along or above these lines
    with a significant
    eccentricity.
%
    This peculiar distribution opens the question 
    on whether it can be an observational bias 
    due to the greater ease of detecting such objects.
    A different explanation can be a dynamical mechanism 
    generated by Venus together with Earth or Mercury. 

    Figure \ref{fig:planet_maps}b  shows the objects in co-orbital motion with the Earth. 
%
    The green and purple horizontal lines in $e=0.28$ 
    and $e=0.52$ 
    correspond to the possible crossings with the orbit of Venus and Mars, respectively. 
%
    Most of the detected asteroids are close to
    the lines corresponding to the singularities 
    generated by the collision with the Earth in the integrable model,
    that is, the line that separates the domain of the QS motion from the HS one. 
%
    Moreover, 
    most of the co-orbital objects
    has eccentricities lower than the ones of the crossing with Venus and Mars.
%
    For the same reason as for the co-orbital objects of Venus,
    such distribution may be due to the easiness of detecting asteroids
    that experience relatively close encounters with the Earth,
    which is the case for QS and HS trajectories.
    
    Figure \ref{fig:planet_maps}c displays 
    the $(\theta,e)$-map of the asteroids in co-orbital motion with Jupiter. 
%
    The green and purple horizontal lines in $e= 0.7$ 
    and $e=0.83$
    correspond to the possible crossing with the Mars' and Saturn's orbit, respectively.
    In particular, it can be seen a very high number of asteroids 
    in the neighborhood of the $L_4$ 
    and $L_5$ Lagrangian equilibrium points with low eccentricities, 
    the very well-known Trojans. 
    But it is possible to notice also some asteroids in QS 
    and HS motion and some TP objects with higher eccentricities.
    
    In the following subsections, we report in detail, for each planet, 
    some significant cases of dynamics observed.


\subsection{Venus}
\label{sec:Ven}

\begin{figure*}
\centering
\includegraphics[width=0.5\textwidth]{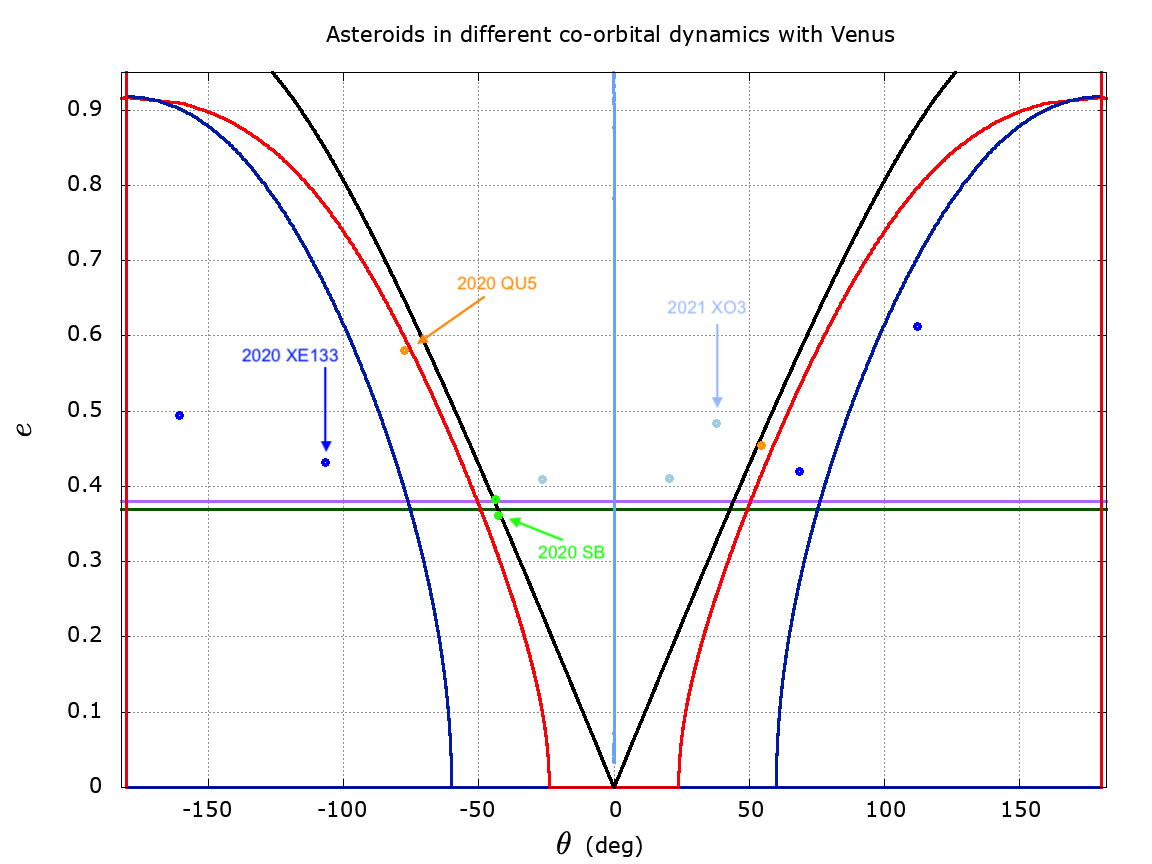}
	\caption{Venus co-orbital objects in the $(\theta,e)$-map are depicted at the current date 
	    and colored according to their dynamics: 
	        QS trajectories in sky blue, 
	        TP motion in blue,
	        compounds orbits in green, 
	        transient behavior in orange. 
	   No object in a stable HS regime (with respect to the considered time frame)
	            has been detected.    
	        The highlighted asteroids are described in detail in the following figures.}
	\label{fig:venus_dyn}
\end{figure*}

\begin{figure*}
\begin{subfigure}[t]{0.49\textwidth}
\centering
\, \, \quad (a)\\
\includegraphics[width=\textwidth]{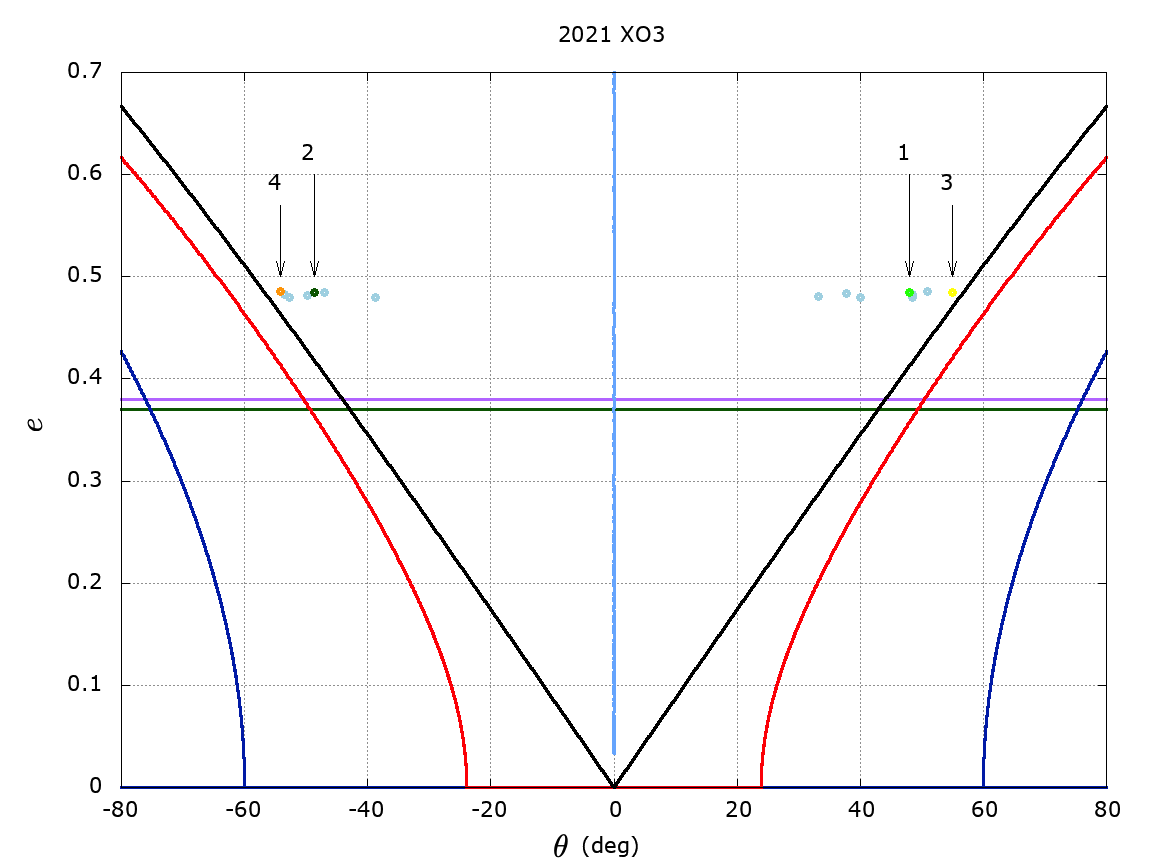}
\label{fig:venus_qsa}
\end{subfigure}
\hfill
\begin{subfigure}[t]{0.49\textwidth}
    \centering
   \quad (b)\\
    \includegraphics[width=\textwidth]{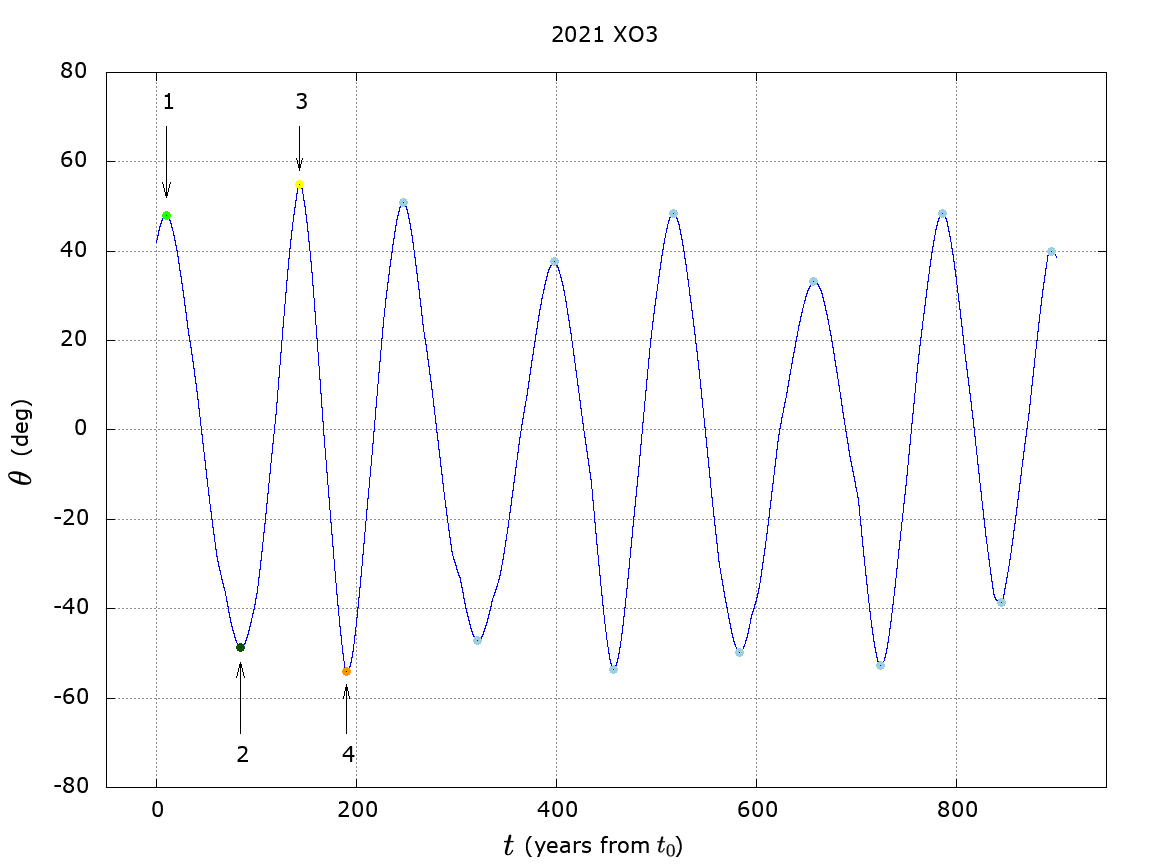}    
   \label{fig:venus_qsb}
\end{subfigure}
\hfill
\begin{subfigure}[b]{0.49\textwidth}
    \centering
  \, \,   \quad (c)\\
    \includegraphics[width=\textwidth]{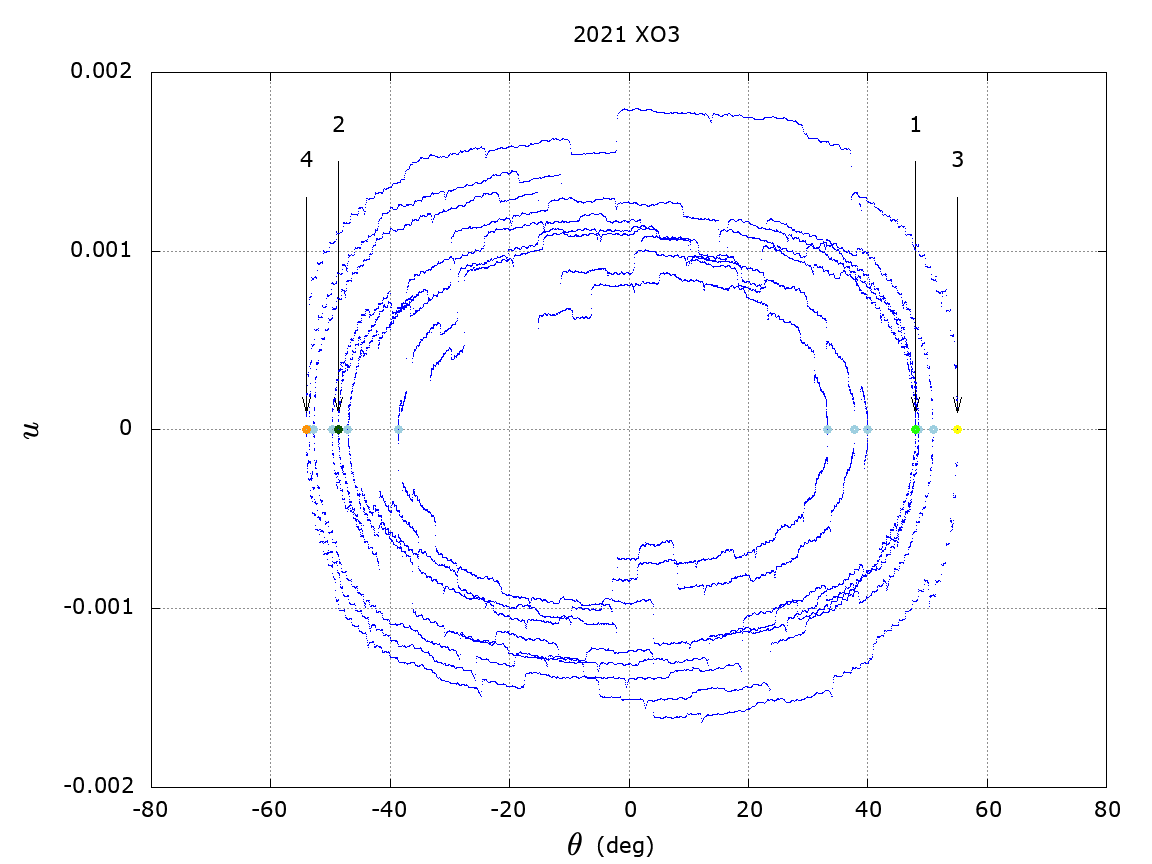}    
   \label{fig:venus_qsc}
   \end{subfigure}
    \hfill
\begin{subfigure}[b]{0.49\textwidth}
    \centering
    \quad (d)\\
    \includegraphics[width=\textwidth]{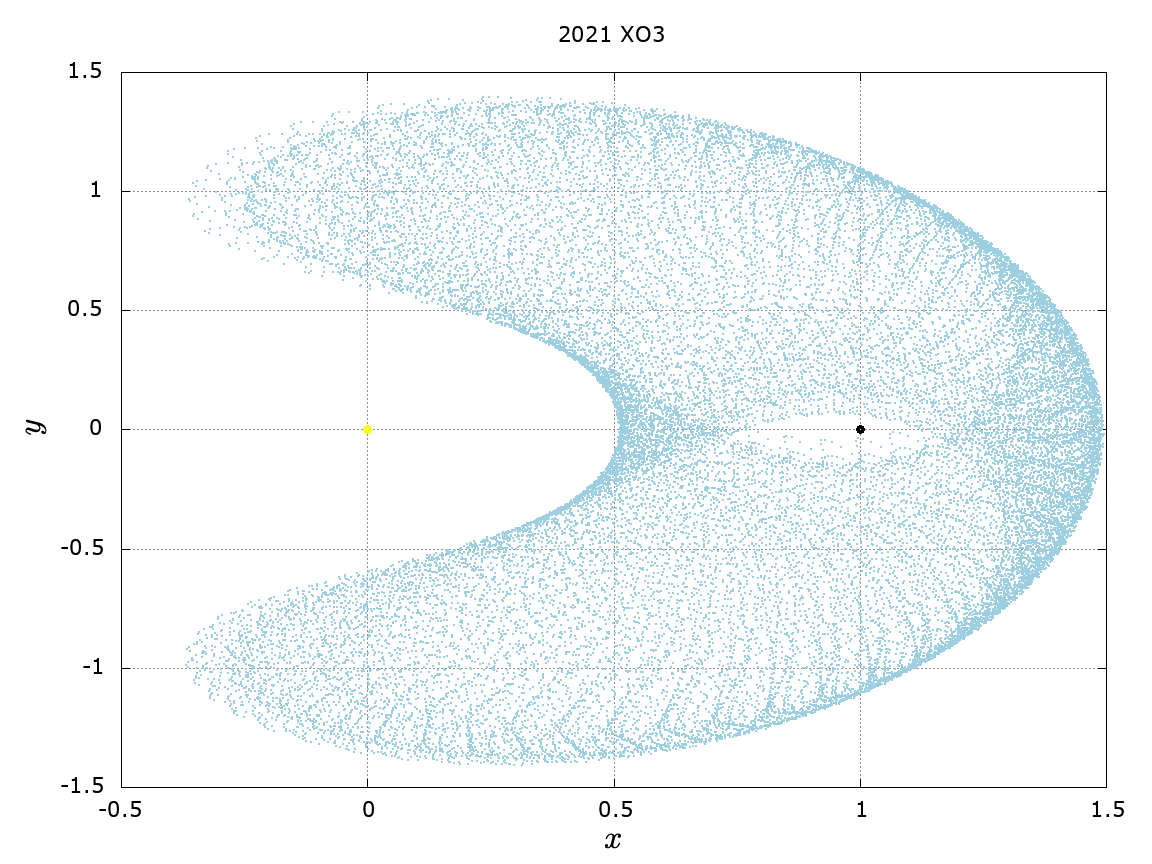}   
   \label{fig:venus_qsd}
\end{subfigure}
    \caption{Asteroid 2021 XO3 in QS motion with Venus. 
    Panel (a):  $(\theta,e)$-map at different times. 
    The numbers represent the chronological order of intersections $u=0$. 
    Panels (b) and (c): respectively, 
    the motion of the asteroid in $\theta$ with respect time,
    and in the  $(\theta,u)$ plane; 
    in each of the three plots the points stand for the intersection of the asteroid 
    with $u=0$; 
    the numbers correspond to the chronological order of intersections.     
    Panel (d): the motion of the asteroid in the heliocentric synodic reference frame 
    colored in sky blue, being  in QS motion; 
    the points $(0,0)$ and $(1,0)$ correspond to Sun and Venus positions, respectively. }
\label{fig:venus_qs}
\end{figure*}

    Let us start by describing the objects in co-orbital motion with Venus. 
%
    In Fig.~\ref{fig:venus_dyn}, 
    groups of asteroids have been highlighted
    according to their dynamics and position on the map at the current date: 
    objects in QS dynamics are plotted in sky blue, 
        while TP asteroids are highlighted in blue.
%
    The objects in green are in {\textit{compound}} motion,
    that is
    a type of dynamics that has not been defined in Sect.\,\ref{sec:av},
    and that results from the composition of two types of dynamics, QS and HS motions in these cases. 
    An example will be given in the following.
%
    The objects in orange are instead \textit{transient} orbits,
    that is objects in a co-orbital motion that in the considered time frame 
    changes in time, passing from a given domain
    to another, for instance from HS to TP, from QS to HS or viceversa.
%
    The highlighted asteroids in Fig.~\ref{fig:venus_dyn} are taken as examples 
    of the main cases that the methodology can model or not.
    To this end, they will be described in detail in the following.

\begin{figure*}
\begin{subfigure}[t]{0.49\textwidth}
\centering
\, \, \quad (a)\\
\includegraphics[width=\textwidth]{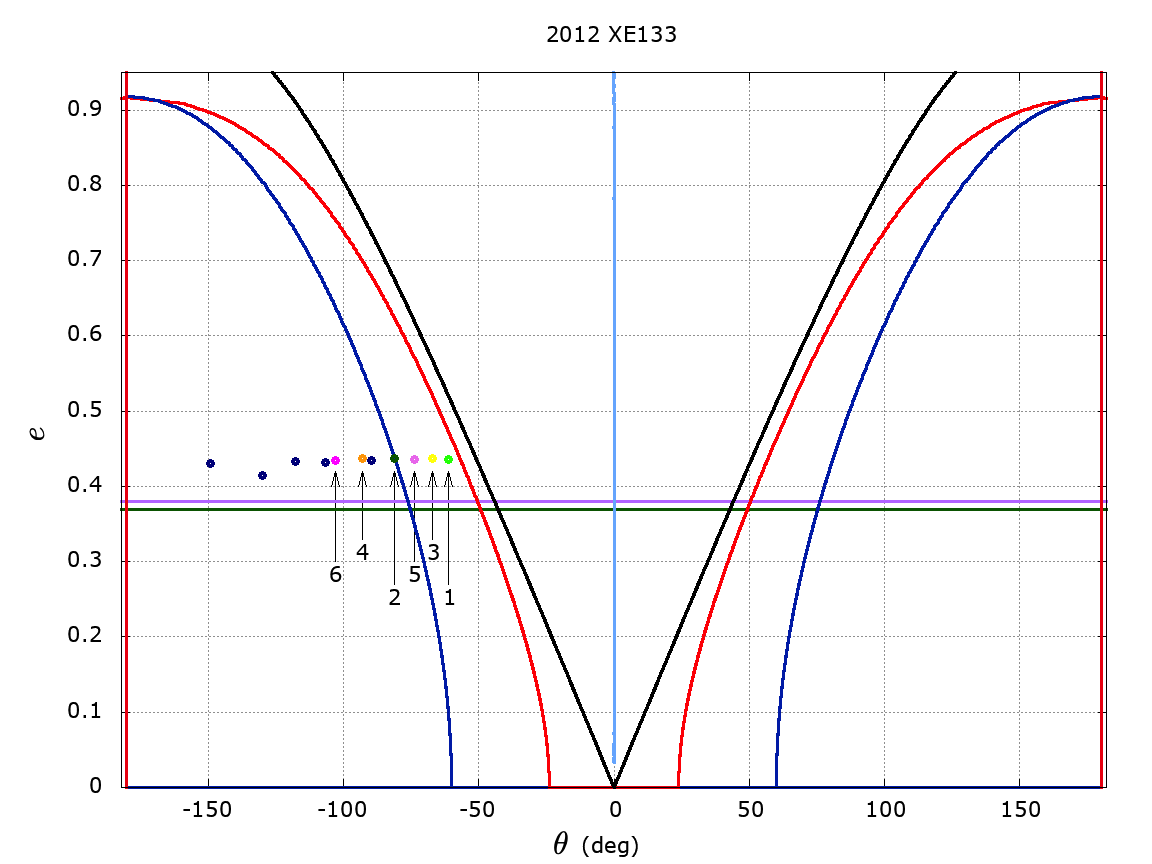}
\label{fig:venus_tada}
\end{subfigure}
\hfill
\begin{subfigure}[t]{0.49\textwidth}
    \centering
     \quad (b)\\
    \includegraphics[width=\textwidth]{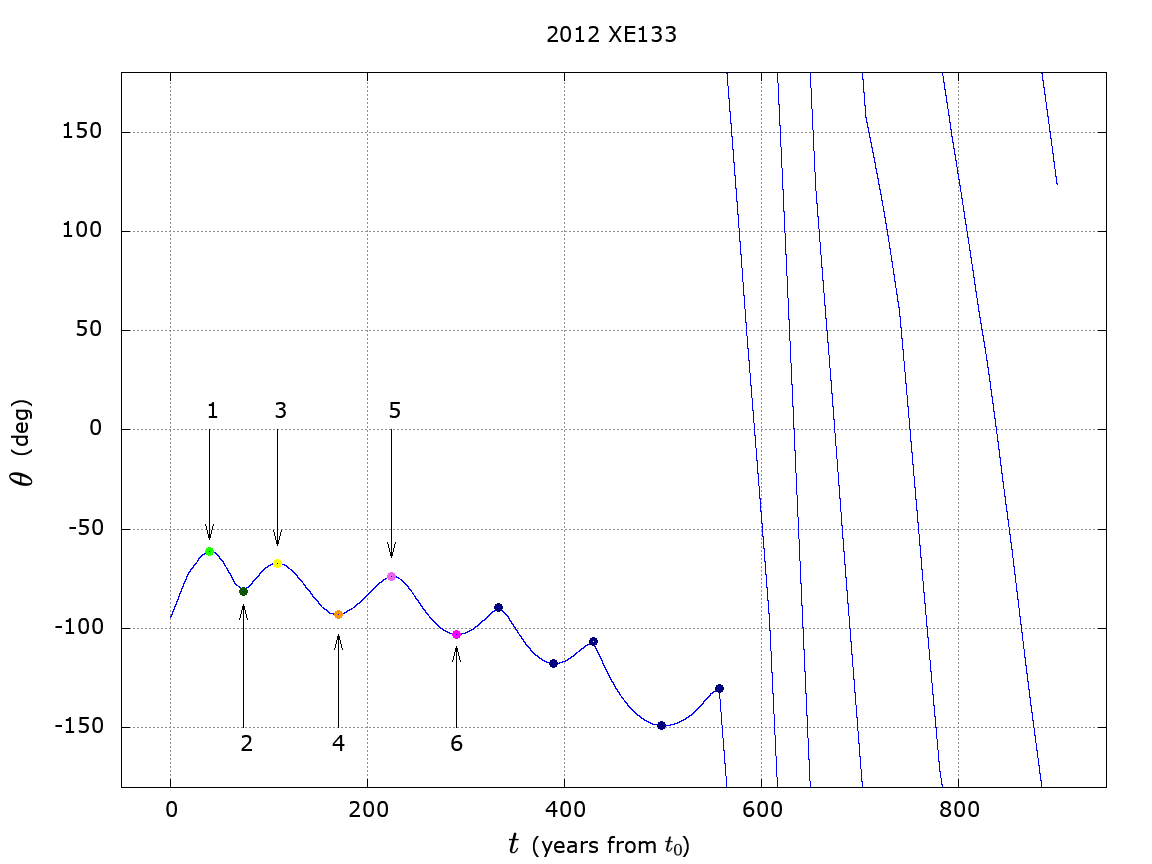}  
   \label{fig:venus_tadb}
\end{subfigure}
\hfill
\begin{subfigure}[b]{0.49\textwidth}
    \centering
    \, \, \quad (c)\\
    \includegraphics[width=\textwidth]{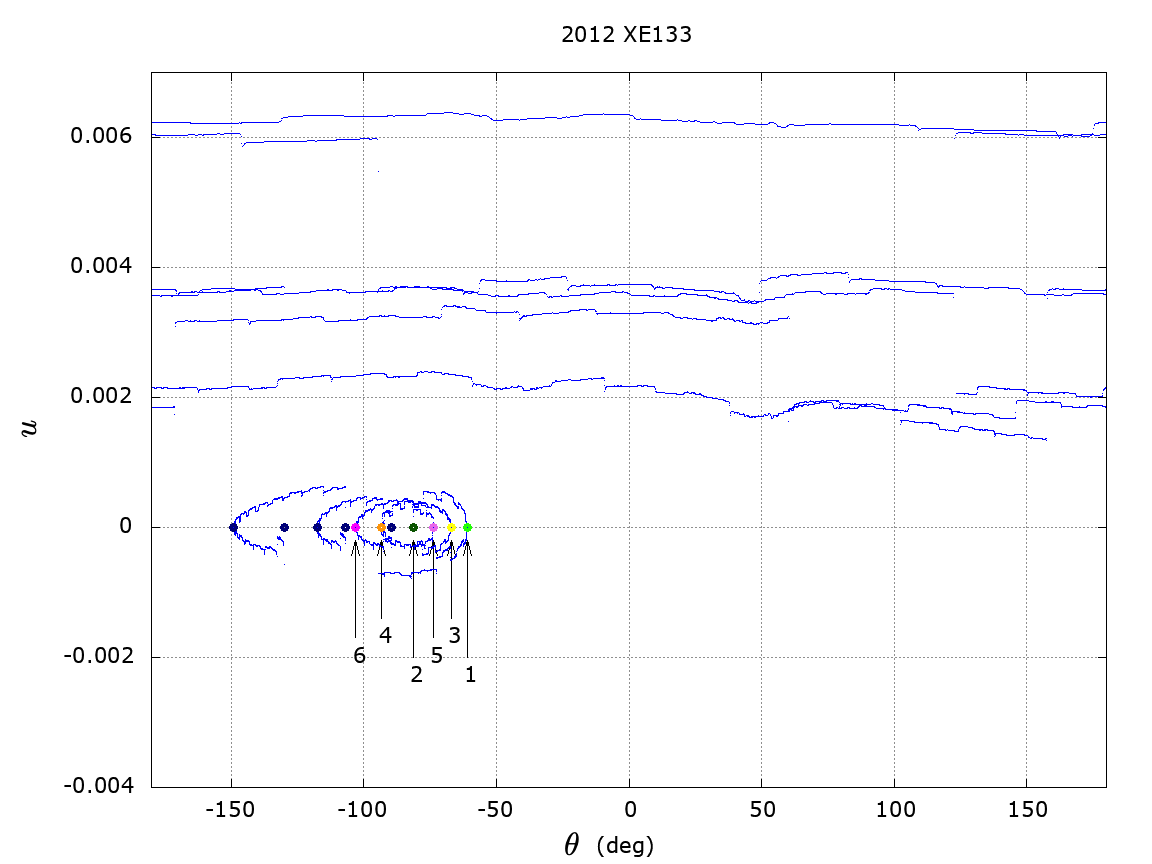}    
   \label{fig:venus_tadc}
   \end{subfigure}
    \hfill
\begin{subfigure}[b]{0.49\textwidth}
    \centering
     \quad (d)\\
    \includegraphics[width=\textwidth]{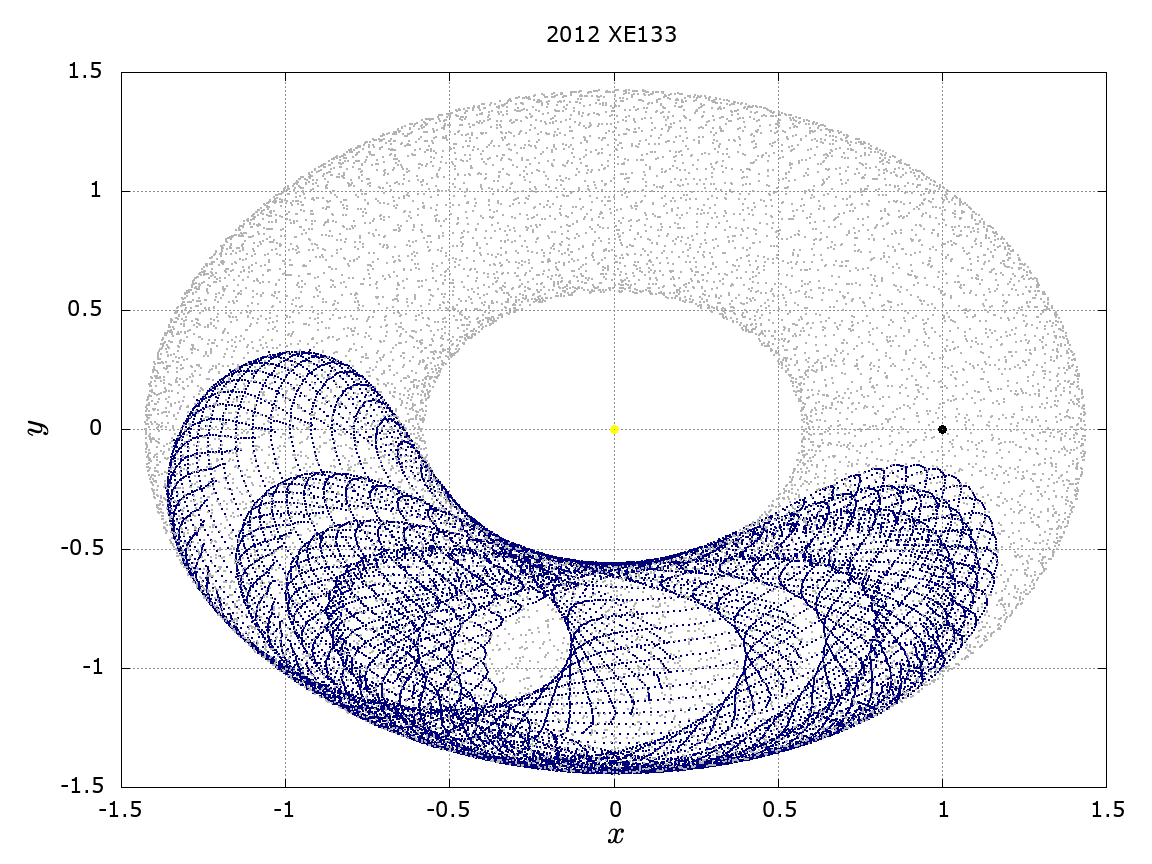}   
   \label{fig:venus_tadd}
\end{subfigure}
\caption{Asteroid 2012 XE133 on a TP orbit of $L_5$ with Venus. 
        Panel (a): $(\theta,e)$-map at different times.  
        The numbers represent the chronological order of intersections $u=0$. 
        Panels (b) and (c): respectively, the motion of the asteroid in the $(t,\theta)$ 
        and $(\theta,u)$ planes; 
        in each of the three plots the points stand for the crossing in $u=0$; 
        the numbers correspond to the chronological order of intersections. 
        Panel (d): the motion of asteroid in the heliocentric synodic reference frame 
            (in blue the TP motion is represented,
            while the interval time where the body is not in resonance with Venus is colored in grey);
        the points $(0,0)$ and $(1,0)$ correspond to the Sun and Venus positions, respectively.}
\label{fig:venus_tad}
\end{figure*}

\begin{figure*}
\begin{subfigure}[t]{0.49\textwidth}
\centering
    \, \, \quad (a)\\
\includegraphics[width=\textwidth]{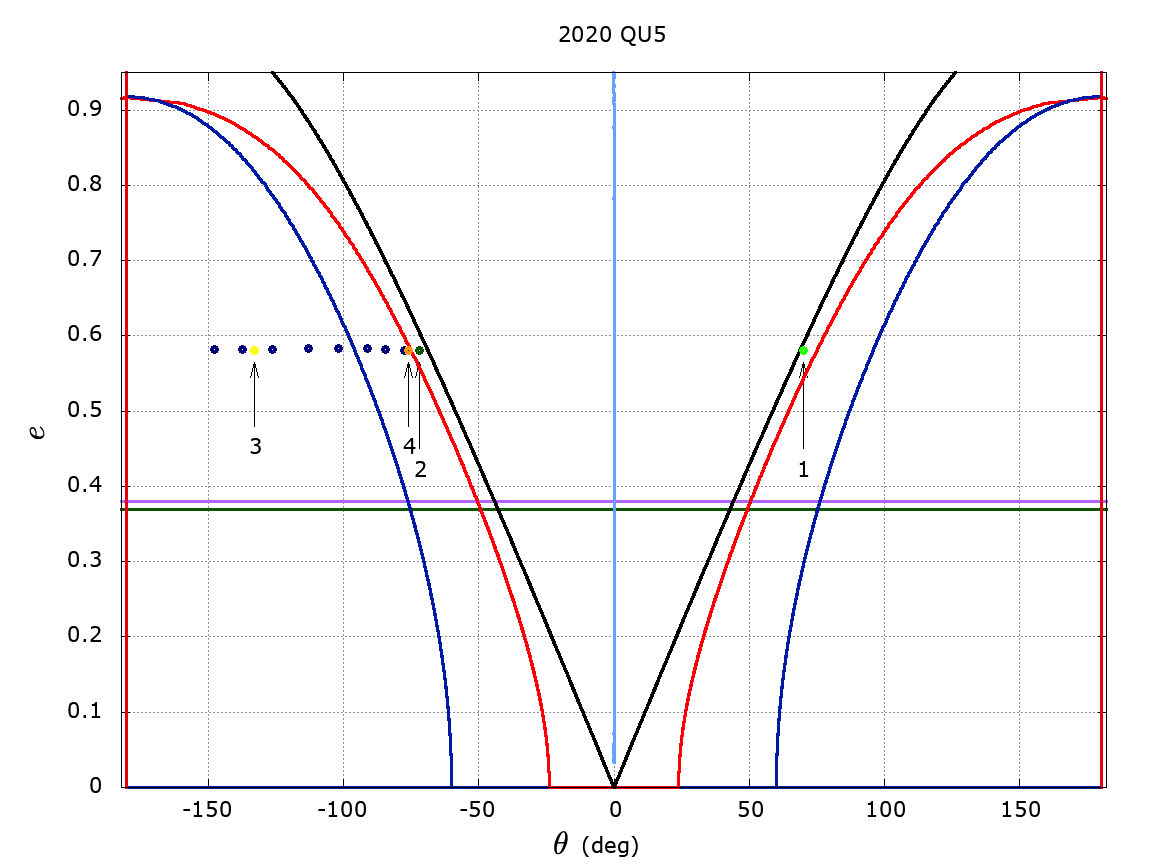}
\label{fig:venus_trana}
\end{subfigure}
\hfill
\begin{subfigure}[t]{0.49\textwidth}
    \centering
         \quad (b)\\
    \includegraphics[width=\textwidth]{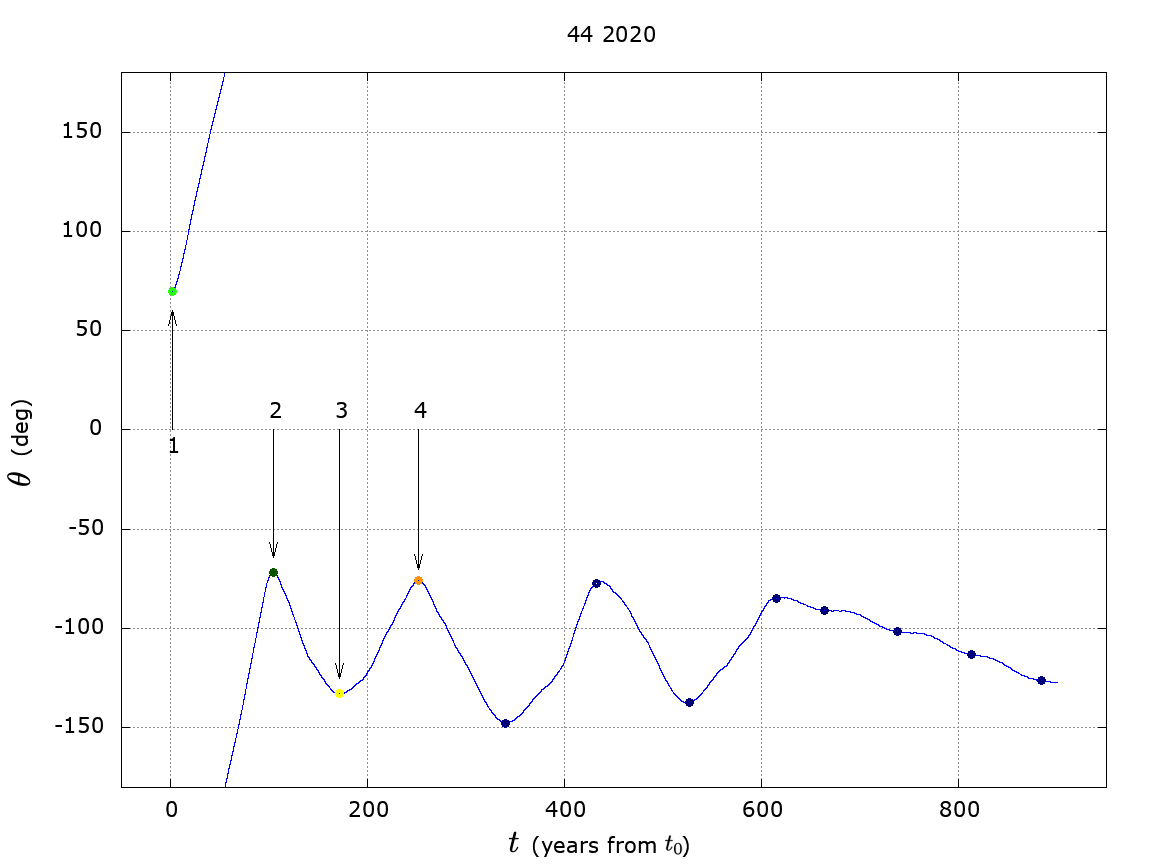}    
   \label{fig:venus_tranb}
\end{subfigure}
\hfill
\begin{subfigure}[b]{0.49\textwidth}
    \centering
     \, \,   \quad (c)\\
    \includegraphics[width=\textwidth]{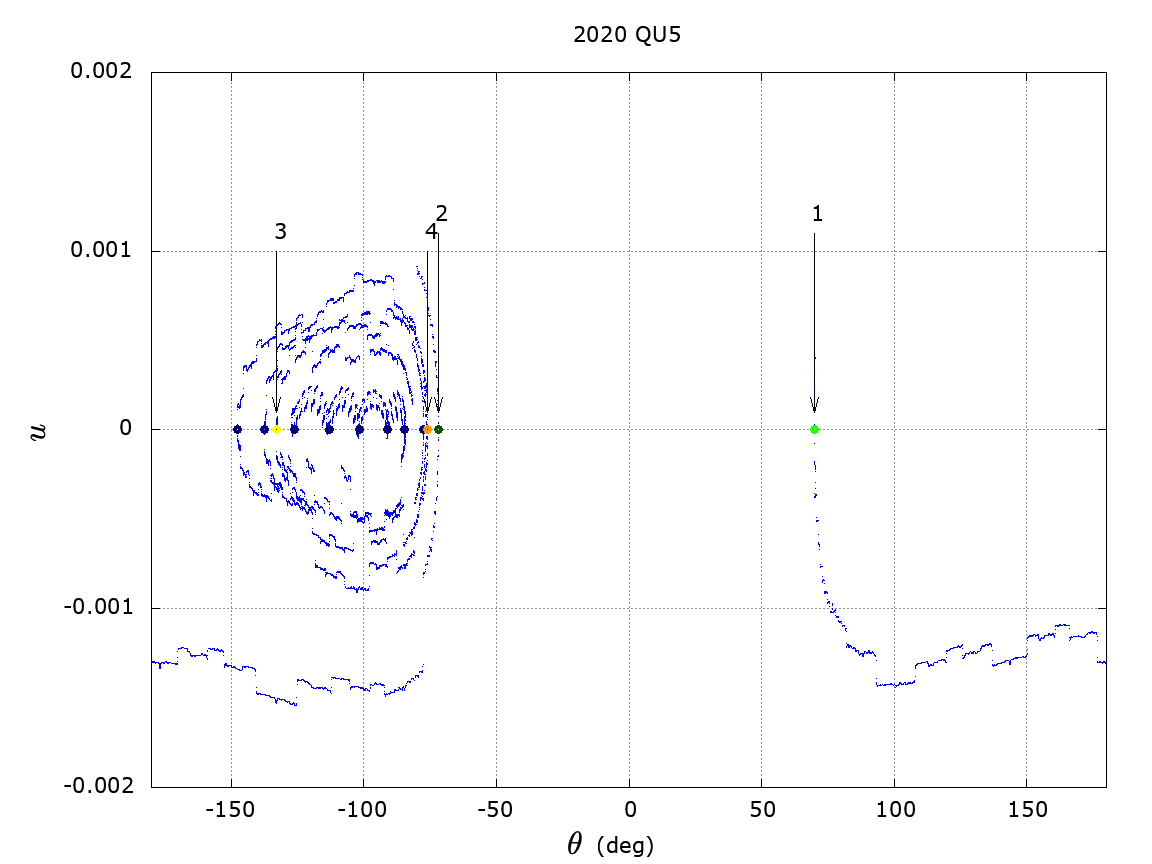}   
   \label{fig:venus_tranc}
   \end{subfigure}
    \hfill
\begin{subfigure}[b]{0.49\textwidth}
    \centering
      \quad (d)\\
    \includegraphics[width=\textwidth]{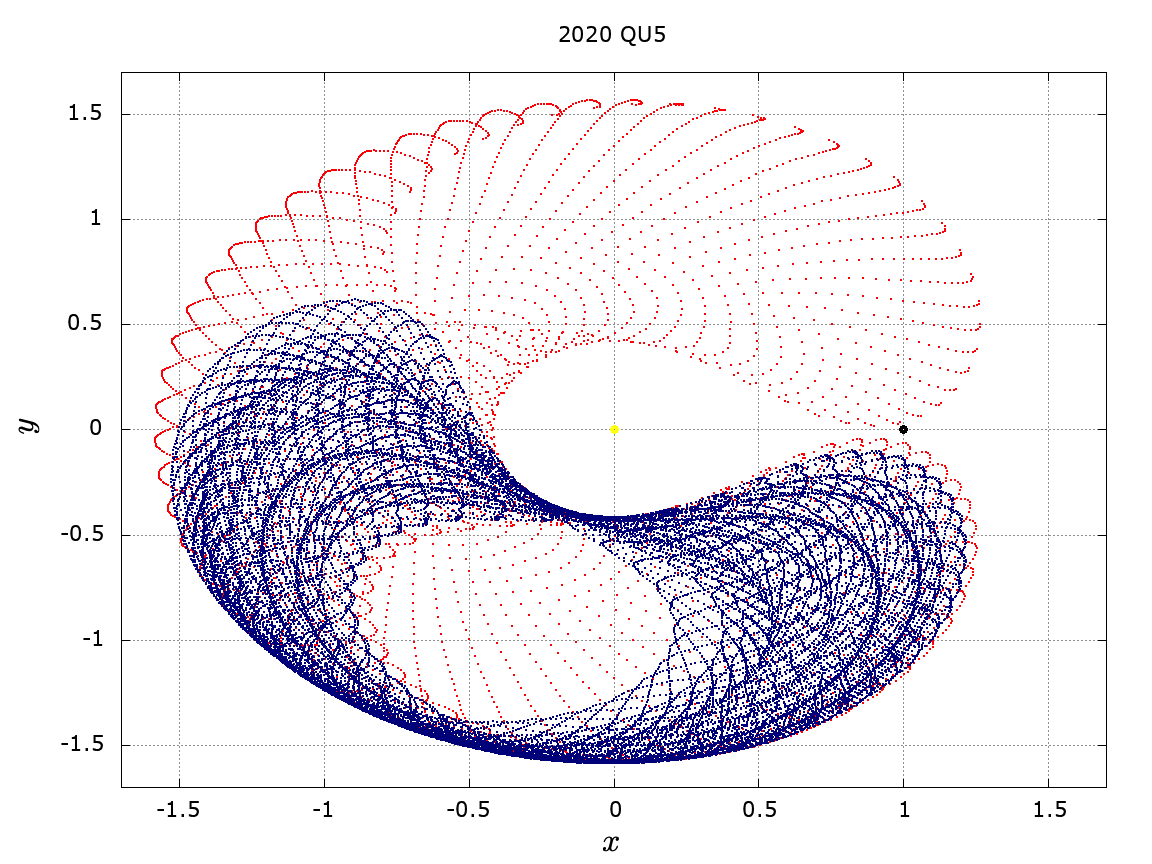}    
   \label{fig:venus_trand}
\end{subfigure}
\caption{Asteroid 2020 QU5 in a co-orbital transient dynamics with Venus. 
        Panels (a), (b), (c) represent, respectively, 
        the crossings with the section $u=0$ in the $(\theta,e)$-map, 
        and the projection of the trajectory in the $(t,\theta), (\theta,u)$ planes;      
        the points represent the  intersections with $u=0$;  
        the numbers indicate the chronological order of part of such intersections. 
        Panel (d) shows the motion of asteroid in the synodic reference frame 
        (in red the HS dynamics and in blue the TP motion); the points $(0,0)$ and $(1,0)$ correspond to the Sun and Venus positions, respectively.
\label{fig:venus_tran}}
\end{figure*}

\begin{figure*}
\begin{subfigure}[t]{0.49\textwidth}
\centering
    \, \, \quad (a)\\
\includegraphics[width=\textwidth]{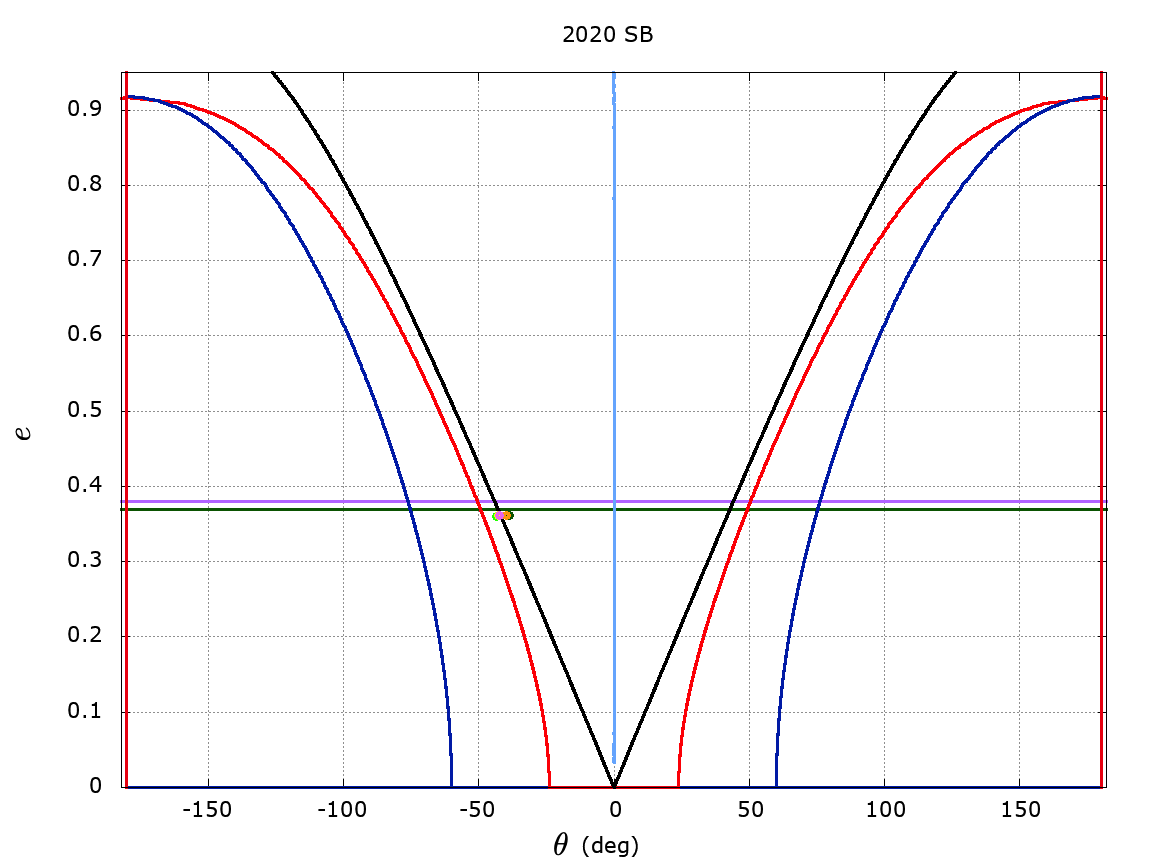}
\label{fig:venus_coma}
\end{subfigure}
\hfill
\begin{subfigure}[t]{0.49\textwidth}
    \centering
      \quad (b)\\
    \includegraphics[width=\textwidth]{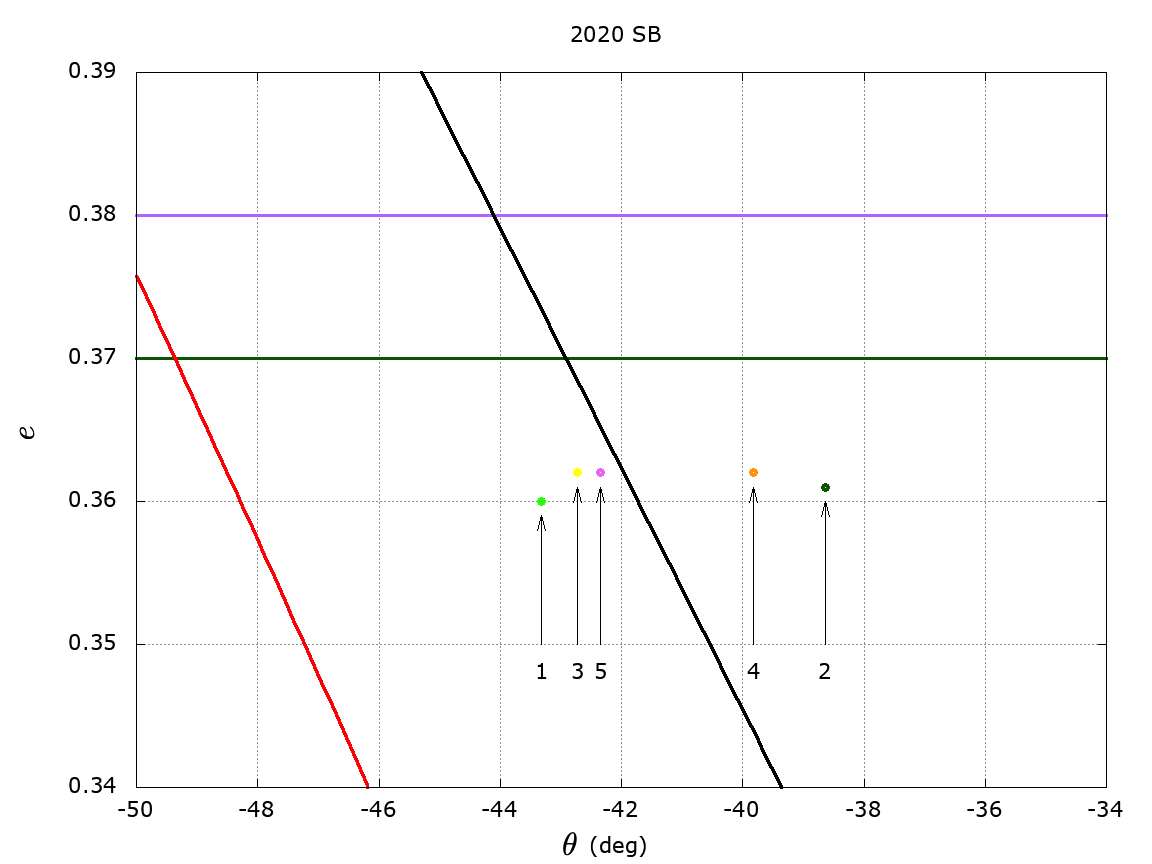}    
   \label{fig:venus_comb}
\end{subfigure}
\hfill
\begin{subfigure}[b]{0.49\textwidth}
    \centering
        \, \, \quad (c)\\
    \includegraphics[width=\textwidth]{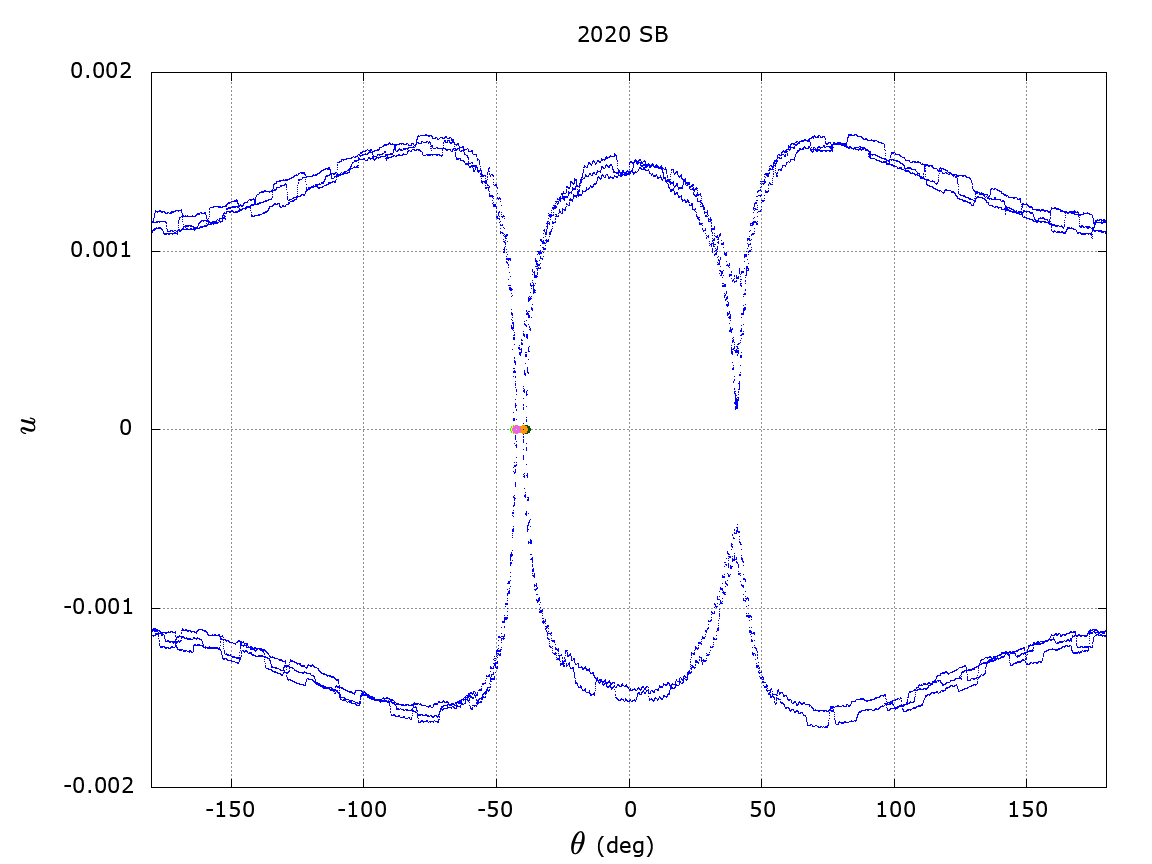}
   \label{fig:venus_comc}
   \end{subfigure}
    \hfill
\begin{subfigure}[b]{0.49\textwidth}
    \centering
     \, \,  \quad (d)\\
    \includegraphics[width=\textwidth]{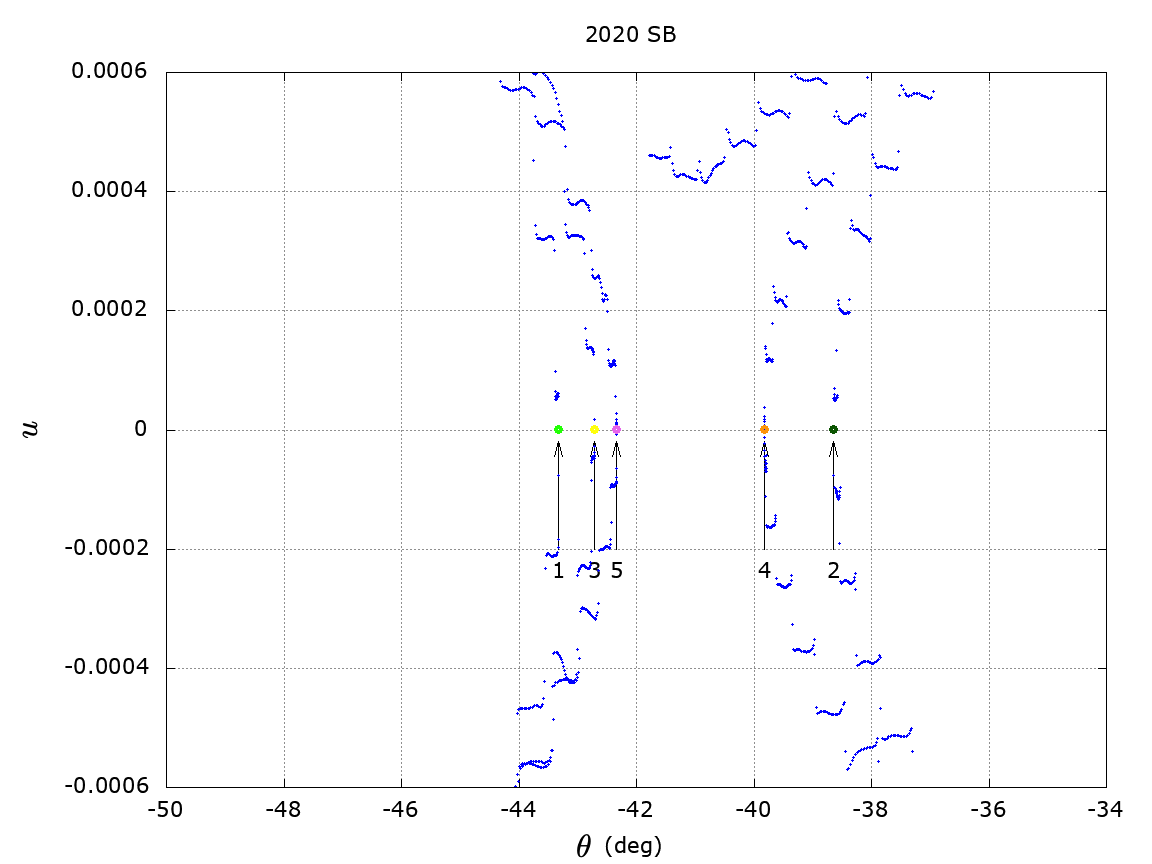}   
   \label{fig:venus_comd}
\end{subfigure}
    \hfill
\begin{subfigure}[b]{0.49\textwidth}
    \centering
        \, \, \quad (e)\\
    \includegraphics[width=\textwidth]{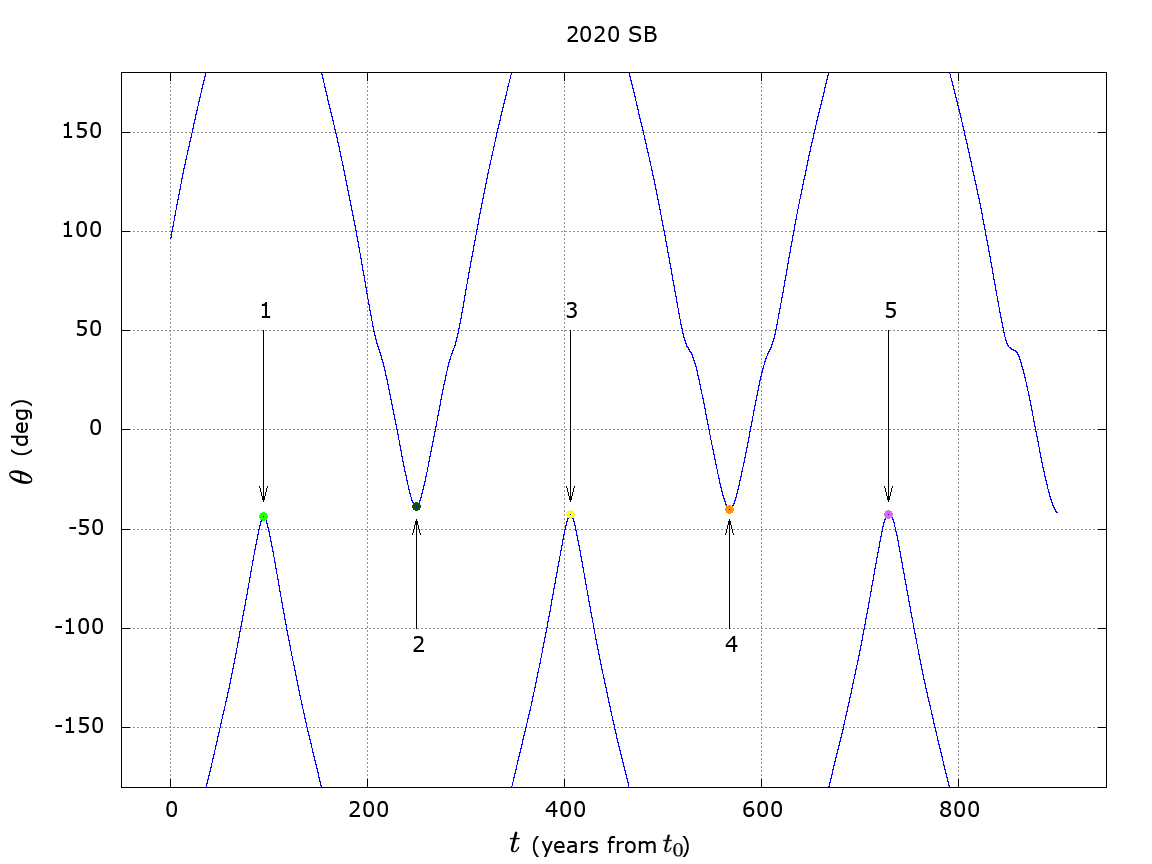}   
   \label{fig:venus_come}
\end{subfigure}
    \hfill
\begin{subfigure}[b]{0.49\textwidth}
    \centering
    \quad (f)\\
    \includegraphics[width=\textwidth]{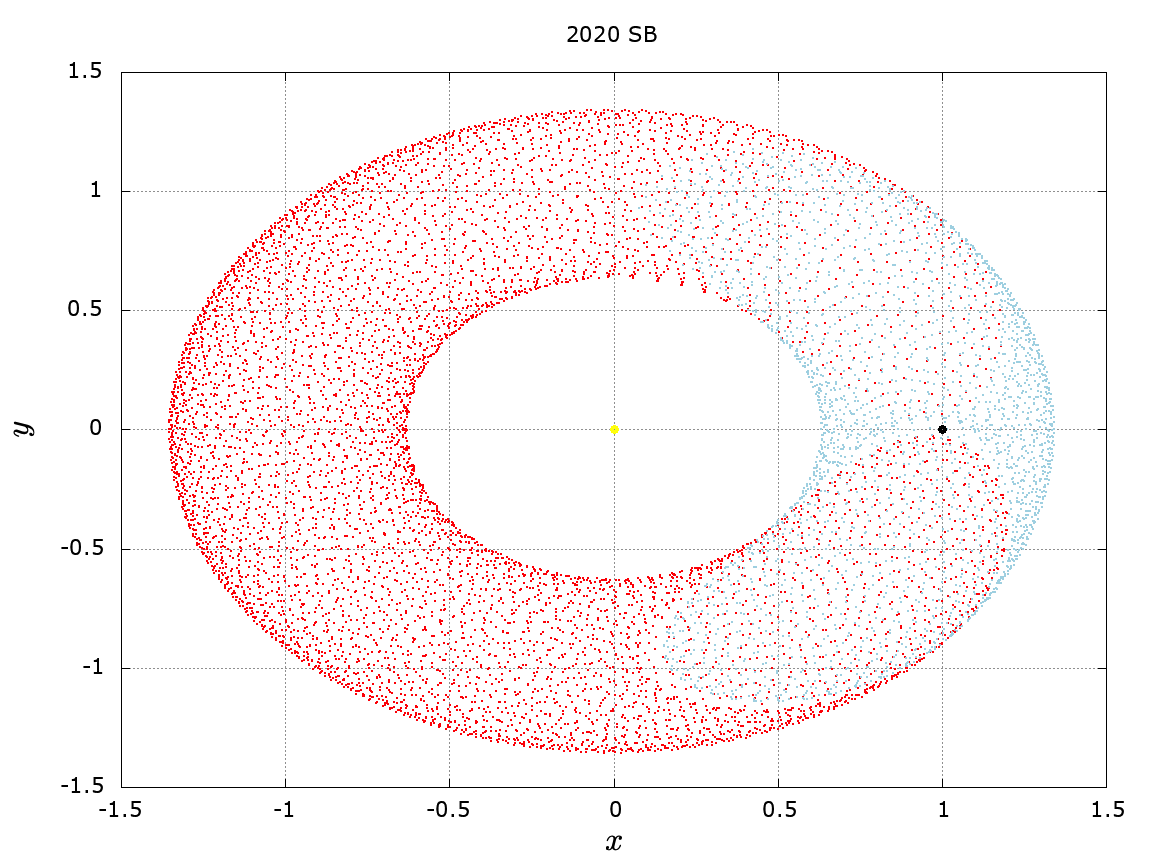}   
   \label{fig:venus_comf}
\end{subfigure}
\caption{Asteroid 2020 SB whose orbit features a compound HS-QS motion.       
        Panels (a), (c), (e) represent, respectively, 
        the crossings of the section $u=0$ in the $(\theta,e)$-map, 
        and the projection of the trajectory in the $(t,\theta), (\theta,u)$ planes; 
        the points stand for the crossings; 
        panels (b), (d) are enlargement of panel (a), (c), respectively; 
        the numbers correspond to the chronological order of crossings. 
        Panel (f) shows the motion of asteroid in the synodic reference frame; the points $(0,0)$ and $(1,0)$ correspond to the Sun and Venus positions, respectively.}
\label{fig:venus_com}
\end{figure*}

    Figure \ref{fig:venus_qs} shows the dynamics of the asteroid 2021 XO3 in QS motion. 
%
    On panel (a), the colored points on the $(\theta,e)$-map represent
    the crossings with the section $u=0$  
    during the considered time frame. 
%
    The numbers represent the chronological order of such intersections. 
%
    It can be useful to compare the first panel (a) with panels (b) and (c), 
    that show the evolution of the variable $\theta$ as a function of time $t$ 
    and the projection of the trajectory in the
    $(\theta, u)$ plane, respectively. 
%
    In the three figures, 
    the first four intersections are highlighted with a color and a number;
    the other intersection points are colored in sky blue,
        being the object in QS motion. 
    Finally, panel (d) shows the motion of the asteroid 
    in the heliocentric synodic reference frame, 
    where the Sun is located at $(0,0)$ and Venus is at $(1,0)$. 

    The asteroid 2021 XO3 has a very low inclination of about $3^\circ$ 
    with respect to the Venus orbital planet,
        therefore it is very close to the domain of validity of the model
        given by the averaged problem in the circular-planar case.
%
    In the framework of the model,
        the crossings should take almost two values of $\theta$ 
        on both sides of the family $f$
        with the same value of eccentricity.
%
    One can observe on panel (a) 
    that the crossings are constant in terms of eccentricity,
            but, in terms of $\modu{\theta}$,
            it extends in the range $[30^\circ, 60^\circ]$.
%
    As it is displayed on panel (b),
        this significant range is the results of the variations
        of the amplitude of libration around the origin of the $(\theta,u)$ plane.
%
    The high eccentricity of the asteroid ($e\simeq 0.5$) 
    indicates that the orbit of the asteroid intersects the orbits of Mercury and Earth 
    at each revolution.
    Hence, the perturbation of the two planets could be the reason of
    the observed variations in the amplitude of libration
    and thus in terms of $\theta$ at the crossing.


    Figure \ref{fig:venus_tad} shows the orbit of the asteroid 2012 XE133. 
%
    It moves on a  TP orbit associated with $L_5$, from the initial time $t_0$ to about $t_0+560$ years. 
%
    After that moment, 
        the object is thrown out from the resonance 
         (the resonant angle $\theta$ circulates),
    and starts to orbit around the Sun not in resonance with Venus,
    although its semi-major axis remains comparable with that of Venus. 
%
    It can be assumed that at that moment some external effects, 
    such as a close approach with a planet 
        (Mercury or Earth) 
    can act in such a way to modify the dynamics. 

    Panels (a), (b), (c) represent, respectively, 
    the crossings in the $(\theta,e)$-map, and the projection of the orbit 
    in the $(t,\theta)$, $(\theta,u)$ planes; 
    as before, the first colored six points represent the six first intersections with $u=0$ 
    and the numbers indicate the chronological order of such intersections. 
%
    The remaining points are plotted in blue 
    as the reference color for the TP motion. 
%
    This asteroid has an inclination with respect to the Venus orbital plane 
    of about $10^\circ$ 
    and its eccentricity is high 
    (about $0.45$). 
%
    As the previous case of the asteroid 2021 XO3, 
    the oscillations of the TP motion phase reflect 
    a sort of instability possibly due to external perturbations:
    in panels (a) and (c), {\it even} numbered intersections do not overlap each other 
    such as the {\it odd} ones do, and in panel (b) 
    the oscillations have always a different amplitude up to disappearing 
    when the motion becomes rotational. 
%
    Notice that the escape from the resonance
        occurs after a small, but notable diminution of the eccentricity
        at the crossing (Fig.~\ref{fig:venus_tad}a).
%
    Finally, panel (d) shows the motion of the asteroid 
    in the heliocentric synodic reference frame:
    in blue the evolution between time $t_0$ 
    and $t_0+560$ years where the asteroids is in TP motion, 
    while the remaining time where no resonances are present is plotted in grey.

    Figure \ref{fig:venus_tran} shows the motion of the asteroid 2020 QU5 
    which follows two types of co-orbital motion. 
%
    At the beginning of the time interval considered, 
    it is temporarily in HS motion and after a certain time it performs a {\it transition} 
    to a TP motion.
%
    As before, panels (a), (b), (c) represent, respectively, 
    the crossings with the section $u=0$ in the $(\theta,e)$-map,
    and the projection of the trajectory in the $(t,\theta), (\theta,u)$ planes; 
    the points represent the  intersections with the plane $u=0$;
    the numbers indicate the chronological order of part of such intersections. 
%
    In particular, numbers 1 and 2 represent a HS dynamics, 
    that could be visualised in all the three figures. 
%
    Numbers 3 and 4 and the remaining blue points stand for the motion in TP associated with $L_5$. 

    In panel (d) the motion of the asteroid in the synodic reference frame is shown 
    for the two different dynamics: 
    before, the red motion (HS dynamics) is plotted 
    from time $t_0$ up to time $t_0 + 130$ years 
    while the blue motion 
        (TP dynamics) 
    is plotted from time $t_0+130$ up to time $t_0 + 950$ years 
        (final time). 
%
    The inclination of the asteroid with respect to the Venus orbital plane 
    is quite low
        (about $2.6^\circ$), 
    while the high eccentricity 
        (almost 0.6) 
    could be responsible of close encounters with Earth or Mercury 
    which would cause the dynamics to change 
    from HS to TP.

    Figure \ref{fig:venus_com} shows a particular case of co-orbital motion
    for which the resonant angle $\theta$ 
    avoids a value close to $-50^\circ$.
%
    Actually, the asteroid 2020 SB is a case of 1:1 resonant motion 
   in  {\it compound} HS-QS dynamics \citep[see, e.g.,][]{2000Ch,2004BrInCo}.
%
    This terminology
        derives from the composition of two different dynamics, 
      -- QS and HS  in this case
        \footnote{In the spatial case of the RTBP,
        other types of compound have been described, such as TP-QS or
        TP-QS-TP \citep[see, e.g.,][]{2000Ch}.} --
    such that the resonant variables $(\theta, u)$  feature periodic oscillations
    and thus the trajectory moves continuously 
    from the HS domain 
    to the QS domain, 
    to then turning back to the HS domain.
%
    Notice that in this case we do not speak of a transition
    from a given temporary type of co-orbital motion to another,
    but we define another orbital regime of the co-orbital resonance.
%
    Indeed, this behavior can be understood by the disappearance 
        of one of the two lines
        corresponding to the singularities of collision 
        in the phase portrait of Fig.\,\ref{fig:PhaseP}, 
        and thus it cannot occur in the framework of the circular-planar case of the RTBP, 
        where such lines bound the QS and the HS domains.

    As a matter of fact,
        according to \cite{2014SiNeAr},
        the compound dynamics appears 
        in the framework of averaged problem of the circular-spatial RTBP. 
    Being the inclination of the asteroid significant, 
        about $9^\circ$ with respect to the Venus orbital plane,
    this is consistent with their theory
    and highlights that a threshold of $10^\circ$ with respect to the orbital plane
    may be too large in order to use the model given by the averaged problem in circular-planar case. This limit should be however verified by considering a larger number of cases and in different Sun-planet systems. 

    In panels (b) and (d), 
        which are enlargements of panels (a) and (c), respectively,
    and in panel (e), five points representing the intersections with $u=0$ are shown 
    and they are numbered in chronological order. 
%
    Odd numbers represent the crossings of the trajectory in the HS domain, 
    while even numbers a crossing in the QS region. 
%
    Panel (b) 
    clearly shows that in the $(\theta,e)$-map 
    the intersections with $u = 0$ cross the collision curve, 
    and thus they belong alternatively to one of the two domains. 
%
    Finally, in panel (f) the motion of the asteroid in the synodic reference frame is shown 
    for two different part of the compound dynamics: 
        the red motion, 
        standing for the HS part of the dynamics, 
    is plotted from time $t_0$ up to time $t_0 + 209$ years 
    and the sky blue motion, 
        standing for the QS part of the dynamics, 
    takes place from time $t_0+209$ up to time $t_0 + 286$ years.
%
    Notice that in the analysis performed a significant number of  objects appear 
        to follow in compound HS-QS motion, especially for the Earth (see next section), 
        and this is one of the reasons 
        why in the near future we aim at analysing the three-dimensional case 
        for a more complete treatment of this dynamics. 

 \begin{table*}
\caption{The co-orbital objects found for the Sun-Venus system on quasi-coplanar orbits.
            The values reported refer to the intersection with $u=0$, 
            that is the closest one to the current date, 
            assumed to be 2021-03-21 00.00.00 (JD 2459294.50).   
            The angular values are reported in degrees and are defined with respect to the orbital plane of the planet.  
            The last column indicates the co-orbital dynamics detected in the given time frame (CP refers to compound, TR to transient).
            These co-orbital configurations are shown also in Fig.~\ref{fig:planet_maps}a
            and Fig.~\ref{fig:venus_dyn}.} \label{venus_outcome}
\begin{tabular*}{\tblwidth}{@{}LL@{}LL@{}LL@{}LL@{}LL@{}LL@{}LL@{}}
\toprule
asteroid & $t$ (JD) & $a$ (au) & $e$ & $I$ & $\theta$ & dynamics\\
\midrule
 2001 CK32  &   2463638.72      &   0.72333158  &   0.382   &   5.68    &   -43.834   & CP \\ 
  2002 LT24 &   2375731.02      &   0.72307523  &   0.494   &   3.84    &  -160.581   & TP  \\
2002 VE68   &   2475698.55      &   0.72333068  &   0.409   &  12.33    &   -26.292  & QS   \\
 2012 XE133 &   2462143.09      &   0.72421895  &   0.432   &   10.03   &  -106.692  & TP   \\
2013 ND15   &   2454656.87      &   0.72343445  &   0.612   &   1.95    &  111.984   & TP   \\
2020 BT2    &   2462807.58      &   0.72333585  &   0.420   &   5.40    &    68.586  & TP  \\ 
  2020 QU5  &   2463543.58      &   0.72333067  &   0.581   &   2.61    &  -77.507  & TR    \\
2020 SB     &   2453862.11      &   0.72338904  &   0.362   &   8.80    &  -42.723  & CP    \\
 2021 XA1   &   2465246.93      &   0.72333198  &   0.410   &   3.02    &   20.629 &  QS   \\
2021 XO3    &   2450832.85      &   0.72363671  &   0.484   &   2.82    &   37.760  & QS    \\
 2022 AP1   &   2509314.03      &   0.72252306  &   0.455   &   1.52    &   54.384  &  TR  \\
\bottomrule
\end{tabular*}
\end{table*}
    In Table~\ref{venus_outcome}, 
    we give the list of the asteroids in co-orbital motion 
    with Venus following the JPL Horizons ephemerides. The data correspond to the intersection with the section $u=0$, 
    that occurs at the closest epoch with respect to the reference epoch. The last column indicates the type of co-orbital motion detected in the given time frame.

\subsection{Earth}

    Let us continue the analysis focusing on two examples of co-orbital motion in the Sun-Earth system.

\begin{figure*}
\begin{subfigure}[t]{0.49\textwidth}
\centering
\,\, \quad (a)\\
\includegraphics[width=\textwidth]{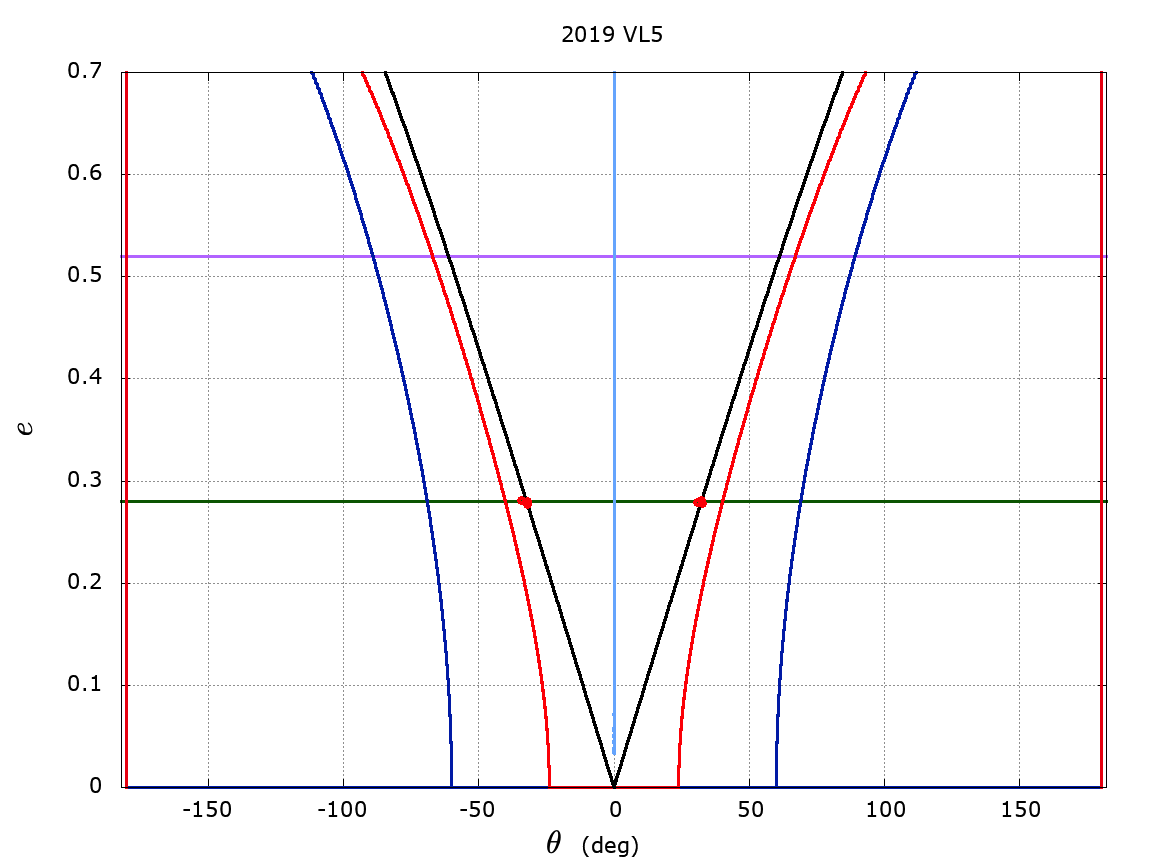}
\label{fig:terra_hsa}
\end{subfigure}
\hfill
\begin{subfigure}[t]{0.49\textwidth}
\centering
 \quad (b)\\
\includegraphics[width=\textwidth]{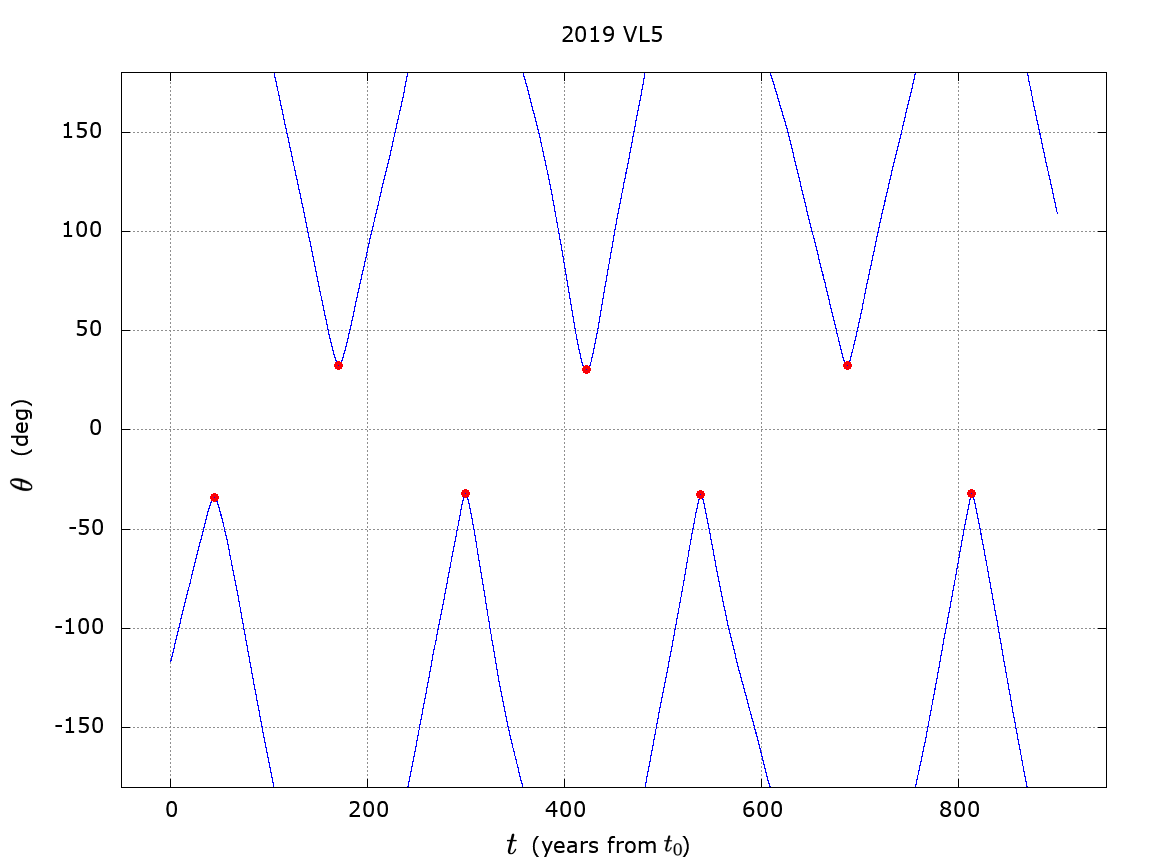}
\label{fig:terra_hsb}
\end{subfigure}
\hfill
\begin{subfigure}[t]{0.49\textwidth}
\centering
\,\, \quad (c)\\
\includegraphics[width=\textwidth]{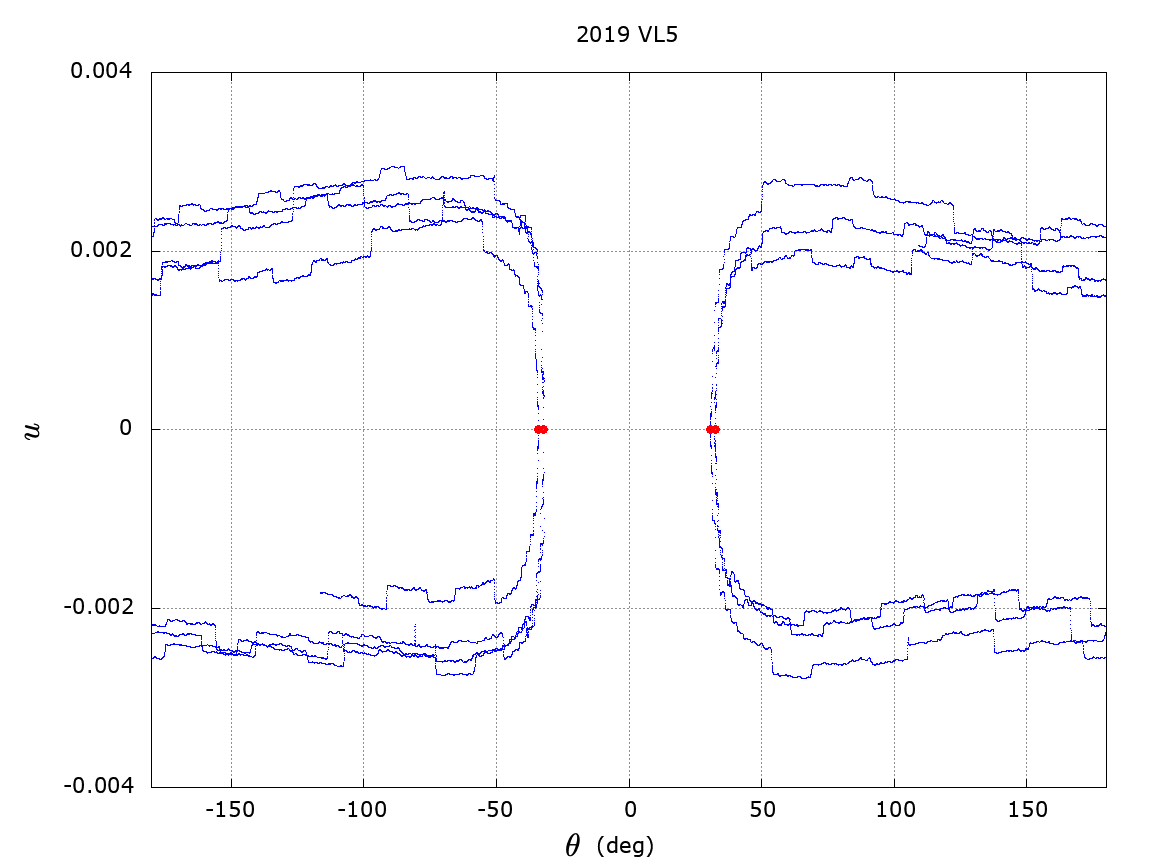}
\label{fig:terra_hsc}
\end{subfigure}
\hfill
\begin{subfigure}[t]{0.49\textwidth}
\centering
 \quad (d)\\
\includegraphics[width=\textwidth]{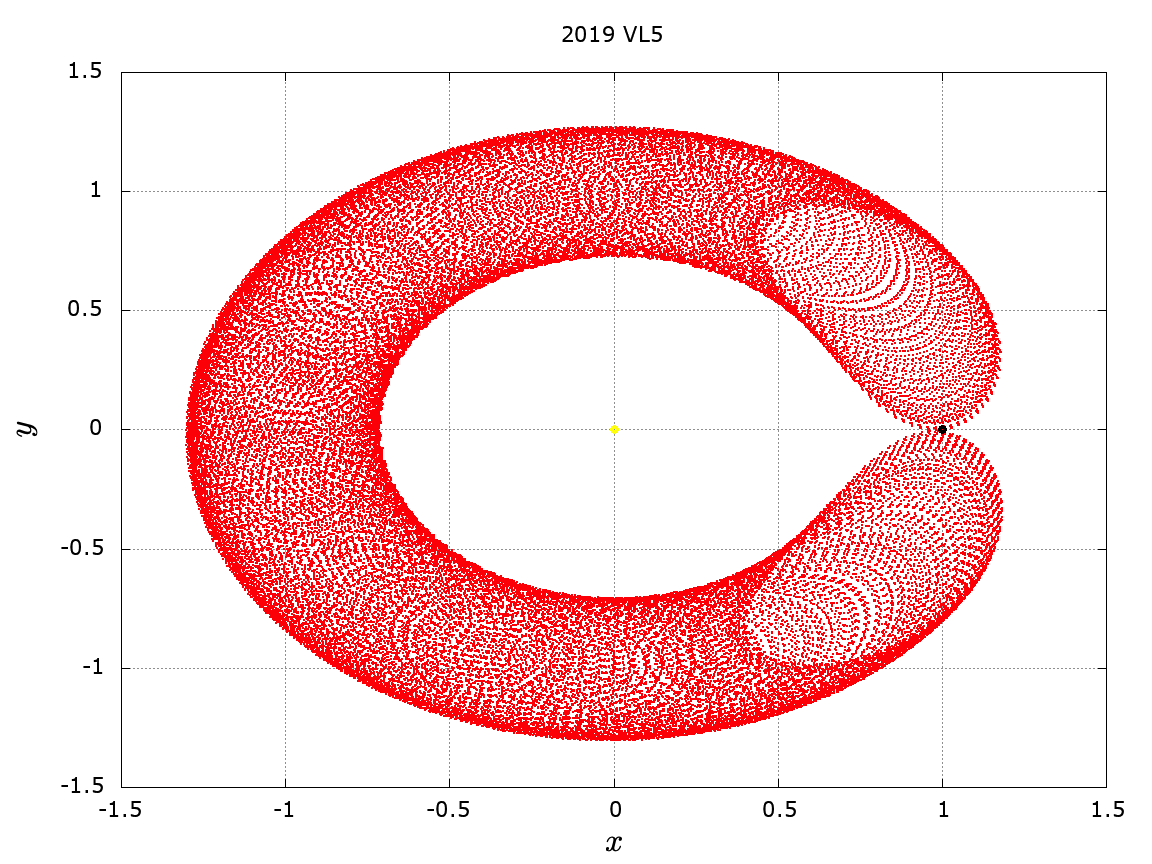}
\label{fig:terra_hsd}
\end{subfigure}
\caption{Asteroid 2019 VL5 in HS dynamics with the Earth. 
        Panels (a), (b), (c) show, respectively, 
        the crossings with the section $u=0$ in the  $(\theta,e)$-map, 
        and the projection of the trajectory in the $(t,\theta)$, $(\theta,u)$ planes; 
        the red dots represent the intersections with $u=0$;  
        panel (d) shows the motion of asteroid in the synodic reference frame; the points $(0,0)$ and $(1,0)$ correspond to the Sun and Earth positions, respectively.}
\label{fig:terra_hs}
\end{figure*}

\begin{figure*}
\begin{subfigure}[t]{0.49\textwidth}
\centering
\,\, \quad (a)\\
\includegraphics[width=\textwidth]{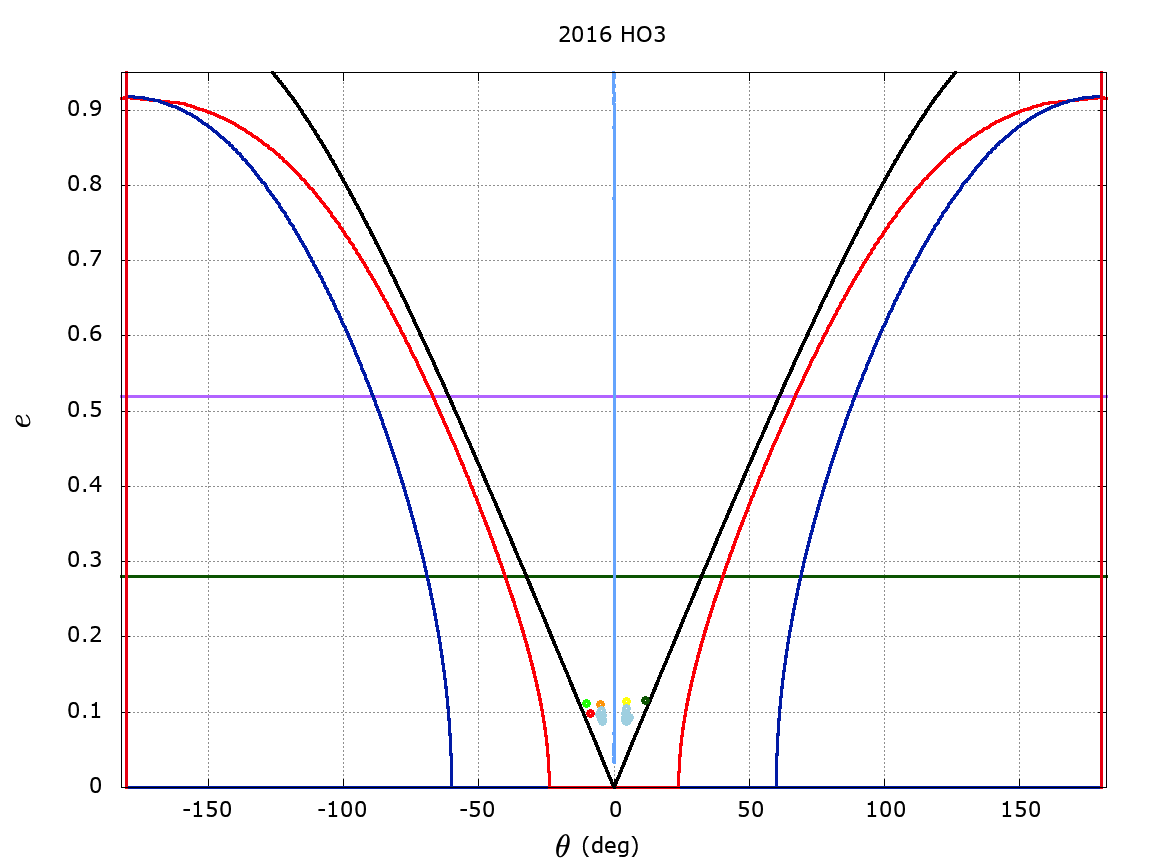}
\label{fig:terra_coma}
\end{subfigure}
\hfill
\begin{subfigure}[t]{0.49\textwidth}
\centering
\quad (b)\\
\includegraphics[width=\textwidth]{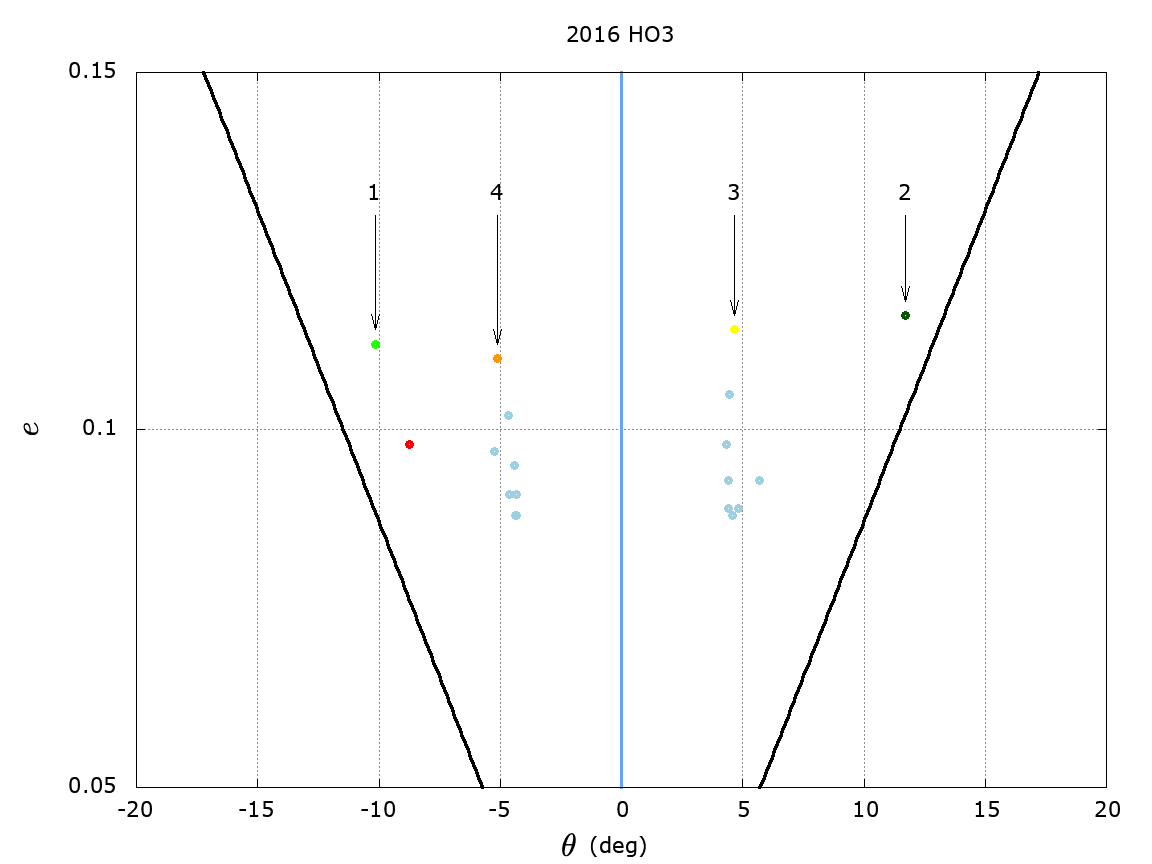}
\label{fig:terra_comb}
\end{subfigure}
\hfill
\begin{subfigure}[t]{0.49\textwidth}
\centering
\,\, \quad (c)\\
\includegraphics[width=\textwidth]{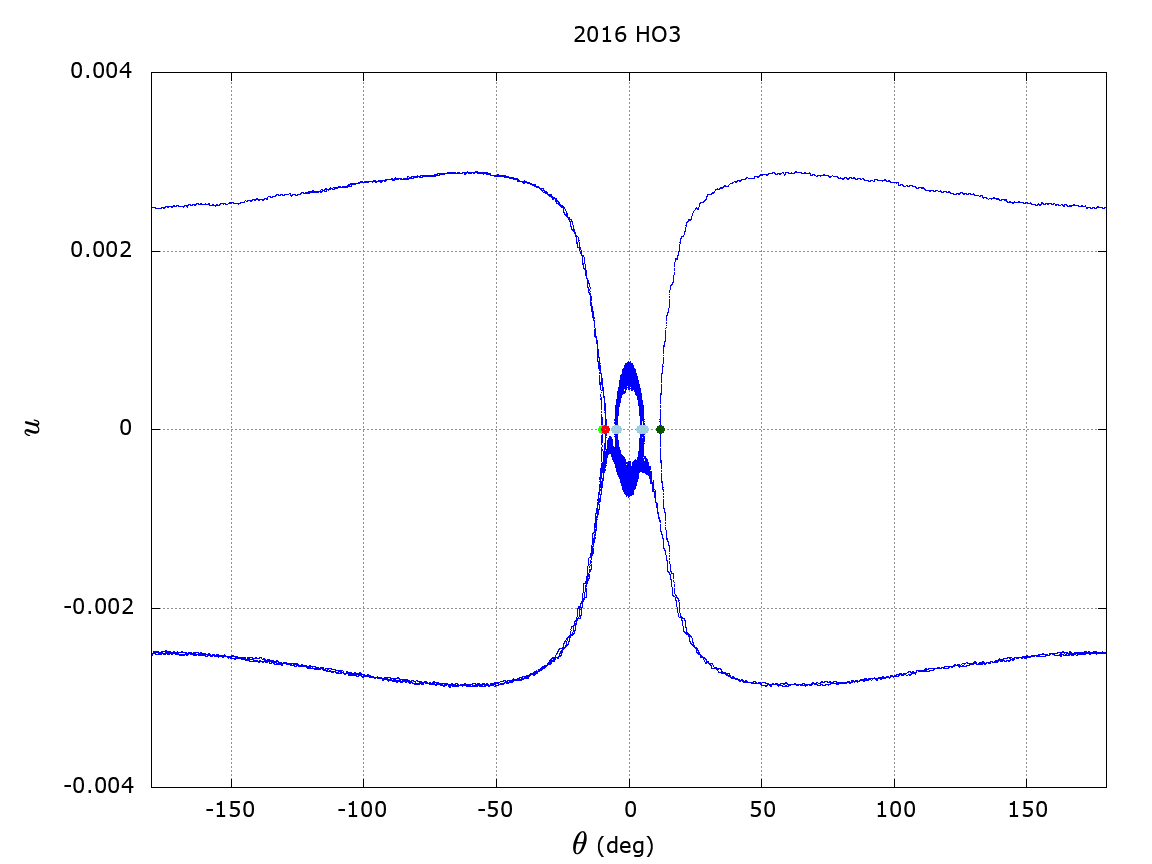}
\label{fig:terra_comc}
\end{subfigure}
\hfill
\begin{subfigure}[t]{0.49\textwidth}
\centering
 \quad (d)\\
\includegraphics[width=\textwidth]{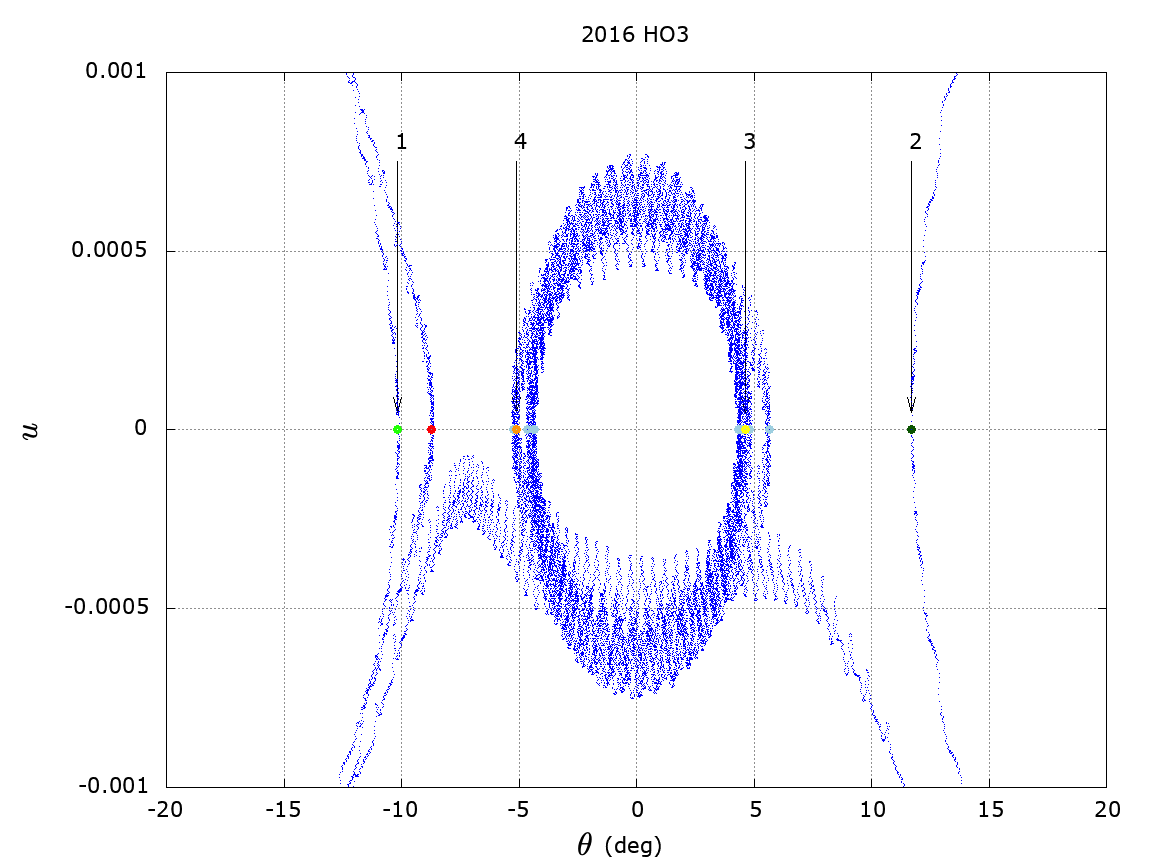}
\label{fig:terra_comd}
\end{subfigure}
\hfill
\begin{subfigure}[t]{0.49\textwidth}
\centering
\,\, \quad (e)\\
\includegraphics[width=\textwidth]{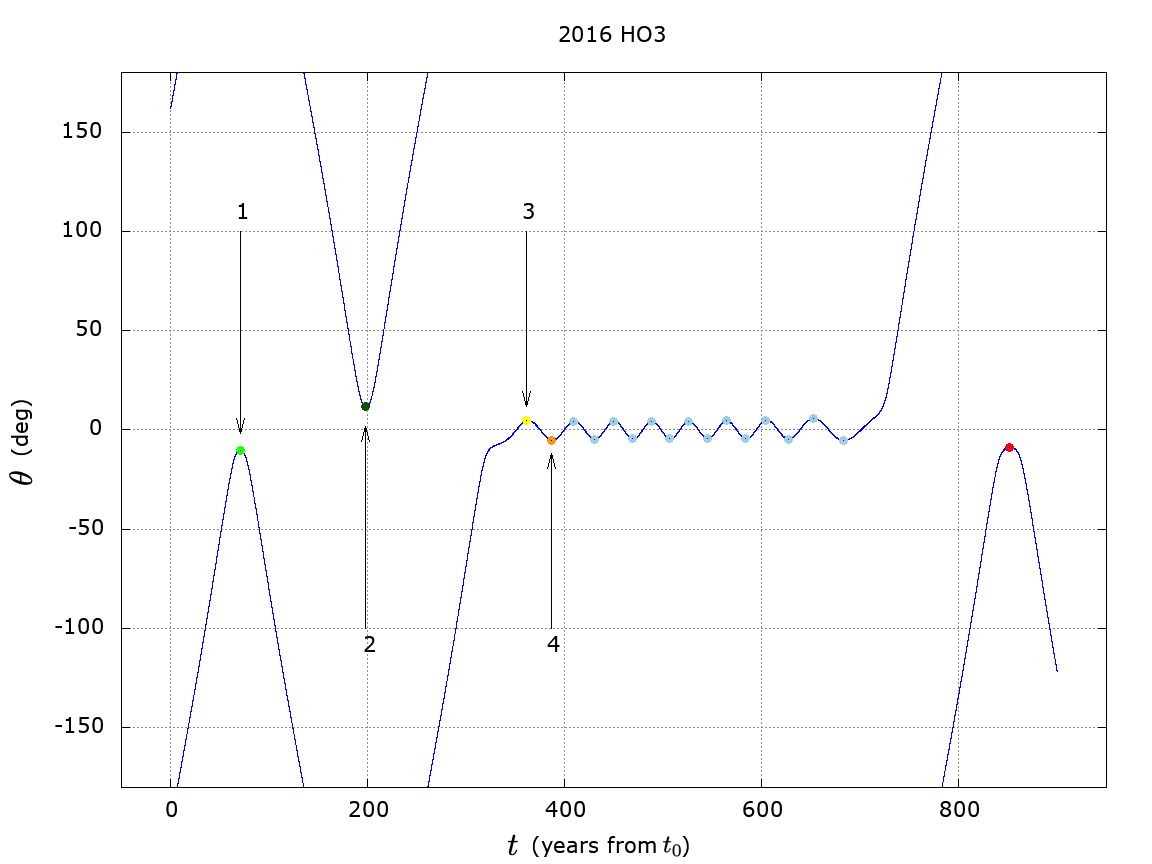}
\label{fig:terra_come}
\end{subfigure}
\hfill
\begin{subfigure}[t]{0.49\textwidth}
\centering
\quad (f)\\
\includegraphics[width=\textwidth]{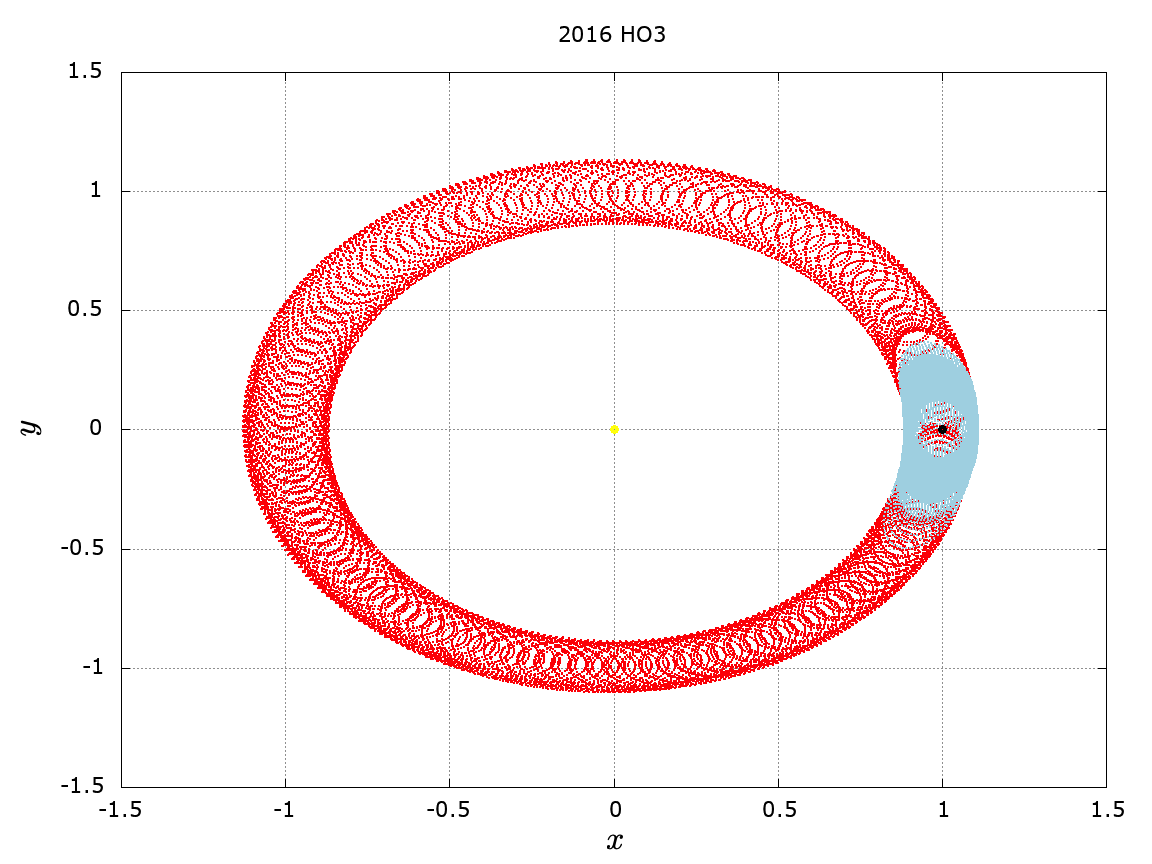}
\label{fig:terra_comf}
\end{subfigure}
\caption{The transient behavior of the asteroid 2016 HO3 with the Earth.        
        Panels (a), (b), (c) show, respectively, 
        the crossings with the section $u=0$ in the  $(\theta,e)$-map, 
        and the projection of the trajectory in the $(t,\theta)$, $(\theta,u)$ planes; 
        the colored points represent the  intersections with $u=0$;    
        panels (b), (d) are enlargements of panels (a), (c), respectively; 
        panel (f) shows the motion of asteroid in the synodic reference frame; the points $(0,0)$ and $(1,0)$ correspond to the Sun and Earth positions, respectively. }
\label{fig:terra_com}
\end{figure*}

\begin{table*}
\caption{The co-orbital objects found for the Sun-Earth system 
    on the quasi-coplanar orbits. 
        The values reported refer to the intersection with $u=0$, 
        that is the closest one to the current date, assumed to be 2021-03-21 00.00.00 
        (JD 2459294.50). 
    The angular values are reported in degrees  and are defined with respect to the orbital plane of the planet.  
            The last column indicates the co-orbital dynamics detected in the given time frame (CP refers to compound, TR to transient).
    These co-orbital configurations are shown also in Fig.~\ref{fig:planet_maps}b.} \label{earth_outcome}
\begin{tabular*}{\tblwidth}{@{}LL@{}LL@{}LL@{}LL@{}LL@{}LL@{}LL@{}}
\toprule
asteroid    & $t$ (JD)  & $a$ (au)  & $e$   & $I$   & $\theta$ & dynamics\\
\midrule
    2000 PH5    & 2452475.70    &    1.00000011  &  0.230  &   1.95  &  -27.004  & HS\\
    2000 EE104  &  2506798.05  &    1.00000011 &   0.293 &    5.10 &    37.777   & HS  \\
    2005 UH6    &  2445884.10   &     1.00000015   & 0.632   &  2.66  &  -94.658 & TR   \\ 
    2005 WK4    &  2545271.43    &     1.00000046   & 0.241   &  9.89  &  -28.449  & CP   \\
    2006 FV35   &  2447416.15   &     1.00000011 &   0.378  &   7.11 &    24.537  & QS  \\ 
    2016 HO3    &  2462506.43    &     1.00000011  &  0.102   &  7.81   &  -4.655 & QS+TR   \\ 
    2001 GO2    & 2452001.06    &     1.00000013  &  0.169   &  4.67   & -18.422 & TR   \\ 
    2002 VX91   &  2395963.64   &     1.00000614 &   0.203  &   2.14  &  -24.799 & CP+HS \\  
    2003 YN107  &  2451752.57  &     1.00000011 &   0.038 &    4.29 &     1.690 & HS \\   
    2007 DD     & 2511123.48     &     1.00000031  &  0.115   &  2.95  &  -12.457 & HS    \\
    2008 NP3    &  2407551.36   &     1.00000012 &   0.334 &    1.47  &   36.868 & TR  \\ 
    2009 HE60   &  2454205.02  &     1.00000011 &   0.265  &   1.52  &  -27.494 & TR   \\
    2009 SH2    &  2455111.84    &     1.00000021  &  0.094   &  6.85   &   9.359  & HS+CP   \\
    2010 CK19   &  2455245.57  &     1.00004886  &  0.153  &   2.24  &   18.978 & HS \\  
    2010 VQ98   &  2512982.25   &     1.00000012  &  0.026 &    1.92  &    0.244  & CP \\   
    2012 BK14   &  2595835.56   &     1.00000188 &   0.190 &    1.79 &   -21.859 & HS  \\  
    2013 BS45   &  2456338.41    &     1.00000015  &  0.086  &   0.82  &   11.336 & HS  \\  
    2013 RZ53   &  2456600.14    &     1.00000011  &  0.050 &    1.52  &    3.856 & HS  \\  
    2014 UR     &  2429919.90     &      1.00000084  &  0.023   &  8.20  &   -4.116   & CP+HS  \\
    2015 SO2    &  2457295.21    &      1.00000011 &   0.108  &   9.20  &   12.145 & HS+TR  \\  
    2015 TC25   & 2563222.35   &      1.00000121 &   0.115  &   4.16 &    11.448 & HS+CP  \\  
    2015 XX169  &  2457734.05   &     1.00000014 &   0.184 &    7.69  &   19.869  & HS+CP \\  
    2015 XF261  & 2484520.70   &     1.00000374  &  0.319 &    0.95  &   38.588  & CP+HS \\  
    2015 XC352  &  2366077.11   &     1.00000040  &  0.140 &    4.87 &    15.793 & HS+CP \\   
    2015 YA     &  2457370.22     &      1.00000075 &   0.280   &  1.78  &  -32.988 & HS+TR  \\ 
    2015 YQ1    & 2457381.44   &      1.00000221   & 0.405  &   2.50  &   46.530 & CP+HS  \\ 
    2016 CO246  &  2458132.16  &      1.00000011 &   0.125  &   6.39 &    16.201 & HS+CP  \\  
    2016 JA     &  2473372.23    &      1.00000201  &  0.270  &   0.09  &   29.382 & HS  \\ 
    2017 DR109  &  2457810.31  &      1.00000020  &  0.243  &   2.99  &  -26.587  & HS+CP \\  
    2017 FZ2    &  2457833.43    &      1.00000264 &   0.265 &    1.73  &   32.256 & HS+TR  \\  
    2017 SL16   &  2458746.73   &      1.00000012 &   0.153  &   8.73 &   -18.738 & HS+CP  \\  
    2018 KS     &  2457896.31    &       1.00000045 &   0.199  &   9.14  &   23.949  & CP \\ 
    2018 PK21   &  2487837.81   &      1.00000011 &   0.077   &  1.54   &  -8.633 & HS    \\
    2018 PN22   &  2456718.64   &      1.00000011   & 0.028  &   4.40   &   3.160 & CP \\   
    2018 UC     & 2530606.45     &      1.00000011  &  0.333  &   0.43  &  -38.332  & HS  \\ 
    2019 DJ1    &  2433930.56    &      1.00000011 &   0.118 &    1.70 &   -15.933  & HS+CP \\  
    2019 GF1    &  2437143.40    &      1.00000011 &   0.044  &   1.84  &    6.954 & HS+TR  \\  
    2019 GM1    &  2462423.17    &      1.00000011  &  0.073   &  6.70   &  -9.689 & HS+CP  \\  
    2019 HM     & 2557019.46     &      1.00000047  &  0.004  &   5.86   &  -2.555 & CP  \\  
    2019 SB6    &  2458763.88    &      1.00000017 &   0.267  &   7.21  &   27.634  & CP \\  
    2019 VL5    &  2459616.44    &      1.00000011  &  0.279  &   1.66  &   30.700 & HS  \\  
    2019 XS     & 2428282.15     &      1.00000014  &  0.325   &  4.20  &  113.230  & TR   \\
    2019 XH2 &  2493516.14    &      1.00000020  &  0.150  &   4.10  &  -18.104  & CP \\ 
    2019 YB4 &  2458851.98    &      1.00000018 &   0.194   &  0.47 &    22.618  & HS \\  
    2020 CB &  2501608.08    &      1.00000038 &   0.287  &   3.82 &   -32.810 & TR  \\ 
    2020 FN3 &  2436661.34    &      1.00000091  &  0.209  &   0.94 &    25.719 & CP  \\  
    2020 GE &  2460575.27    &      1.00000551 &   0.048 &    2.32 &    -7.112 & CP  \\ 
    2020 HO5  & 2482334.95    &      1.00000064 &   0.023 &    4.73 &     3.727 & CP  \\  
    2020 KJ4 &  2459000.79   &      1.00028859 &   0.137 &    2.74 &    16.421 & HS+TR  \\ 
    2020 PN1 &  2459453.40    &      1.00000011 &   0.125  &   4.95 &   -14.883  & HS+CP \\  
    \bottomrule
\end{tabular*}
\end{table*}

\begin{table*}
\caption*{(End of Table \ref{earth_outcome})}
\begin{tabular*}{\tblwidth}{@{}LL@{}LL@{}LL@{}LL@{}LL@{}LL@{}LL@{}}
\toprule
asteroid    & $t$ (JD)  & $a$ (au)  & $e$   & $I$   & $\theta$ & dynamics\\
\midrule
    2020 PP1 &  2441019.81   &       1.00000011  &  0.059  &   5.80  &    6.594  & HS+TR   \\
    2021 CC7 &  2609920.84   &       1.00000046 &   0.039  &   7.35 &    -4.577 & CP   \\ 
    2021 EY1 &  2517726.61   &       1.00000014  &  0.456  &   5.97  &   54.569   & CP \\
    2021 GM1 &  2424782.01   &       1.00000256 &   0.043 &    1.53 &     2.700 & CP  \\  
    2021 JE1 &  2471610.72   &       1.00007781  &  0.264  &   0.45  &  -31.182 & CP  \\  
    2021 LF6 &  2459359.91   &       1.00000013 &   0.022  &   2.99  &    1.271 & CP  \\  
    2021 RZ3 &  2563353.39   &       1.00000083 &   0.054  &   2.18  &    7.606 & HS+CP  \\  
    2021 XV  & 2459547.63    &       1.00000567 &   0.323  &   6.79 &   -38.580  & CP+HS  \\ 
\bottomrule
\end{tabular*}
\end{table*}

    Figure \ref {fig:terra_hs} shows the HS dynamics of the asteroid 2019 VL5. 
%
    Differently from the previous cases analysed for Venus, 
    here, a greater regularity in the dynamics is observed: 
    in the considered time span 
    the asteroid moves in the $(t,\theta)$ plane 
        (panel (b)) 
    with regular oscillations with a same amplitude 
    and in panels (a) and (c) 
    the different points referring to the intersection with $u=0$ overlap perfectly. 
%
    The inclination of such asteroid is $1^\circ$ 
    with respect to the Earth's orbital plane, 
    so that the averaged problem in circular-planar fits very well in this case. 
%
    Moreover, the eccentricity of the orbit of this asteroid 
        (about 0.279) 
    is such that its orbit is close to the intersection with the orbit of Venus. 
%
    It can be argued that this configuration locks the regular motion of the asteroid.
%
    Finally, panel (d) shows the motion of asteroid in the synodic reference frame 
    for the whole interval time.

    Figure \ref{fig:terra_com} analyses the motion of the well-known asteroid 2016 HO3, 
    named Kamo'oalewa. 
    At the current date, 
    it turns out to follow a QS dynamics, 
    but, as it can be inferred from the figures, 
    this dynamics is temporary:
    its orbit will switch to the HS regime in the future,
    in the same way it did in the past, going from the HS to the QS regime.
%
    In other words, as in the case of asteroid 2020 QU5 for Venus, 
    the asteroid has a transient behavior.
%
    Observing panel (b), 
    which is an enlargement of the $(\theta,e)$-map represented in panel (a), 
    all the dots corresponding to the intersections with $u=0$
    belong to the QS domain although some of them, 
    namely the ones numbered 1 and 2 and the not-numbered red dots lie in the HS realm. 
%
    The dots numbered 2 and 4 and all the other sky blue dots belong instead to the QS dynamics, following 
    the $(\theta,e)$-map issued from the circular-planar model. 
%
    Also in this case, we conclude that due to the non-negligible inclination of the orbit of this asteroid 
        (about $9^\circ$ with respect to the Earth's orbital plane), 
    the planar model is not entirely satisfactory to describe the dynamics, 
 especially close to the boundary associated with the collision curve,
    and a three-dimensional model is needed to explain this kind of motion. 
%
    In panels (b), (d), (e) the four numbered dots stand for 
    the first four intersections $u=0$ in chronological order. 
    Panel (f) shows the motion of asteroid in the synodic reference frame 
    and in particular the red part represents the HS motion between time $t_0$ 
    and time $t_0 + 317$ years and the sky blue represents the QS motion between time $t_0 + 317$  and time $t_0 + 720$ years.

    In Table\,\ref{earth_outcome},
    the list of the co-orbital objects on quasi-coplanar orbits with the Earth 
    is given. 
    The objects correspond to the data depicted in Fig.~\ref{fig:planet_maps}b.
    The last column of the Table indicates the detected co-orbital dynamics. It refers to the dynamics in the whole time span, except for the cases where the asteroid escapes from the co-orbital regime or if it experiences a too close approach with the Earth and thus the ephemerides are not reliable. In these cases we report the closest dynamics to what is considered as the current date.  
        
    Note that to be able to identify the objects in co-orbital motion with the Earth 
    at a given epoch as a function to their heliocentric orbital elements 
    is of particular relevance both for planetary science and space engineering purposes. 
    Indeed, asteroids that can orbit in the neighborhood of the Earth 
    are not necessarily a treat for the Earth, 
    but they can represent unique opportunities for science and resource retrieval. 
    In this regards, the map in Fig.~\ref{fig:planet_maps}b, 
    can be enriched by adding the information on 
    the minimum distance to the Earth 
\citep{2022PoAl},
    the corresponding Jacobi constant
\citep{2021PoAl},
    the fundamental frequencies of the quasi-periodic approximation of the co-orbital trajectory, 
    or the taxonomy of asteroids (e.g., C-type), if known. 
    The map depicted shows the co-orbital configuration at a date that it is not necessarily 
    in the future, following the approach of this study, 
    but a different choice can be made. 
    In this way, the map can represent a valuable tool in the preliminary phases of the mission analysis.

\begin{table*}
\caption{The co-orbital objects on quasi-coplanar orbits, 
    not in a regular TP regime, 
    found for the Sun-Jupiter system. 
    The values reported refer to the intersection with $u=0$, 
    that is the closest one to the current date, assumed to be 2021-03-21 00.00.00 (JD 2459294.50).
    The angular values are reported in degrees and are defined with respect to the orbital plane of the planet.  
            The last column indicates the co-orbital dynamics detected in the given time frame (CP refers to compound, TR to transient).   
    These co-orbital configurations are shown also in Fig.~\ref{fig:planet_maps}c.} \label{jupiter_outcome}
\begin{tabular*}{\tblwidth}{@{}LL@{}LL@{}LL@{}LL@{}LL@{}LL@{}LL@{}}
\toprule
asteroid & $t$ (JD) & $a$ (au) & $e$ & $I$ & $\theta$ & dynamics\\
\midrule
2005 TS100  &  2472880.96  &    5.20366671  &   0.298  &   13.35  &    34.121 & HS \\  
2014 SE288  &  2333777.10  &    5.20336302  &   0.358  &    8.94  &    26.695 & HS \\   
2006 SA387  &  2436624.78  &    5.20336306  &   0.194  &    2.79  &    29.869 & HS \\   
1998 WK44   &  2448225.35  &    5.20336303  &   0.057  &    5.69  &    24.585 & HS \\   
2000 QV233  &  2453185.37  &   5.20336306  &   0.258  &    4.61  &   -32.581 & HS \\  
2002 GE195  &  2468467.67  &    5.20336514  &   0.191  &    6.53  &   -24.533 & HS \\   
2004 AE9  &  2448352.65  &      5.20336301  &   0.642  &    2.08  &   -45.845 & QS \\    
2007 EV40  &  2435078.29  &     5.20336472  &   0.634  &    3.72  &    78.947 & QS \\  
2009 KE31  &  2447987.01  &     5.20336303  &   0.172  &    6.43  &   -23.593 & HS \\ 
2009 KQ31  &  2462775.62  &     5.20336306  &   0.089  &    3.77  &   -24.875 & HS \\  
2009 SV412  &  2440099.70  &    5.20336340  &   0.416  &   10.85  &   -39.691 & CP \\   
2009 UK120  &  2446482.25  &    5.20336319  &   0.634  &    0.89  &    74.976 & HS \\   
2009 WQ109  &  2476705.06  &    5.20336301  &   0.712  &    2.06  &   -52.277 & QS \\    
2010 ST19  &  2503334.19  &     5.20336330  &   0.110  &    6.18  &    -1.800 & HS \\   
2012 TT139  &  2457889.34  &    5.20336304  &   0.166  &    8.44  &   -30.646 & TR \\  
2013 NF15  &  2486427.97  &     5.20336329  &   0.736  &    1.78  &   -93.401 & TR \\   
2014 EC59  &  2460325.34  &     5.20336315  &   0.163  &    7.28  &    29.940 & TR \\   
2014 EN60  &  2465603.12  &     5.20336317  &   0.175  &    6.88  &    29.454 & HS \\   
2014 EL63  &  2461772.06  &     5.20336301  &   0.144  &   11.39  &    33.633 & TR \\   
2014 ES72  &  2493233.04  &     5.20336308  &   0.244  &    5.94  &   -25.510 & CP \\  
2014 ED74  &  2450750.08  &     5.20336301  &   0.100  &    7.41  &    24.886  & HS \\   
2014 EM82  &  2448567.13  &     5.20336301  &   0.047  &    7.62  &    22.004 & HS \\   
2014 EB85  &  2449302.04  &     5.20336355  &   0.129  &    9.28  &    20.801 & HS \\   
2014 ET119  &  2449327.92  &    5.20336305  &   0.145  &    5.94  &    19.498 & HS \\   
2014 EM120  &  2460943.70  &    5.20336317  &   0.136  &    4.01  &    28.669 & HS+TR \\   
2014 EB132  &  2461602.75  &    5.20336301  &   0.107  &    8.03  &    27.361 & HS \\   
2014 EF133  &  2419283.15  &    5.20336792  &   0.283  &    3.50  &    30.812 & HS \\   
2014 EO135  &  2449156.61  &    5.20336304  &   0.091  &    7.93  &    19.949 & HS \\   
2014 EG141  &  2449712.64  &    5.20336305  &   0.085  &    9.10  &    21.163 & HS \\  
2014 EP147  &  2448562.62  &    5.20336314  &   0.094  &    7.27  &    25.307 & HS \\   
2014 EA155  &  2449429.26  &    5.20336311  &   0.110  &    7.55  &    20.669 & HS \\   
2014 ES157  &  2449443.15  &    5.20336311  &   0.164  &    6.58  &    24.907 & HS \\   
2014 EY166  &  2449528.38  &    5.20336306  &   0.164  &    7.18  &    26.464  & HS\\  
2014 EP169  &  2448561.02  &    5.20336302  &   0.145  &   11.14  &    25.141 & HS \\  
2014 EG177  &  2449864.21  &    5.20336309  &   0.102  &    8.34  &    18.930 & HS \\   
2014 EX180  &  2449411.76  &    5.20336312  &   0.121  &    8.44  &    23.814 & HS \\   
2014 EA187  &  2449243.72  &    5.20336326  &   0.101  &    8.38  &    20.206 & HS \\   
2014 EL209  &  2449705.65  &    5.20336314  &   0.169  &    7.03  &    23.661 & HS \\   
2014 ET220  &  2449629.93  &    5.20336307  &   0.093  &    7.51  &    19.888 & HS \\   
2014 EX220  &  2448756.35  &    5.20336302  &   0.069  &    8.88  &    23.884 & HS \\   
2014 EJ226  &  2447679.67  &    5.20336301  &   0.138  &    9.08  &    28.139 & HS \\   
2014 ED239  &  2450381.92  &    5.20336302  &   0.082  &   12.00  &    19.677 & HS \\   
2014 EY243  &  2459294.68  &    5.21540765  &   0.135  &    8.50  &    33.971 & HS+TR \\   
2014 EK245  &  2448618.35  &    5.20336302  &   0.051  &    7.79  &    22.866 & HS \\   
2014 FP59  &  2462035.57  &     5.20339839  &   0.395  &   11.30  &    47.257 & HS+TR \\  
2014 OY165  &  2494332.56  &    5.20336356  &   0.714  &    4.52  &   -85.671 & CP  \\   
2015 BB555  &  2438741.70  &    5.20336301  &   0.680  &   10.77  &    63.329 & QS \\   
2015 YJ22  &  2462187.93  &     5.20336303  &   0.558  &    4.00  &   -71.883 & HS+TR \\   
2016 CE150  &  2457268.76  &    5.20336314 &   0.379  &    1.13  &    48.996 & HS \\  
2016 GJ51  &  2461599.19  &     5.20336314  &   0.737  &    1.55  &   102.801 & HS \\  
   \bottomrule
\end{tabular*}
\end{table*}

\begin{table*}
\caption*{(End of Table \ref{jupiter_outcome})}
\begin{tabular*}{\tblwidth}{@{}LL@{}LL@{}LL@{}LL@{}LL@{}LL@{}LL@{}}
\toprule
asteroid    & $t$ (JD)  & $a$ (au)  & $e$   & $I$   & $\theta$ & dynamics\\
\midrule
2016 UY220  &  2492599.21  &    5.20336358  &   0.531  &    2.26  &    66.141 & HS \\  
2016 UG221  &  2473905.66  &    5.20339692  &   0.650  &    2.15  &   -73.971 & HS \\   
2016 UE223  &  2440247.96  &    5.20336303  &   0.660  &    3.66  &    87.896 & HS \\  
2017 QO100  &  2383735.18  &    5.20336338  &   0.515  &    8.01  &   -63.341 & TR \\  
2017 WJ30  &  2433862.33  &     5.20337222  &   0.542  &    8.96  &    60.345 & HS+CP \\ 
2018 PZ27  &  2474081.46  &     5.20336304  &   0.657  &    7.78  &    69.511 & QS \\ 
2019 AS35  &  2464610.62  &     5.20336301  &   0.734  &    1.51  &   -85.384  & QS \\  
2019 QU75  &  2442259.38  &     5.20336376  &   0.223  &    8.69  &   -24.977  &  QS \\   
2020 MM5  &  2457519.22  &      5.20336301  &   0.576  &    7.27  &    29.890 & QS \\  
2021 PM66  &  2518743.99  &     5.20339257  &   0.163  &   17.71  &   -22.188 & HS \\ 
\bottomrule
\end{tabular*}
\end{table*}


\subsection{Jupiter}


    As very well known and also evident from Fig.\,\ref{fig:planet_maps}c, 
    Jupiter has a huge number of asteroid orbiting in 1:1 mean-motion resonance with it. 
%
    The peculiarity of the case of Jupiter lies in the {\it regularity} of the motions of such objects. 
%
    In this case the JPL Horizons ephemerides covers a time span of 600 years. 
%
    Four cases are analyzed here: the first three are chosen in order to show such regularity 
    which is not present in the cases of Venus and Earth.

\begin{figure*}
\begin{subfigure}[t]{0.49\textwidth}
\centering
\, \, \quad (a) \\
\includegraphics[width=\textwidth]{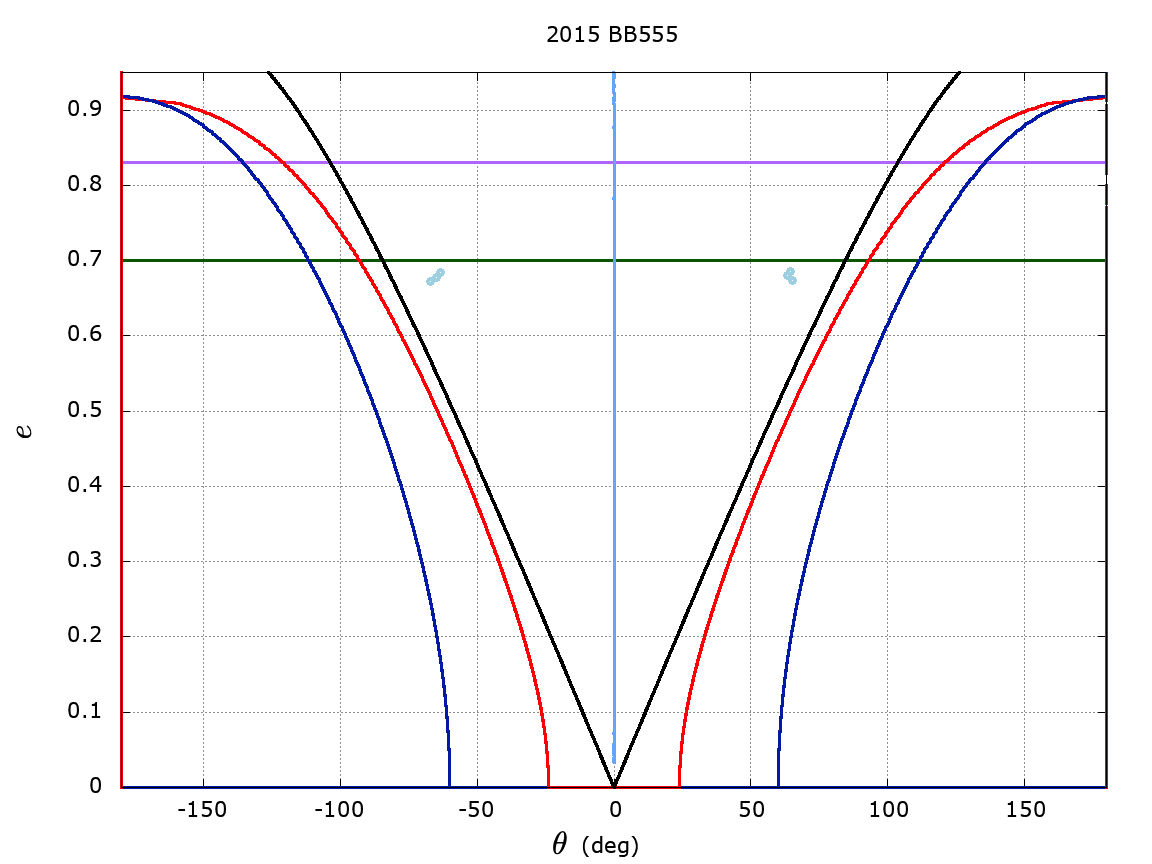}
\label{fig:giove_qsa}
\end{subfigure}
\hfill
\begin{subfigure}[t]{0.49\textwidth}
\centering
 \quad (b) \\
\includegraphics[width=\textwidth]{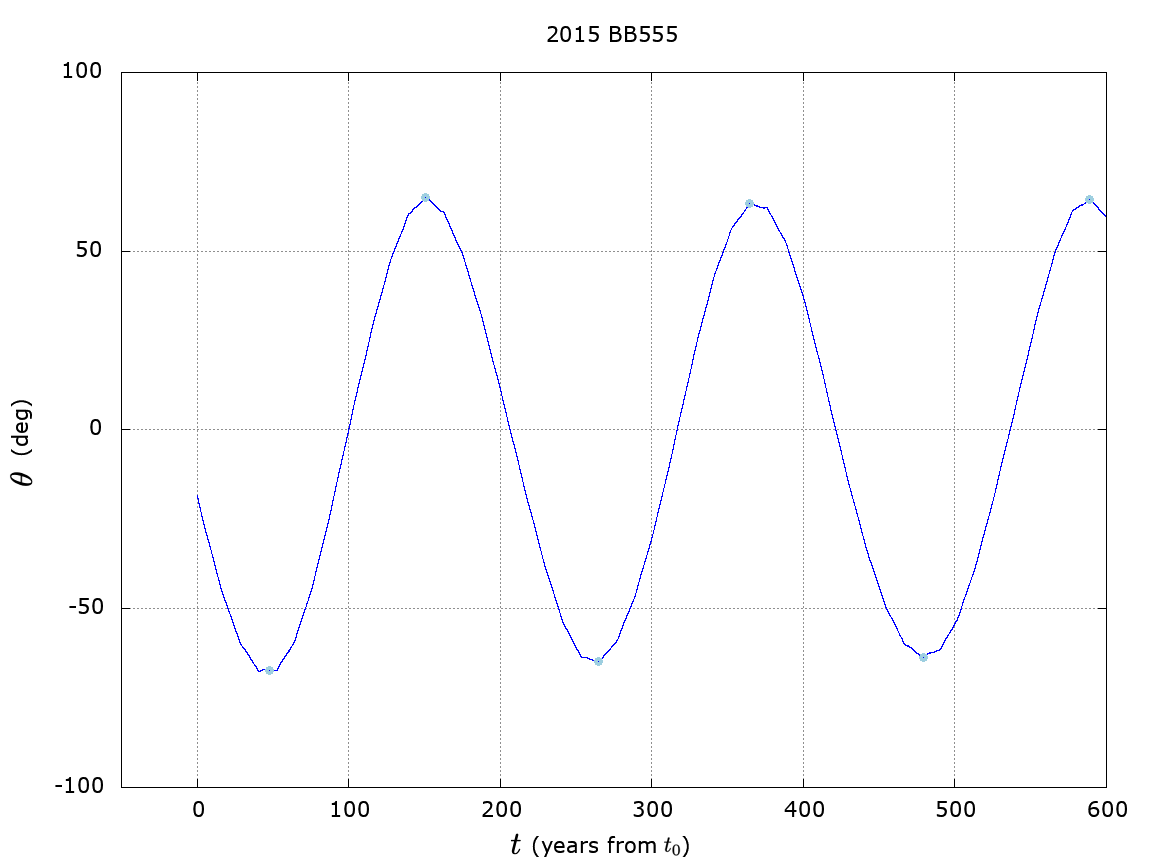}
\label{fig:giove_qsb}
\end{subfigure}
\hfill
\begin{subfigure}[t]{0.49\textwidth}
\centering
\, \, \quad (c) \\
\includegraphics[width=\textwidth]{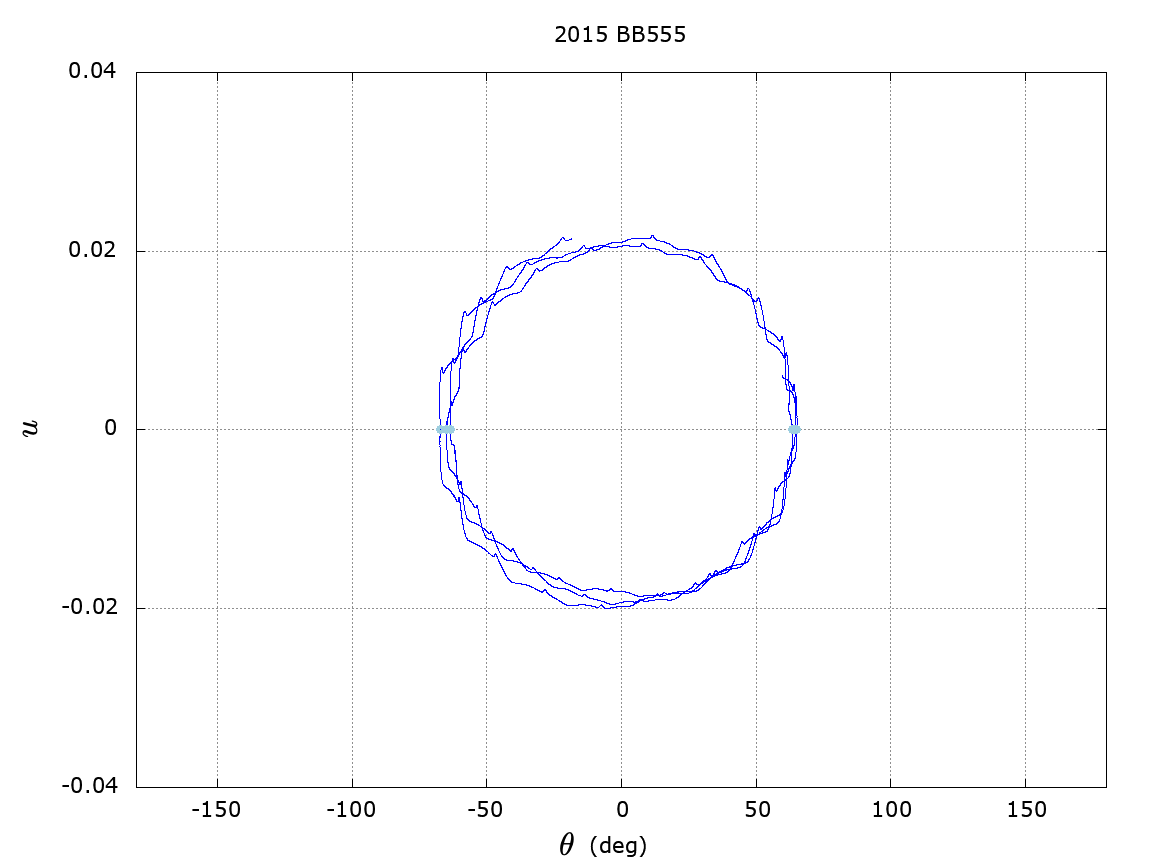}
\label{fig:giove_qsc}
\end{subfigure}
\hfill
\begin{subfigure}[t]{0.49\textwidth}
\centering
 \quad (d) \\
\includegraphics[width=\textwidth]{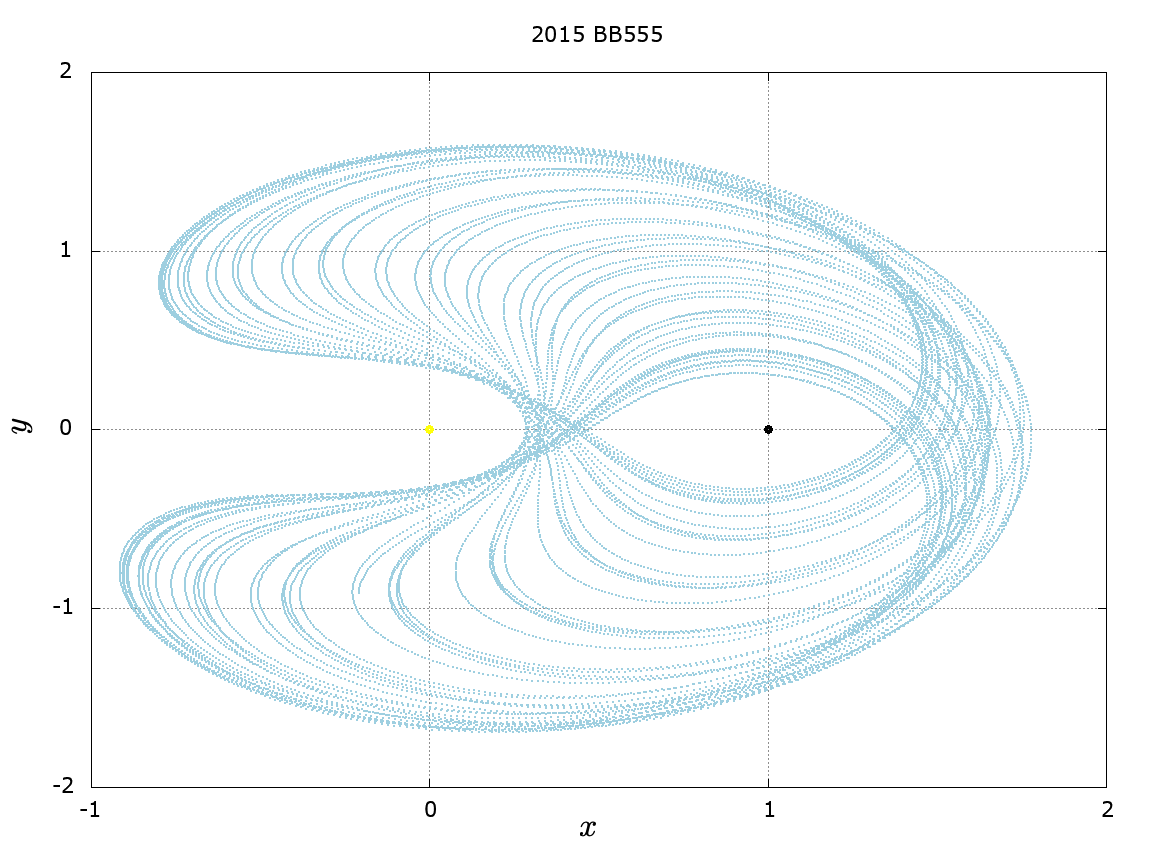}
\label{fig:giove_qsd}
\end{subfigure}
\caption{Asteroid 2015 BB555 
            in QS motion with Jupiter. 
         Panels (a), (b), (c) show, respectively, 
        the crossings with the section $u=0$ in the  $(\theta,e)$-map, 
        and the projection of the trajectory in the $(t,\theta)$, $(\theta,u)$ planes; 
            the sky blue dots represent the  intersections with the plane $u=0$;          
            panel (d) shows the motion of asteroid in the synodic reference frame; the points $(0,0)$ and $(1,0)$ correspond to the Sun and Jupiter positions, respectively.}
\label{fig:giove_qs}
\end{figure*}

    Figure \ref{fig:giove_qs} shows the asteroid 2015 BB555 in a QS dynamics. 
    Although the eccentricity and inclination of the orbit of this body are high, $0.68$ and $10^\circ$ 
    (with respect to the Jupiter orbital plane), respectively,
    panels (a), (b) and (c) show that the amplitudes of oscillations of $\theta$ are regular 
    for the whole interval of time. 
    These figures can be compared with the case of a QS dynamics found for Venus, the one 
    shown in Fig.~\ref{fig:venus_qs} 
    where each oscillation in the $(t,\theta)$ plane is different from the others, 
    due probably to external perturbations 
    (as, for example, close encounters with Earth or Mercury).

\begin{figure*}
\begin{subfigure}[t]{0.49\textwidth}
\centering
\, \, \quad (a) \\
\includegraphics[width=\textwidth]{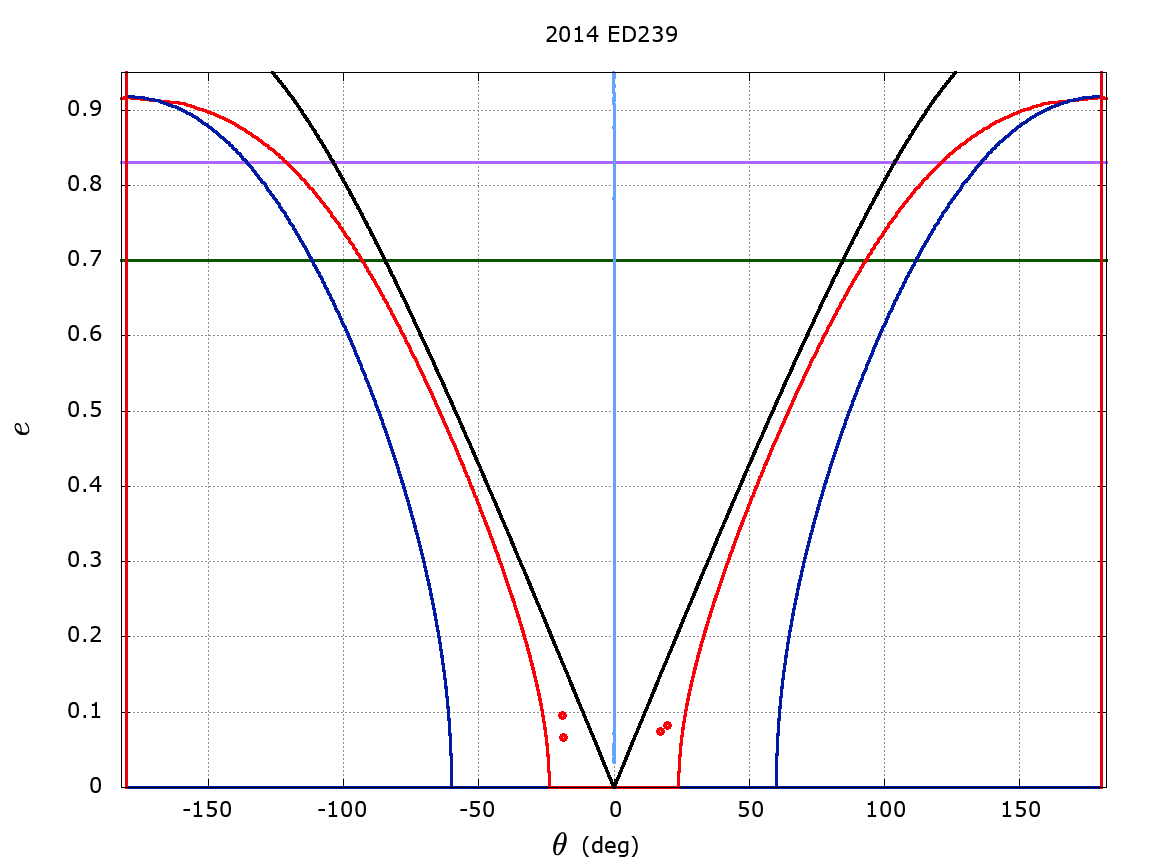}
\label{fig:giove_hsa}
\end{subfigure}
\hfill
\begin{subfigure}[t]{0.49\textwidth}
\centering
\quad (b) \\
\includegraphics[width=\textwidth]{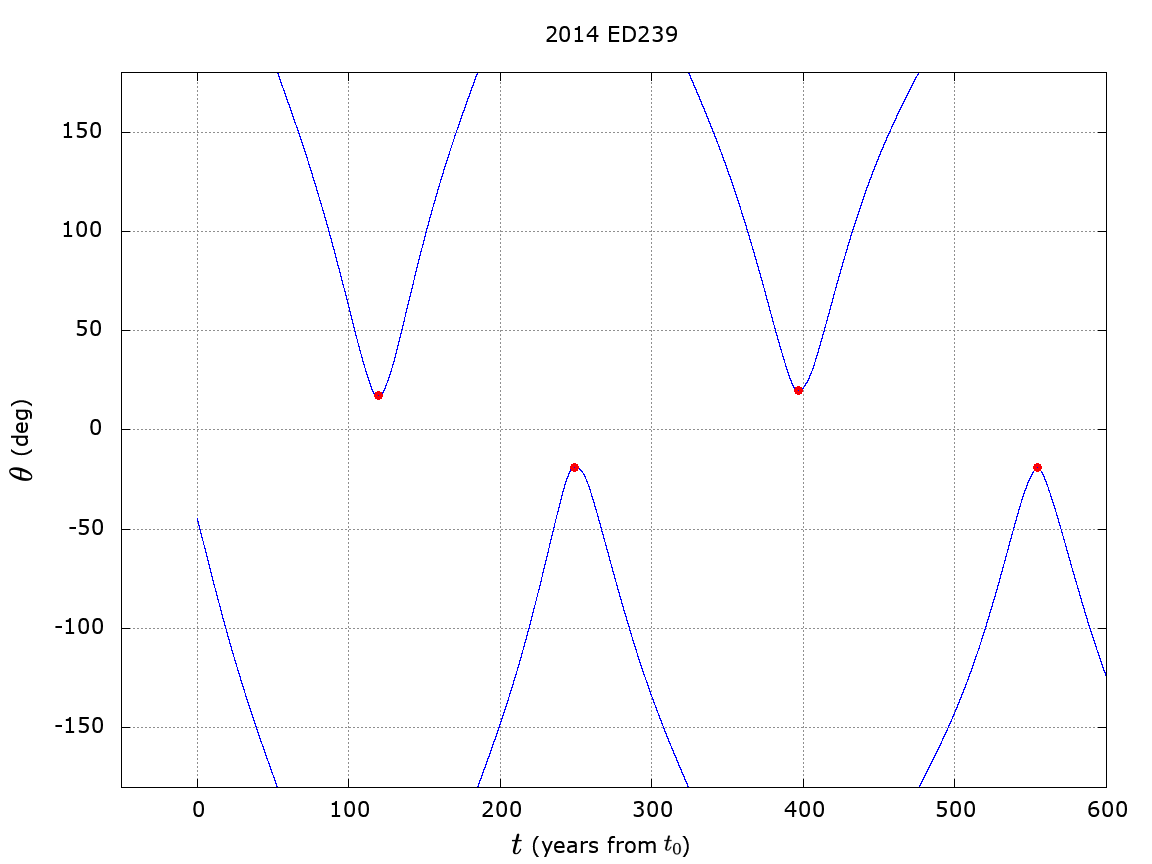}
\label{fig:giove_hsb}
\end{subfigure}
\hfill
\begin{subfigure}[t]{0.49\textwidth}
\centering
\, \, \quad (c) \\
\includegraphics[width=\textwidth]{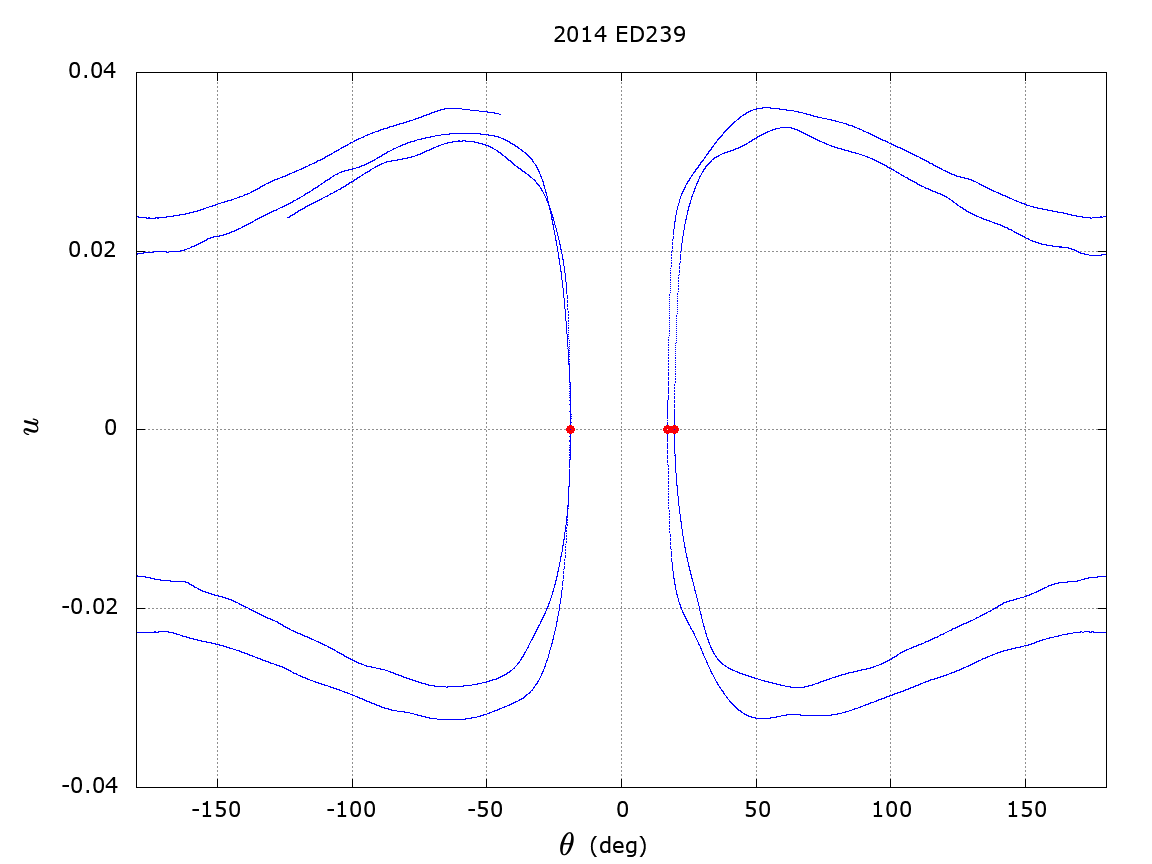}
\label{fig:giove_hsc}
\end{subfigure}
\hfill
\begin{subfigure}[t]{0.49\textwidth}
\centering
\quad (d) \\
\includegraphics[width=\textwidth]{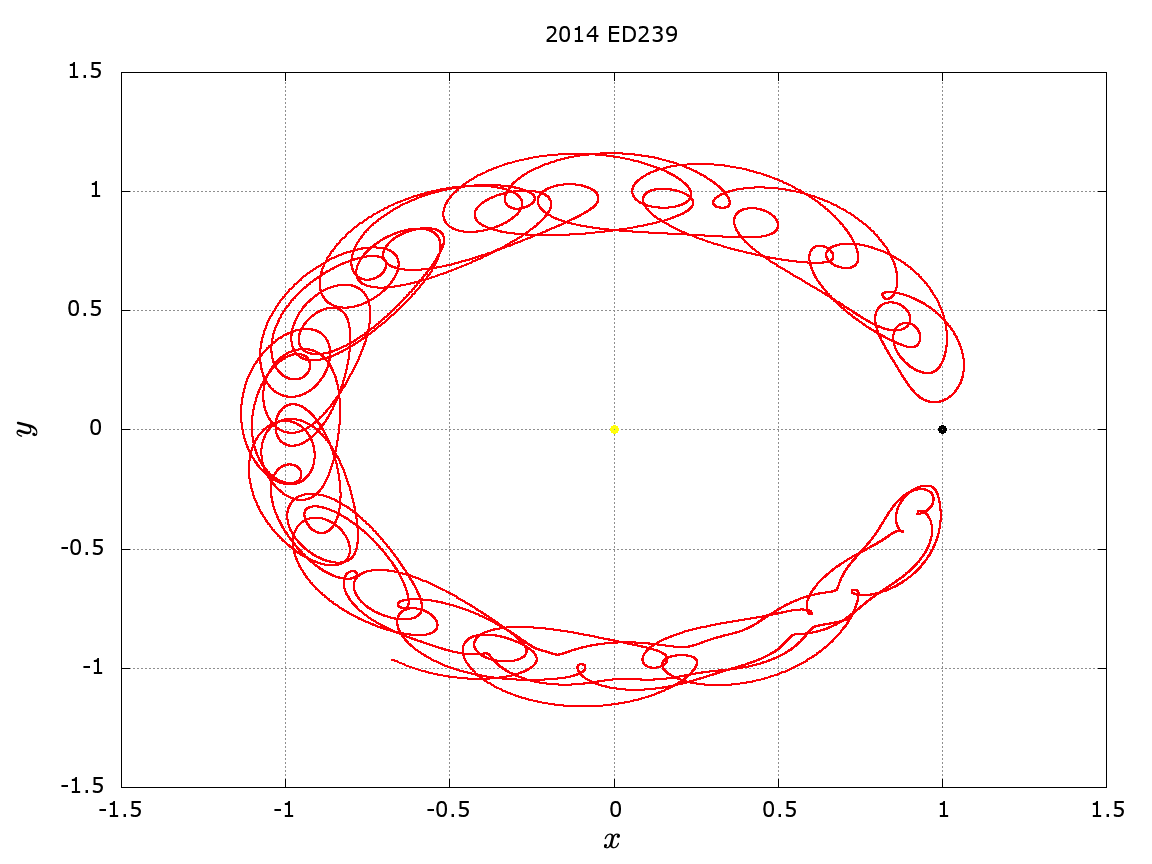}
\label{fig:giove_hsd}
\end{subfigure}
\caption{Asteroid 2014 ED239 in HS motion with Jupiter. 
        Panels (a), (b), (c) show, respectively, 
       the crossings with the section $u=0$ in the  $(\theta,e)$-map, 
        and the projection of the trajectory in the $(t,\theta)$, $(\theta,u)$ planes; 
    the red dots represent the  intersections with the plane $u=0$;  
    panel (d) shows the motion of asteroid in the synodic reference frame; the points $(0,0)$ and $(1,0)$ correspond to the Sun and Jupiter positions, respectively.}
\label{fig:giove_hs}
\end{figure*}
\begin{figure*}
\begin{subfigure}[t]{0.49\textwidth}
\centering
\, \, \quad (a) \\
\includegraphics[width=\textwidth]{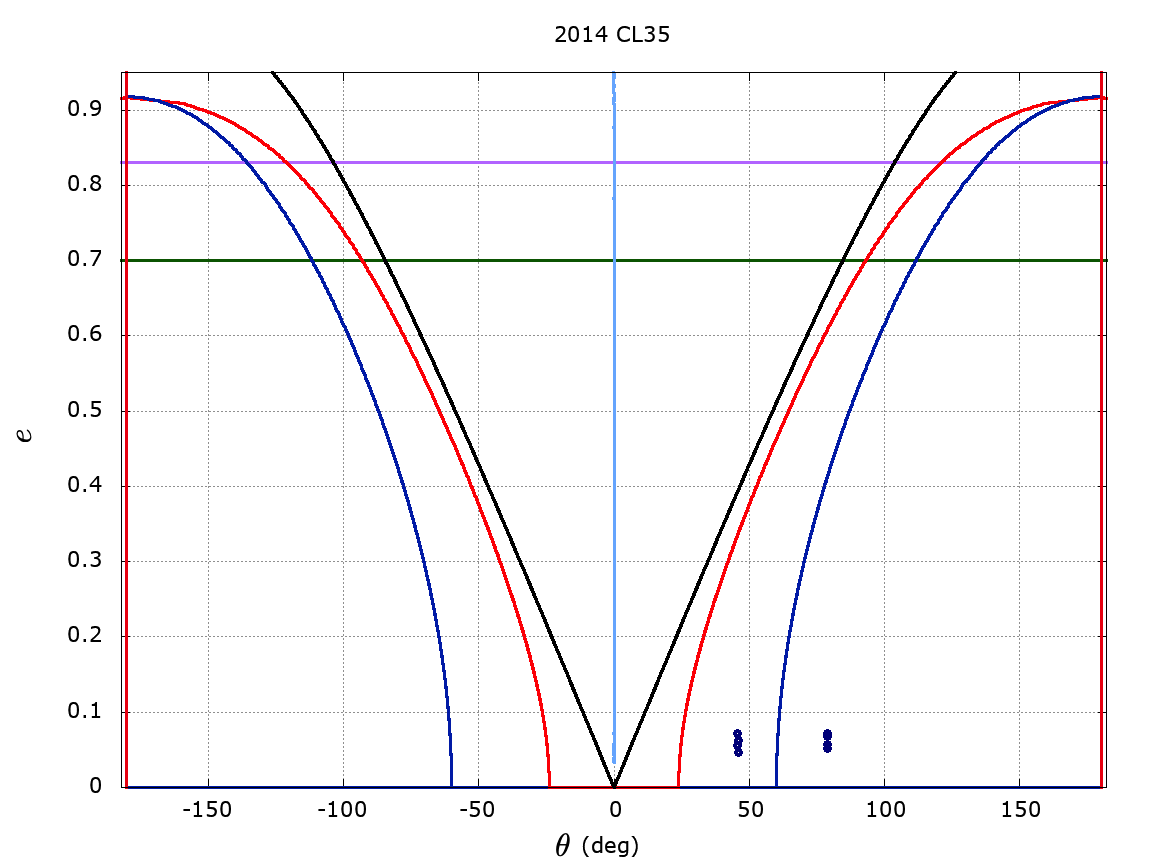}
\label{fig:giove_L4a}
\end{subfigure}
\hfill
\begin{subfigure}[t]{0.49\textwidth}
\centering
 \quad (b) \\
\includegraphics[width=\textwidth]{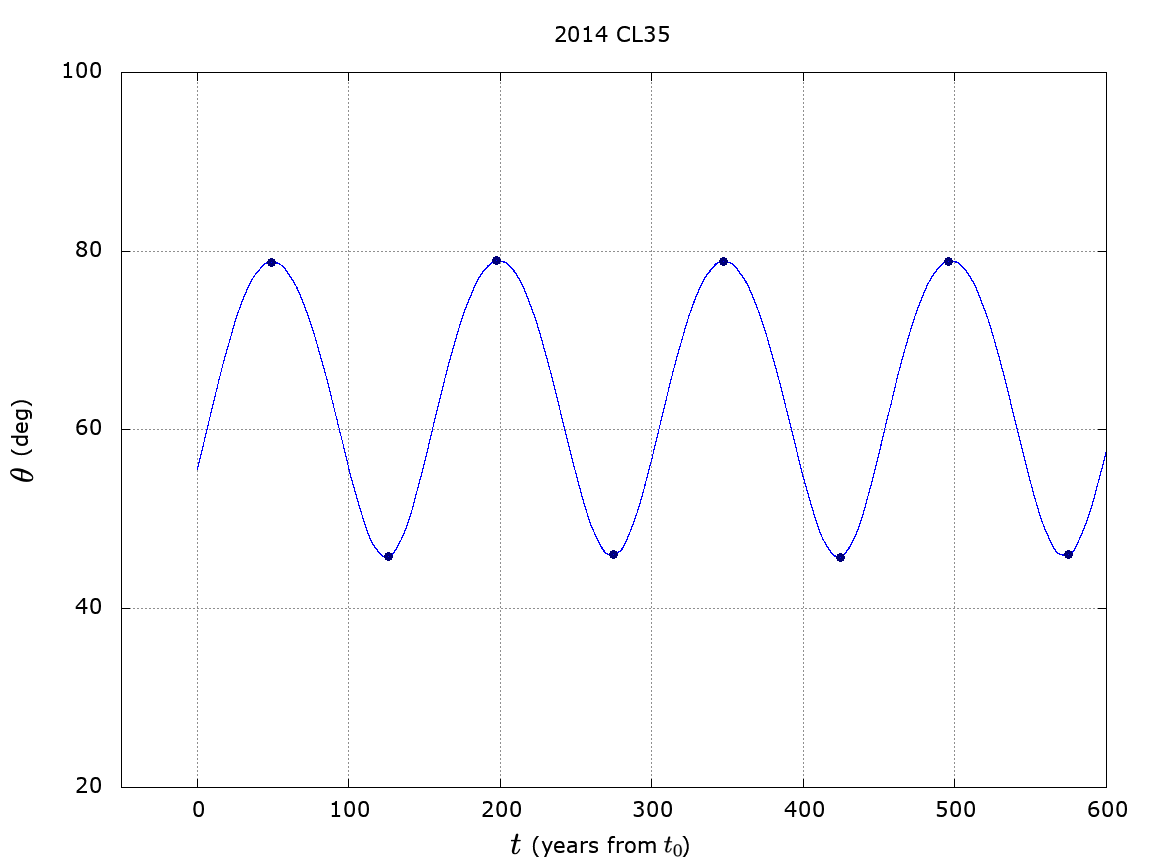}
\label{fig:giove_L4b}
\end{subfigure}
\hfill
\begin{subfigure}[t]{0.49\textwidth}
\centering
\, \, \quad (c) \\
\includegraphics[width=\textwidth]{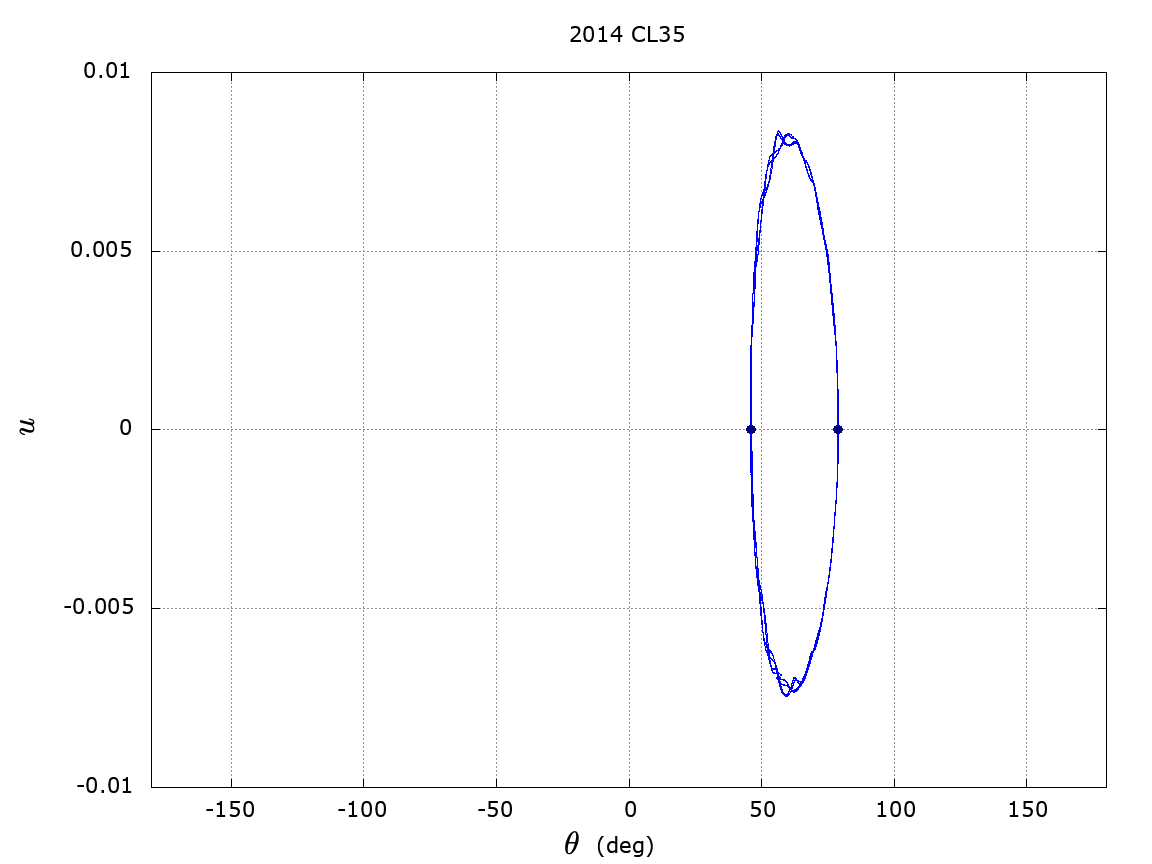}
\label{fig:giove_L4c}
\end{subfigure}
\hfill
\begin{subfigure}[t]{0.49\textwidth}
\centering
\, \, \quad (d) \\
\includegraphics[width=\textwidth]{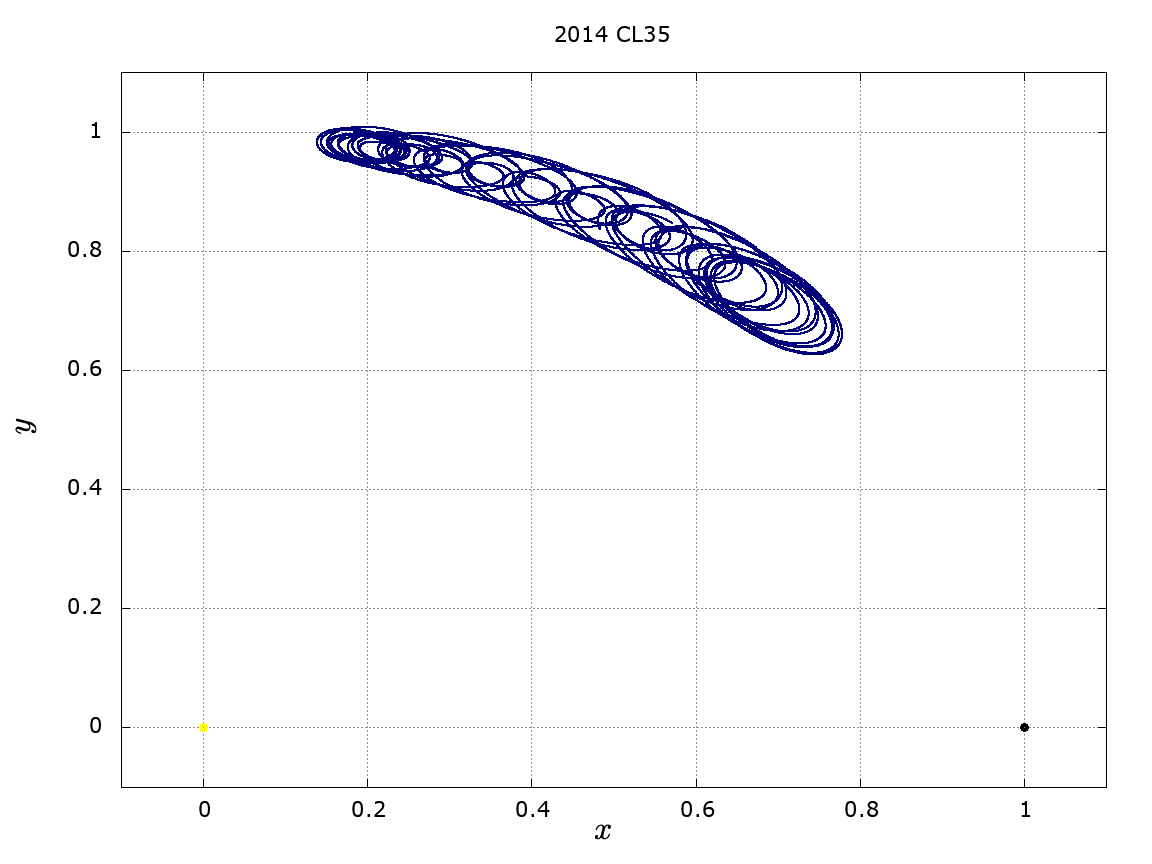}
\label{fig:giove_L4d}
\end{subfigure}
\caption{Asteroid 2014 CL35 in TP motion of $L_4$ with Jupiter. 
         Panels (a), (b), (c) show, respectively, 
        the crossings with the section $u=0$ in the  $(\theta,e)$-map, 
        and the projection of the trajectory in the $(t,\theta)$, $(\theta,u)$ planes; 
    the blue dots represent the  intersections with $u=0$;     
    panel (d) shows the motion of asteroid in the synodic reference frame; the points $(0,0)$ and $(1,0)$ correspond to the Sun and Jupiter positions, respectively.}
\label{fig:giove_L4}
\end{figure*}
\begin{figure*}
\begin{subfigure}[t]{0.49\textwidth}
\centering
\, \, \quad (a) \\
\includegraphics[width=\textwidth]{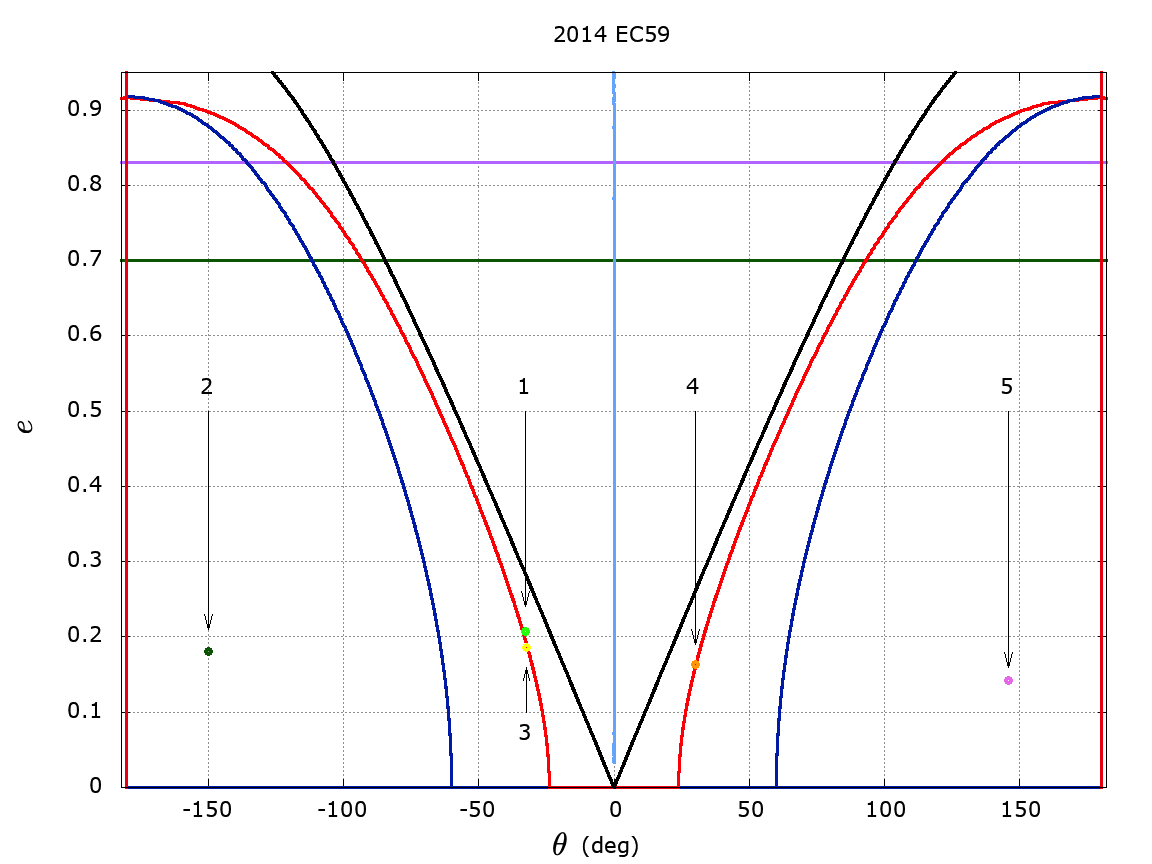}
\label{fig:giove_trana}
\end{subfigure}
\hfill
\begin{subfigure}[t]{0.49\textwidth}
\centering
\quad (b) \\
\includegraphics[width=\textwidth]{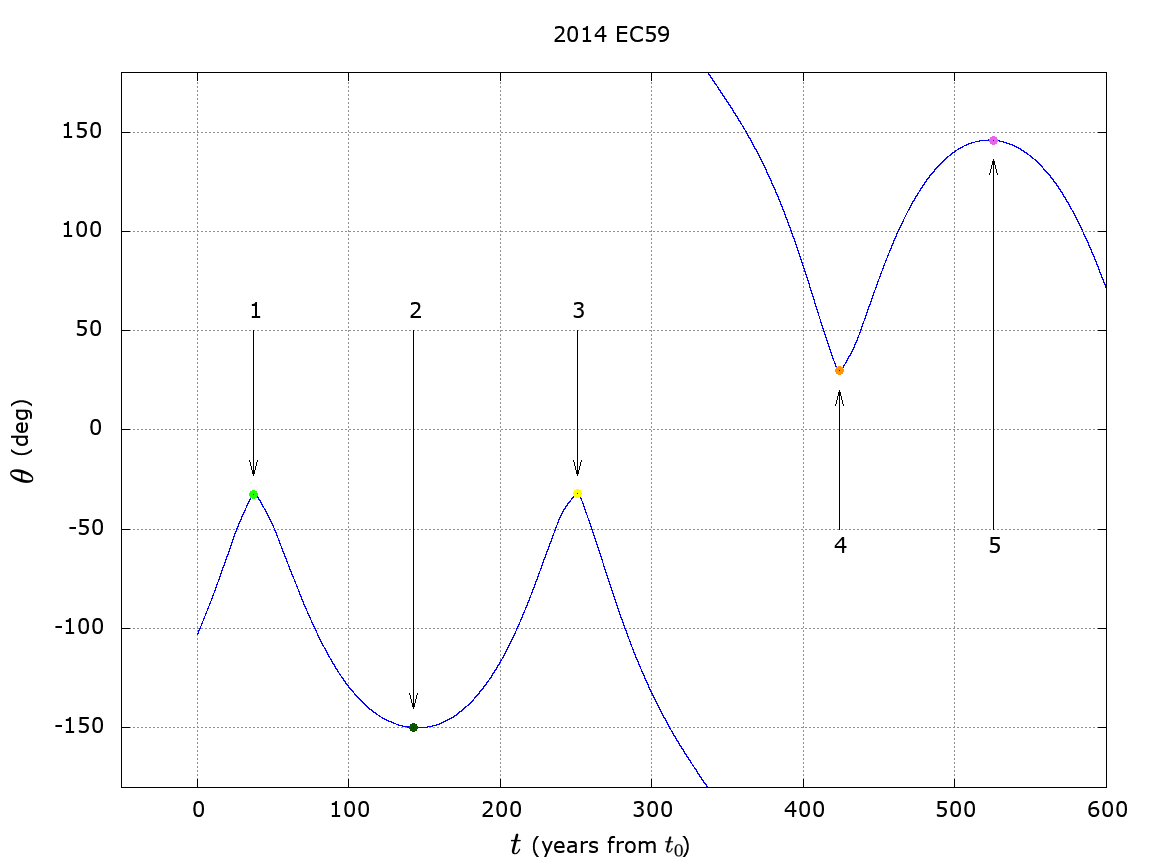}
\label{fig:giove_tranb}
\end{subfigure}
\hfill
\begin{subfigure}[t]{0.49\textwidth}
\centering
\, \, \quad (c) \\
\includegraphics[width=\textwidth]{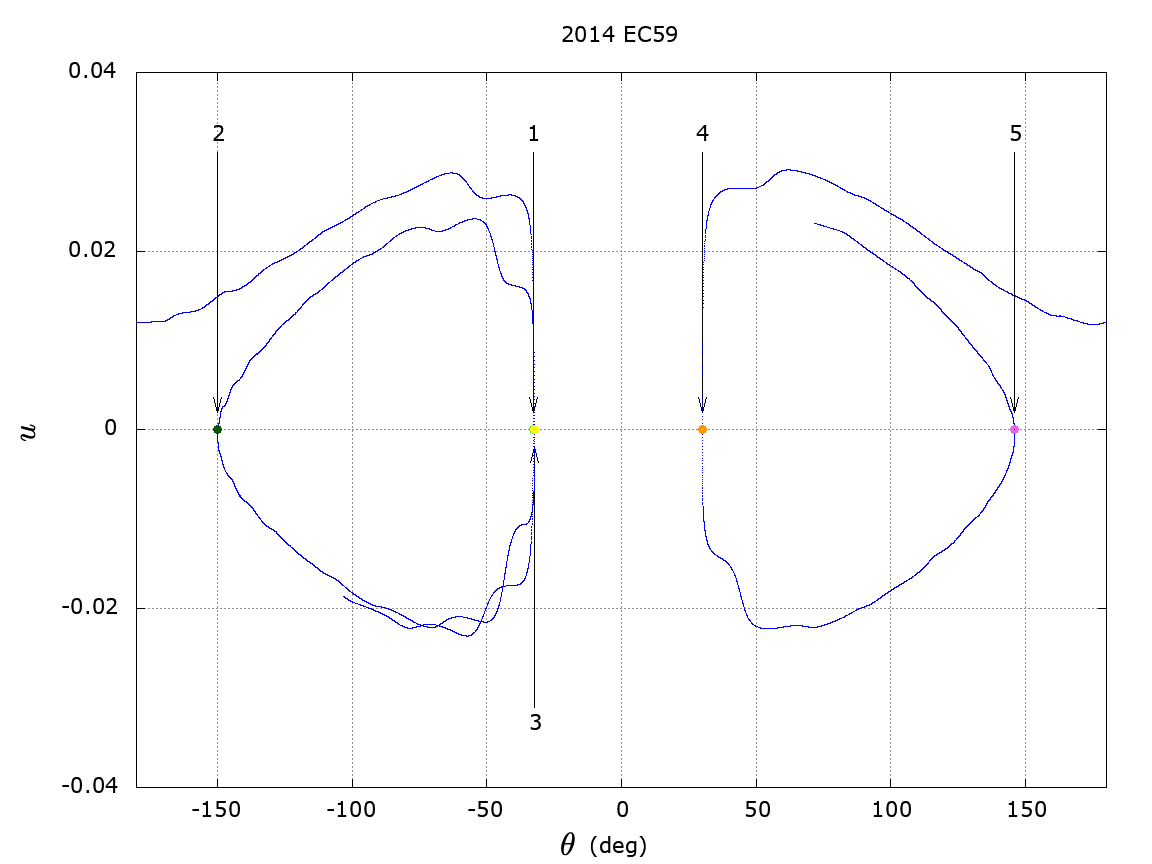}
\label{fig:giove_tranc}
\end{subfigure}
\hfill
\begin{subfigure}[t]{0.49\textwidth}
\centering
\quad (d) \\
\includegraphics[width=\textwidth]{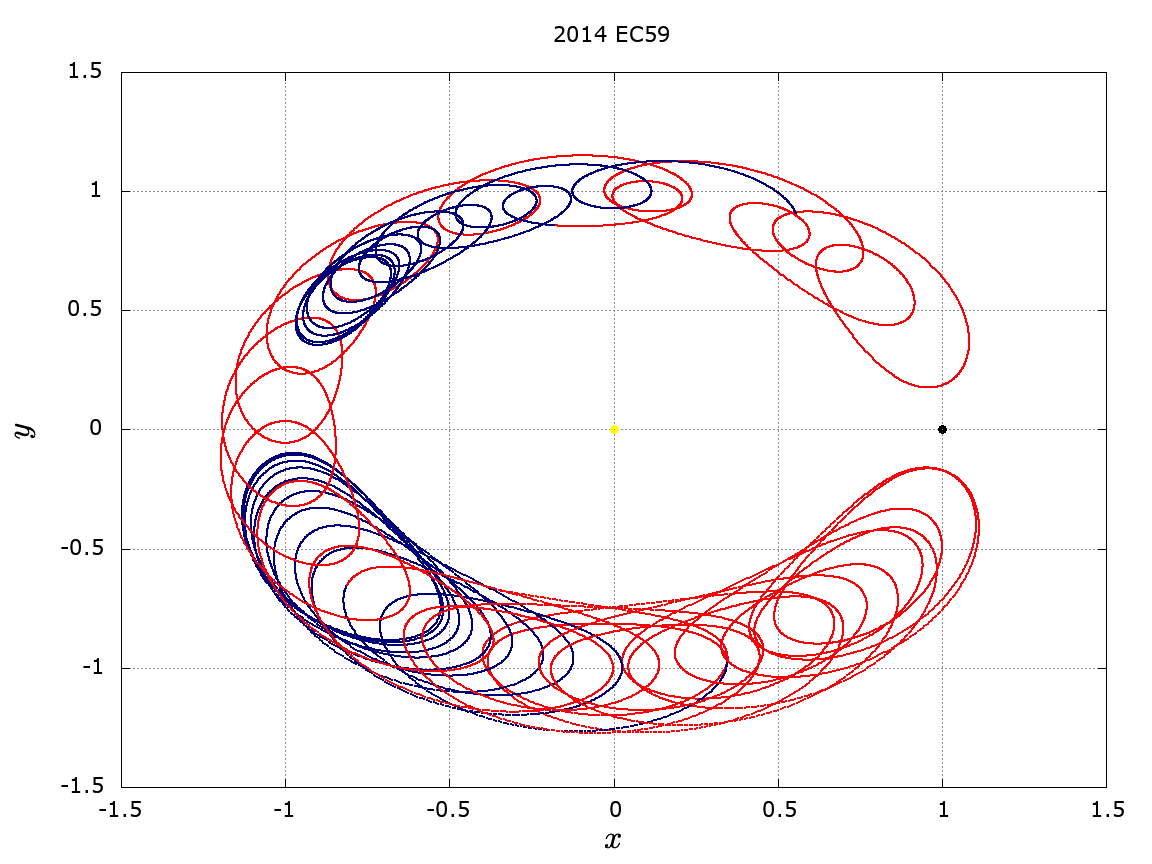}
\label{fig:giove_trand}
\end{subfigure}
\caption{Asteroid 2014 EC59 in a transient dynamics from HS to TP motion with Jupiter.
    Panels (a), (b), (c) represent, respectively, 
    the motion in the $(\theta,e)$, $(t,\theta)$, $(\theta,u)$ planes; 
    the colored dots represent the intersections with $u=0$; 
    panel (d) shows the motion of asteroid in the synodic reference frame; the points $(0,0)$ and $(1,0)$ correspond to the Sun and Jupiter positions, respectively.}
\label{fig:giove_tran}
\end{figure*}

    The same analysis can be done for the asteroid 2014 ED239, 
    as shown in Fig.~\ref{fig:giove_hs}. 
%
    In this case, the orbit follows a HS dynamics, characterized by a small eccentricity 
        (less than 0.08) 
    and a high inclination 
        (about $12^\circ$). 
    In the considered time span the asteroid remains trapped in its dynamics. 
    Nevertheless, it is expected that 
    after a certain amount of time, the asteroid will
    switch to another dynamics or escape from the co-orbital resonance.
    Indeed, 
    according to the numerical simulations of 
\cite{2012CuHaHo},
    all the HS orbits become unstable
    if the mass of the third body 
    is at least about one Jupiter mass.

    Even more striking is the case of the Trojan 2014 CL35, shown in Fig.\,\ref{fig:giove_L4}. 
%
    With its quite small eccentricity (about 0.05) 
    and quite high inclination (about 7$^\circ$), 
    this asteroid looks like to be in a perfect TP motion associated with $L_4$. 
    In panel (c) the intersection points with $u=0$ coincide perfectly.
    As panel (a) shows, only the eccentricity at the crossing experiences small variations,
    that are probably due the eccentricity-inclination exchanges
    generated by the Lidov-Kozai-type first integral 
    that
    exists in the framework of the averaged problem of the 1:1 mean-motion resonance 
    in circular-spatial case 
    \citep[see, e.g.,][]{2022PoAl}.

    The last case of Jupiter has some other peculiarities. 
%
    In Fig.~\ref{fig:giove_tran}, the orbit of the asteroid 2014 EC59 performs 
    many transitions in the considered time span: 
    it begins to be in an HS dynamics and then moves into a  TP orbit associated with $L_5$, 
    then moves back to a HS dynamics 
    and again into a TP orbit associated with $L_4$. 
%
    Although the orbit is characterized by a quite high inclination (more than $7^\circ$) 
    and a moderate eccentricity (about $0.2$),
    the $(\theta,e)$-map shown in panel (a) 
        fits quite well with the model. 
%
    The numbers indicate the chronological order of the $u=0$ intersections 
    and it is possible to notice that number 1 is in the HS domain, 
    number 2 is in the $L_5$ TP domain, 
    again number 3 and 4 are in the HS domain and, 
    finally, number 5 is in the $L_4$ TP domain.
%
    The same is well represented with the same numbers in panels (b) and (c). 
%
    Panel (d) is split in four time intervals: 
        the first is from $t_0 = 0$ to $t_0 + 75$ years 
        and the third is from $t_0 +200 $ years to $t_0 + 460$ 
        and they are plotted in red representing the HS motion; 
        the second is from $t_0 +75$ years to $t_0 + 200$ years
        and the fourth is from $t_0 +460 $ years to $t_0 + 600$ 
        and they are plotted in blue representing  TP motion of $L_5$ the former 
        and  
        TP motion of $L_4$ the latter.

    In Table~\ref{jupiter_outcome}, 
    the objects in the Sun-Jupiter system 
    that in the given time span follow a co-orbital motion, 
    that is not a regular TP orbit, are listed. 
    In other words, only the asteroids in QS, HS, compound or transition regime 
    (under the quasi-coplanar approximation) 
    are given.

    Finally, the list of the objects, 
    that following the algorithm explained in Section~\ref{sec:JPL} can be considered Trojans, 
    have been compared with the ones given by the \citet{MPC}, 
    in the considered inclination range. 
    It turns out that the objects 2000 QV233, 2009 WF218, 2014 ES72, 2014 EY243 are co-orbitals according to the analysis presented here but not properly in a TP motion, 
    while the MPC considers them as Trojans. 
%
    On the other hand, the MPC's list does not contain the following asteroids, 
    that according to our analysis are Trojans, 
        2008 YL115, 
        2014 ES70, 
        2014 EL76, 
        2014 EJ135, 
        2014 EJ169, 
        2014 EL244, 
        2014 RV60, 
        2015 HF178, 
        2015 YB2, 
        2016 UP208, 
        2016 UN244, 
        2017 SF121. 
%
    Notice that among these objects, 
        2014 ES70, 
        2014 EL76, 
        2014 EJ135, 
        2014 EJ169, 
        2014 EL244 are instead listed as Trojans by the \citet{NASASDB}. These discrepancies will be addressed in the future.


\section{Conclusions and future directions}\label{sec:concl}



    This work has focused on the medium-term ephemerides of the asteroids
    computed by the JPL Horizons system  
    to look for those objects that move on a co-orbital configuration. 
%
    The  time frame constraints of the given ephemerides,
    as well as the expected period of the resonant behavior,
    led to limit the investigation 
    to the Sun-Venus, 
    Sun-Earth and 
    Sun-Jupiter systems.
    
    The analysis of the real data has aimed at a possible application of
        the integrable model given by the averaged problem of the RTBP in circular-planar case,
        for which the different types of co-orbital motion
        are neatly defined. 
%
    In this framework,
        the type of orbital regime
        can be identified by 
        a simple bi-dimensional map 
        that has the remarkable feature
        to be independent on the mass ratio of the system, 
        as long as this can be considered as a small parameter.
%
    Being this condition fulfilled
        for all Sun-planet systems of the solar system,
    once the tool was computed
        (i.e., crossings of the separatrices, collision curve and equilibrium points)
        then it has been applied equally 
        for the co-orbitals of Venus, Earth and Jupiter.

    In order to be close to the limit of validity of the model,
        the data has been restricted in inclination
        with respect to the orbital plane of the given planet,
        to what we called the ``quasi-coplanar" orbits.

 The procedure developed
    has allowed to 
        highlight, 
        represent in a synthetic way,
        and classify
        the co-orbital objects of the solar system
        in terms of their dynamics,
        whether being stable or temporary 
    in the considered time frame.
    According to the
        study of the trajectories
        in the resonant variables
        as well as in the synodic reference frame,
    the results show 
    that under some assumptions of low inclination
    with respect to the Sun-planet reference system 
    the bounds of each domain of motion defined by the map
    fit very well with the observed behavior. 
  
    An improvement of the condition of quasi-coplanarity
        is necessary in order to apply the method
        in a systematic way.
%
    The compound dynamics,
        that does not exist in the framework of the circular-planar case,
    could be a possible future research focus in order 
    to refine the criteria,
    for instance
        through the investigation of the phase space 
        of the averaged problem in the circular-spatial case,
        to show the possible existence of a hyperbolic structure
        that divides the horseshoe, quasi-satellite and compound domains.

    Another advantage of the map
        is that it allows to         follow the transitions
        that are generated by different perturbations, namely, effects that are not included in the averaged approximation of the RTBP and effects that are external to the RTBP itself.
    In this regards, the map
        can be enriched with information on 
        the minimum distance to the considered planet, 
        the crossing with the orbit of another planet,
        the Jacobi constant,
        or the evolution of the fundamental frequencies 
        in the different orbital regimes.
    In other words,
        a cartography of the co-orbital resonance could be drawn
        for each planet of the solar system
        in order to identify phenomena 
        that can generate instabilities, transitions 
        or escapes from the co-orbital resonance.
  
    Finally,
        another long-term research focus 
        is to remove the constraints in inclination
        in order to obtain a systematic method of 
        detection and classification of the co-orbital objects in the solar system.
    To this end,
        a first step would be a complete exploration of the
        phase space of the averaged problem in circular-spatial case,
        (4-dimensional and not integrable),
        and the identification of the invariant structures
        that may help to define
        the dynamics and characterize
        their domains in the phase space.

\section*{Acknowledgements}
The authors are indebted with Tommaso Del Viscio from IMATI-CNR for the support he provided for the data processing and with Federica Spoto from CFA Harvard for the suggestions on the data retrieval.

A.P. and E.M.A. acknowledge the support received by the project entitle ``\textit{co-orbital motion and
three-body regimes in the solar system}",
funded by Fondazione Cariplo through the program ``\textit{Promozione dell'attrattivit\`a e competitivit\`a
dei ricercatori su strumenti
dell'European Research Council -- Sottomisura rafforzamento}".


\printcredits

\bibliographystyle{cas-model2-names}
\newpage
\bibliography{DiRuzza_Pousse_Alessi_2022}



\end{document}